\documentclass[smallextended]{svjour3}       % onecolumn (second format)
\smartqed  % flush right qed marks, e.g. at end of proof

\usepackage{cite}
\usepackage[dvipsnames]{xcolor}
\usepackage{lipsum}
\usepackage{amsmath,amssymb,amsfonts}
\usepackage{algorithmic}
\usepackage{graphicx}
\usepackage[title]{appendix}
\usepackage{booktabs,array}
\usepackage{booktabs}

\usepackage{textcomp}
\usepackage{xcolor}
\usepackage{balance}
\usepackage{tabularx,ragged2e}                            
\usepackage{balance}     
\usepackage{multirow}      
\usepackage{multicol}            
\usepackage{fancyvrb}       
\usepackage{booktabs}
\usepackage{hyperref}
\usepackage{amsmath}
\usepackage{color, colortbl}

\usepackage{framed}
\usepackage{wrapfig}
\include{macro}
\usepackage{pifont}
\usepackage{xspace}
\usepackage{multirow}
\usepackage{xcolor,colortbl}
\usepackage{hhline}
\usepackage{adjustbox}
\usepackage{booktabs} 
\usepackage{tabularx} 
\usepackage{amssymb}
\usepackage{balance}
\usepackage[most]{tcolorbox}
\usepackage{url}
\usepackage{booktabs}
\usepackage{tikz}
\usepackage{float}
\usepackage{mdframed}

\usepackage{noindentafter}
\NoIndentAfterEnv{itemize}
\NoIndentAfterEnv{description}
\NoIndentAfterEnv{enumerate}

\newcommand{\framework}{SDC-Scissor\xspace}
\newcommand{\frameworklong}{\textbf{SDC-Scissor} (\textbf{SDC} co\textbf{S}t-effe\textbf{C}t\textbf{I}ve te\textbf{S}t \textbf{S}elect\textbf{OR})\xspace}

\newcommand{\FIX}{\textsc{FIX}\xspace}
\newcommand{\REACH}{\textsc{REACH}\xspace}

\definecolor{table_colour}{rgb}{0,128,0}

\definecolor{mid}{HTML}{B8DBDB}
\newtcolorbox{resultbox}{colback=mid, arc=0.5mm, top=1mm, bottom=1mm, left=1mm, right=1mm}

\definecolor{white}{HTML}{FFFFFF}
\newtcolorbox{observationbox}{colback=white, arc=0.5mm, top=1mm, bottom=1mm, left=1mm, right=1mm}

\newcounter{HypothesisCounter}
{\end{tcolorbox}}

\newcounter{ObservationCounter}
{\end{tcolorbox}}

\newcounter{ConclusionCounter}
{\end{tcolorbox}}

\usepackage{xcolor}             
\include{macro}  
\usepackage{xspace}
\usepackage{cite}       
\usepackage{graphicx}               
\usepackage{paralist}        
\usepackage{tabularx,ragged2e}          
\usepackage{url}          
\usepackage{balance}          
\usepackage{multirow}         
\usepackage{multicol} 
\usepackage{lipsum}
\usepackage{listings}
\usepackage{booktabs}
\usepackage{xspace}
\usepackage{amsmath}
\usepackage{xcolor}
\usepackage{comment}
\usepackage{colortbl}

\newcommand{\ie}{i.e.,\xspace}
\newcommand{\eg}{e.g.,\xspace}

\newcommand{\etal}{et al.\xspace}

\newcommand{\minor}[1]{\color{black}{#1}\color{black}\xspace}

\newcommand{\major}[1]{\color{black}{#1}\color{black}\xspace}
\newcommand{\minorrevision}[1]{\textcolor{black}{#1}}

\begin{document}

%\title{An Empirical Investigation of Cost-effective Test Selection Strategies for Self-driving Cars in Virtual Environments}

%\title{Empirically Investigating the Cost-effectiveness of Machine Learning-based Test Selection Strategies for Simulation-based Testing of Self-driving Cars Software}
 
 \title{Machine Learning-based Test Selection for Simulation-based Testing of Self-driving Cars Software}
 \titlerunning{ML-based Test Selection for Simulation-based Testing of SDC Software}

%\title{Automated Test Cases Selection for Autonomous Systems in Virtual Environments}

\author{Christian Birchler, Sajad Khatiri, Bill Bosshard, Alessio Gambi, Sebastiano~Panichella}
\authorrunning {C.Birchler, S.Khatiri,B.Bosshard, A.Gambi, S.Panichella}
\institute{
	Christian Birchler and Sebastiano Panichella \at
	Zurich University of Applied Science, Switzerland  \\
	\email{birc@zhaw.ch, panc@zhaw.ch} \\\\
	Sajad Khatiri \at
	Zurich University of Applied Science \& Software Institute - USI, Lugano, Switzerland  \\
	\email{mazr@zhaw.ch, mazras@usi.ch} \\\\
	Bill Bosshard \at
	Meier Planungsdienste GmbH, Switzerland. \\
	\email{bill.bosshard@outlook.com } \\\\
	Alessio Gambi \at
	IMC University of Applied Science Krems, Austria, University of Passau, Germany\\
	\email{alessio.gambi@fh-krems.ac.at}\\\\
% 	    Sebastiano Panichella \at
% 	Zurich University of Applied Science, Switzerland \\
% 	\email{panc@zhaw.ch} \\\\
}

\date{Received: date / Accepted: date}

\maketitle

\begin{abstract}
\minor{
Simulation platforms facilitate the development of emerging Cyber-Physical Systems (CPS) like self-driving cars (SDC) because they are more efficient and less dangerous than field operational test cases. Despite this, thoroughly testing SDCs in simulated environments remains challenging because SDCs must be tested in a sheer amount of long-running test cases.
Past results on software testing optimization have shown that not all the test cases contribute equally to establishing confidence in test subjects' quality and reliability, and the execution of ``safe and uninformative" test cases can be skipped to reduce testing effort.
However, this problem is only partially addressed in the context of SDC simulation platforms.
In this paper, we investigate test selection strategies to increase the cost-effectiveness of simulation-based testing in the context of SDCs. We propose an approach called \frameworklong that leverages Machine Learning (ML) strategies to identify and skip test cases that are unlikely to detect faults in SDCs before executing them.
%
% Specifically, \framework extracts features concerning the characteristics of the test scenarios being executed in the simulation environment and via ML strategies predict test cases that lead to faults before executing them. 
%Our evaluation shows that \framework achieved high classification accuracy 
%%(up to 93.4\%)
%\minorrevision{(up to 96\%)}
%in classifying test cases leading to a fault and improved testing cost-effectiveness.

\minorrevision{
Our evaluation shows that \framework outperforms the baselines.
With the Logistic model, we achieve an accuracy of 70\%, a precision of 65\%, and a recall of 80\% in selecting tests leading to a fault  and improved testing cost-effectiveness.
}
\minorrevision{
Specifically, \framework avoided the execution of 50\% of \textit{unnecessary} tests as well as
%reduced 
%\seba{@Christian, please check the following number}
%(of ca. 50\%) the time spent in running ``uninformative'' test cases and 
outperformed two  
%(80\% of the total amount of) failure triggering test cases  compared to 
baseline strategies.
}
%(that identified  between 42.6\% and 60\% of them).
% without introducing significant computational overhead in the SDCs testing process. 
% Finally, 
%\alex{Is the following really necessary? Do we evaluate SDC in this context? It does not really seem to fit with the paper... but maybe it is just a matter of wording}
Complementary to existing work, 
we also integrated \framework into the context of an industrial organization in the automotive domain to demonstrate how it can be used 
%to generate signals compatible with their physical protocol, which is critical to SDC development 
in industrial settings. 
}
\end{abstract}


\maketitle

%\IEEEpeerreviewmaketitle  

\section{Introduction}  
\label{sec:intro} 
% Are those commands really necessary
%\newcommand{\rqone}{\textit{To what extent is possible to identify safe and unsafe test cases for SDCs before executing them?}}
%\newcommand{\rqtwo}{Does \framework reduce resources allocated in SDC virtual test cases compared to baseline approaches?} 

Cyber-Physical Systems (CPSs) leverage physical capabilities from hardware components as well as computational and artificial intelligence from software components to operate in complex and dynamic environments, potentially involving humans~\cite{baheti2011cyber}.
\minor{
Specifically, CPSs continuously collect sensor data from the surrounding environment and analyze them to control physical actuators at run-time~\cite{baheti2011cyber,national201721st}.
}

\minor{
CPSs find application in many domains ranging from Robotics and Transportation to Healthcare and are expected to drastically improve the quality of life of citizens and the economy~\cite{S2424862217500129}.
%
% Among various and emerging CPS application domains, 
% For instance, self-driving cars (SDCs), an emerging application of CPS in transportation, are expected to impact our society profoundly. Human errors cause more than 90\% of driving accidents (e.g., driving while under the influence of alcohol, fatigue, and other distractions) \cite{KalraPaddock:2016}; hence, automated driving systems such as SDCs have the potential to reduce such errors and eliminate most accidents.
For instance, self-driving cars (SDCs), an emerging application of CPS in transportation, are expected to impact our society profoundly by drastically reducing human errors that currently cause more than 90\% of driving accidents, improving passenger comfort, and limiting pollution~\cite{KalraPaddock:2016}.
}
%
% However, the recent fatal crashes involving self-driving cars suggest that the advertised large-scale adoption of SDCs appears optimistic and premature~\cite{baheti2011cyber,TheGuardian-2018}.
% One of the main factors limiting the usage of autonomous driving solutions is the lack of adequate testing. Consequently, the risk of releasing SDCs equipped with defective software, which might become erratic and lead to fatal crashes, is still quite high~\cite{TheGuardian-2018}.
\minor{Currently, one of the main factors limiting the widespread usage of SDCs is the lack of adequate testing. Releasing SDCs equipped with defective software poses the risk that they might become erratic, which has already led to some fatal crashes~\cite{baheti2011cyber}\cite{TheGuardian-2018}.}

Testing automation is crucial for ensuring the safety and reliability of software, including the one controlling SDCs~\cite{KalraPaddock:2016,Kim2019}. However, most developers rely on human-written test cases %(at unit and system levels) 
to assess SDCs' behavior. This practice has several limitations and drawbacks: 
\minorrevision{
\begin{inparaenum}[(i)]
%\item limited possibility to repeat tests under the same conditions~\cite{Kim2019};
\item difficulty in testing SDCs in representative and safety-critical scenarios~\cite{TheGuardian-2018,washingtonpost:2019,Ingrand19};
\item difficulty in assessing SDC's behavior in different environments and execution conditions~\cite{KalraPaddock:2016}.
\end{inparaenum}
}
As a consequence, SDC practitioners in the field are facing a fundamental development challenge: \textit{observability, testability, and predictability of the behavior of SDCs are highly limited}~\cite{TheGuardian-2018,washingtonpost:2019,Ingrand19}. Thus, new testing practices and tools are needed to find SDC faults earlier during development and, eventually, support the widespread usage of autonomous driving.

\minor{Simulation environments can potentially address several of the challenges mentioned above~\cite{beamNG,BondiDKPSFDHIJT18,DosovitskiyRCLK17,nvidia_drive} since simulation-based testing is more efficient than and can be as effective as traditional field operational testing~\cite{afzal2020study,DosovitskiyRCLK17}.
Additionally, simulation-based testing results are easier to replicate and can support %and complement well-
established model-in-the-loop (MiL), software-in-the-loop (SiL), and hardware-in-the-loop (HiL) development strategies.
Consequently, an increasingly large number of commercial and open-source simulation environments have been delivered to the market to conduct testing in the autonomous driving domain~\cite{DosovitskiyRCLK17,beamNG} as well as other CPS domains~\cite{Shin2018}.
For such reasons, our work focuses on %leveraging 
simulation-based testing %technologies 
in the context of SDCs.
}

\minor{\subsection{Problem Statement and Research Questions}
% The usage of 
Simulation environments enable automated test generation and execution~\cite{Gambi2019}. However, the potential size of the testing space of simulation environments is, in principle, infinite, which poses several  challenges and questions (What SDC test cases to select to identify faults efficiently? Is it possible to characterize safety-critical SDC tests?) in exercising the SDC behaviors adequately~\cite{3533818,Birchler2022Cost,Gambi2019,abdessalem2018testing}.}
\minorrevision{The time budget and computational resources devoted to testing activities are usually limited, making the identification of faults particularly challenging in the SDC domain since the execution of simulation-based tests is considerably slower compared to other forms of tests (e.g., unit and system tests of traditional software systems).
%~\cite{YohanandhanEMM20,Flores-GarciaKY20}.
%\alex{Do those two references talk about testing time? From the title is not evident that this is the case.}
}

\minorrevision{
%\major{
For instance, testing how an ego-car handles a driving scenario can easily take several minutes~\cite{SBST:2021,3533818,Birchler2022Cost}; in contrast, running a unit or system test of a traditional software system takes some (milli)seconds.
%\alex{running a system tests might take more than milliseconds also in traditional systems...}
It is important to point out that simulation-based testing tests the subject on the system level, which involves all components and not just a unit, and simulates the environment from which the test subject takes its inputs.
%}
}
Therefore, it is paramount that developers test SDCs cost-effectively, for example, by using test suites optimized to reduce testing effort or \major{by improving existing automated test generators' efficiency without affecting their ability to identify faults~\cite{Yoo:2010, DBLP:journals/tse/NucciPZL20,abdessalem2018testing}}.
%using simulations both in nominal operating conditions and corner cases
 % in SDCs~\cite{Yoo:2010, DBLP:journals/tse/NucciPZL20,abdessalem2018testing}}.

\major{In this paper, we investigate %test case selection (TCS) 
techniques to improve the cost-effectiveness of simulation-based testing in the context of SDCs. Specifically, we focus on techniques that employ Machine Learning (ML) models for supporting test case selection (TCS), addressing the following main challenges:
%
%The main challenges we focuses on while designing such ML-based test case selection strategies for SDCs are as follow:
\begin{inparaenum}[(i)]
\item to leverage test case characteristics as well as ad-hoc SDC test case metrics to characterize best \emph{unsafe} (fault revealing) and \emph{safe} (not fault revealing) SDC test cases;
\item to identify suitable ML models that can reliably predict the SDCs' behavior before executing those test cases;
\item to experiment with the usage of such ML strategies to effectively distinguish unsafe test cases from safe ones;
\item to integrate the proposed ML-based approach into the context of an industrial organization in the automotive
domain, thus demonstrating its applicability in industrial settings.
\end{inparaenum}
}

% ALESSIO: I cannot make sense of this. In general, no one can know prior their execution the outcome of tests, no matter the domain or the test subject. Note that one can argue that SDC are easier to test because the SDC are CPS they must obey to some extent to the physical laws, the are more stable and predictable than software.
%The reason why this is a more challenging task compared to traditional systems is that SDC simulation scenarios do not consist of sequences of method calls as in traditional tests~\cite{abdessalem2018testing,Gambi2019}, but by more complex static and dynamic objects (e.g., number of vehicles, road shape, length and width of the road, weather conditions)~\cite{abdessalem2018testing,Gambi2019}. 
%Hence, typical test data available for traditional systems such as test distance based on signal~\cite{arrieta2018multi} and fault-detection capability~\cite{DBLP:journals/jss/ArrietaWSE19} is not known \textit{upfront}, i.e., without executing the actual SDC tests.
We are interested in testing the safety of SDCs; therefore, we deem as relevant those scenarios that expose a fault (e.g., an SDC drives out off the road). We call those scenarios \emph{unsafe}. Consequently, our TCS techniques exploit ML models to classify SDC test cases that are unsafe (i.e., likely to expose a fault) or safe. 

\major{To address the aforementioned challenges, in this paper, we seek to answer the following research questions:}

\begin{itemize}
\major{
\item \textbf{RQ$_1$}: \emph{To what extent is it possible to identify safe and unsafe SDC test cases before executing them?} \\
% The primary goal of \textit{RQ$_1$} is to propose approaches able to identify (or predict) safe and unsafe test cases for SDCs before executing them, while in \textit{RQ$_2$} we focus more on comparing such an approach with baseline strategies.
% Hence, in the context of \textit{RQ$_1$}, we focus on using  driving scenarios \emph{input features}, i.e., test case characteristics that concern the SDC simulation-based tests and can be extracted before their execution. 
% Then, we propose \framework, a framework that leverages the aforementioned features to train ML models that classify test cases as safe or unsafe (see Section \ref{sec:sdc-scissor}, describing our approach). 
% Specifically, to distinguish between safe and unsafe test cases we focus on \textit{lane-keeping} functionalities in which unsafe scenarios cause a self-driving car to depart its lane~\cite{Gambi2019} and investigate features that describe the geometry of a road as a whole (i.e., \textit{road features}). 
% It is important to specify that in our work we focus on the \textit{lane-keeping} functionalities to automatically label safe and unsafe tests, but with \framework is possible to easily integrate and experiment with further safety criteria, relevant for future research.
%Finally, we investigate the accuracy of \framework in classifying safe and unsafe test cases of SDCs.
Answering \textit{RQ$_1$} is important to understand whether, and to what extent, it is possible to classify test cases for SDCs before executing them and by only considering \emph{static input features} (i.e., referred to as \textit{Road Characteristics}).
We investigate the use of ML models for classifying test cases and study their application in the context of \textit{Lane Keeping}, the fundamental requirement in autonomous driving. 
Specifically, in testing lane-keeping systems, unsafe scenarios cause self-driving cars to depart their lane~\cite{Gambi2019,3533818,Birchler2022Cost}, and input features describe the geometry of a road as a whole (i.e., \textit{Road Features}).

\item \textbf{RQ$_2$}: \textit{Does \framework improve the cost-effectiveness of simulation-based testing of SDCs?} \\ 
\textit{RQ$_2$} investigates whether \framework improves the cost-effectiveness of simulation-based testing of SDCs, compared to baseline approaches.
Hence, in the context of \textit{RQ$_2$}, we investigated whether \framework reduces the time dedicated to executing irrelevant (safe) tests without affecting testing effectiveness. %(i.e., its ability to identify unsafe tests) compared to such baselines.
% We study \framework's behavior 
% %in two opposite setups and contextualize our findings by comparing the results against a random baseline approach (i.e., the scenarios are randomly generated, selected, and executed). In the first study, 
% by comparing its results against two baselines approach, with \framework leveraging ML models trained on off-line data (i.e., trained on a large \textit{static dataset}).} This setup lets us evaluate the application of the proposed technique for regression testing. 

\item \textbf{RQ$_3$}: \textit{What is the  actual upper bound on the precision and recall of ML techniques in identifying SDC safe and unsafe test cases when using static SDC features?}  
%\alex{@All: The following requires a bit of love: we never spoke about ad-hoc metrics, maybe features, and never mentioned hyper-parameter tuning strategies. }
In RQ$_1$ and RQ$_2$, we focused on investigating the feasibility and cost-effectiveness of using SDC \textit{Road Characteristics} as features for the problem of classifying SDC test cases before executing them.
In RQ$_3$, we explore a complementary aspect, which is investigating whether there is an actual upper bound on precision and recall of ML techniques in identifying SDC safe and unsafe test cases when using static SDC features (available before executing the tests).
Hence, once we identified the best ML models for classifying safe and unsafe test cases when compared to baseline approaches (in RQ$_1$ and RQ$_2$), we focus on answering RQ$_3$ by (i) designing additional SDC test case features, called \textit{Diversity Metrics} (compared to the previous features used in RQ$_1$ and RQ$_2$ for training the ML models, these metrics  are more complex than just computing simple road characteristics of SDC test cases); and (ii) leveraging hyperparameter tuning strategies to find the optimal configurations of the most promising ML models (as observed in RQ$_1$ and RQ$_2$).
%\seba{@christian, do we generate new test cases as reported below? If not, jus comment the following sentences.}
%To perform such experiments, we generate new test cases by increasing the level of realism of generated simulation, this by including  obstacles in the generate tests. This was done to observe the behavior of the SDCs as well as the ability of \framework in identifying safe and unsafe test cases, in the context of more articulated test cases.
}
\end{itemize}

\major{
%\seba{@Christian, please update in the following paragraph what we used in our study. Include basically background information on BeamNG.tech, and test generators from the SBST tool competition...}
We conducted our investigation using the freely available SDCs simulator BeamNG.tech~\cite{beamNG} (elaborated in Section \ref{sec:background}).
%, and the open-source tools AsFault~\cite{Gambi2019}, and Frenetic~\cite{CastellanoCTKZA21}. 
\minorrevision{We selected BeamNG.tech because it can execute procedurally generated driving scenarios, and it was recently adopted as the reference simulator in the ninth and tenth editions of the Search-Based Software Testing tool competition\footnote{https://sbst21.github.io/tools/}\cite{SBST:2021}.}
%We selected AsFault because it can automatically generate test cases to assess SDCs' lane-keeping and is compatible with BeamNG.tech.
}

%\alex{ALESSIO: The following requires some rephrasing, maybe something like: is \framework beneficial in an industrial setting? Can \framework be effectively applied in practice? Is SDC that generates the CAN signal... isn't this out of its scope (predicting test cases)?}
%\textit{To what extent simulation-based tests generated by \framework can be integrated into the context of an industrial organization in the automotive domain?}  
%previous studies in the state-of-the-art, 
\major{
Complementary to the investigation of the aforementioned research questions, we investigate the extent to which \framework can be integrated into the context of industrial organizations in the automotive domain.
%(\textit{To what extent can \framework be integrated into an industrial setting?}).
% Hence, compared to previous studies in the state-of-the-art, this research question provides to this article a practical focus, with \framework  integrated SDC-Scissor into the context of an industrial organization in the automotive domain, named AICAS \footnote{https://www.aicas.com/wp/}.
Specifically, to perform such an investigation, we generate SDC test cases and assess the ability of \framework to generate signals compatible with the CAN Bus protocol \cite{canbus-history,8923315,GunduM22} used in the AICAS organization (details  about the AICAS company, their protocol, as well as the design and results of our integration study, are provided in Section~\ref{sec:integration}).
}

%\alex{Consider merging together summary of results and paper contributions to save some space}
\major{\subsection{Summary of Results \& Paper contributions} 
%\seba{@Christian, please, update the following part as soon as all results are in}
\major{
%\minorrevision{Our results (of RQ$_1$) show that \framework achieved high accuracy (between 72\% and 96\% in predicting unsafe test cases) and a F$_1$ score up to 70\%}.
\minorrevision{
%Moreover,  
\framework avoided the execution of 50\% of \textit{unnecessary} tests as well as identified more 
%(80\% of the total amount of) 
failure triggering test cases  compared to 
%\seba{@Christian, please check the following sentence (did you integrate the second baseline?)} 
two baseline strategies. 
}
%(that identified  between 42.6\% and 60\% of them).
}

\minorrevision{
SDC-Scissor outperformed the baseline across all test pools; with the Logistic model, we achieved an accuracy of 70\%, a precision of 65\%, and a recall of 80\% (Table~\ref{tab:pre_adaptive_compare}) in selecting unsafe tests.
}

Our assessment of \framework shows that \framework successfully selects test cases independently from the AI engine used or different driving styles, with the Logistic model providing the more stable results.
Our results also show that the knowledge is not transferable from one AI engine to another one, i.e., \framework performed worse when training ML models on data from a specific AI engine and testing on data from a different AI engine.
%Our findings demonstrate that \framework can predict the outcome of SDCs test cases before executing them and reduce testing costs by applying ML-based test case selection strategies. 
%\seba{@Christian, please, update the following part as soon as we know if an additional baseline was considered..}
%Interestingly, \framework does not introduce significant computational overhead in the SDCs testing process compared to baseline approaches (RQ$_2$), which is critical to SDC development in industrial settings. 
However, from the discussion of our results (in RQ$_3$), we also observed that there is an upper bound for the extent to which static SDC features can be used to predict SDC testing outcomes.
Finally, the integration of  \framework into the AICAS use case allowed us to demonstrate that the proposed approach can automate the testing process of such a large automotive company, coping with the need to complement their hardware-based simulation (based on the Can Bus protocol) with simulation-based testing automation.
}
%\textcolor{blue}{SEBA TODO: add some qualitative findings from results section}

\major{
%\subsection{Paper contributions}
%\seba{Check the following part as soon as all changes are in}
The contributions of this paper can be summarized as follows:
\begin{itemize}
	\item \textbf{Selection of SDCs test cases (RQ$_1$)}: We investigated new methods in the area of SDCs for test case selection.
	We first compute SDC features that can be used to characterize safe and unsafe test cases before executing them.	
	Hence, we introduced \framework that leverages ML models to support test case selection for SDCs, to enhance testing cost-effectiveness. 
%	\item \textbf{Offline v.s. Real-time Training} (RQ$_1$): We investigated two opposite setups for SDC test case selection that leverage ML models trained on off-line data (i.e., trained on a large \textit{static dataset}) and real-time data (i.e., dynamically generated tests).
	\item \textbf{\framework's Cost-effectiveness} (RQ$_2$):	
	%\seba{@Christian, please, update the following part as soon as we know if an additional baseline was considered..}
	We compared the proposed approach against two distinct baseline approaches to demonstrate the testing cost-effectiveness of \framework. % via test case selection.
	The first one is a random baseline approach that selects tests randomly.
	The second baseline selects tests based on their road length, which means that test cases with long roads are preferred based on the intuitive assumption that long roads have a higher probability of being unsafe.
	\minorrevision{\item \textbf{Offline v.s. Real-time Training} (RQ$_2$): We investigated two opposite setups for SDC test case selection that leverage ML models trained on offline data (i.e., trained on a large \textit{static dataset}) and real-time data (i.e., dynamically generated tests).}
	\item \textbf{Upper-bound of SDC static features} (RQ$_3$): We empirically investigated whether there is an actual upper-bound on the precision and recall of ML techniques in identifying SDC safe and unsafe test cases when using static SDC features (available before executing the tests).
	\item \textbf{Integration of \framework in an Industrial Use Case} (analysis detailed Section \ref{sec:integration}): We integrated \framework into the development context of the AICAS use case, demonstrating that the proposed tool can automate the testing process of such a large automotive company.
	
\end{itemize}
}

\major{
%\seba{@Christian, we should add here also the link to the link to \framework... and edit the following part.. }
To foster the replicability of our study, we built a large dataset of labeled test cases~\cite{RP2021} that can be used for replicating our results and promoting further research. 
Furthermore, \framework is publicly available on GitHub~\footnote{\url{https://github.com/ChristianBirchler/sdc-scissor}}, which can be used with the data to replicate our results.
}

\major{
\textbf{Paper structure}.
 %\seba{Double-check the following part as soon as all sections are in}
The paper proceeds as follows: Section~\ref{sec:background} provides some background about CPS simulation technologies, regression testing, a discussion of the simulation-based testing (of Lane Keeping) systems used in the context of our study,  a discussion on automated test generation in the context of SDCs, and a summary of the main terminology used in our study. 
Section~\ref{sec:sdc-scissor} presents the  approach proposed in this paper.
Section~\ref{sec:design} describes the empirical study design, while Section~\ref{sec:results} presents its main results.
Section~\ref{sec:integration} provides a brief background on AICAS, the industrial organization involved in our study, details on the Can Bus (i.e., their signal-based protocol), and elaborates on the design and results of \framework's integration within the AICAS organization. 
Section~\ref{sec:discussion} reflects on the results reported in Section~\ref{sec:results} and Section~\ref{sec:integration}, providing complementary insights and providing a discussion on future work for researchers and SDC developers. 
Section~\ref{sec:related} discusses related work, while Section~\ref{sec:threats} discusses the threats that could affect the validity of our results. Finally, Section~\ref{sec:conclusions} concludes the paper and outlines future research directions. 

}

\section{Background}
\label{sec:background}
\major{This section introduces background elements to make this paper self-contained. It presents the main approaches to SDC simulation (\ref{sec:cps-simulation}) and discusses automated testing of Lane Keeping systems (\ref{sec:cps-testing}). Finally, it concludes with a recap of the terminology used in the rest of this paper (\ref{sec:terminology}).
% it describes AICAS, the industrial organization involved in our study, as well as details on the Can Bus, i.e., their signal-based protocol.
}
% The point here is to say: we use soft-body because is more accurate, but it takes more time to execute. So our ML techniques are very much needed.
% The second point is about existing automated/search based test generation tools, which will benefit from such an approach.

\subsection{CPS Simulation Technologies} 
\label{sec:cps-simulation}
\major{Several simulation technologies have been developed to support developers in various stages of the design and validation of CPSs. Those technologies provide various levels of accuracy and realism at different execution costs, i.e., more accurate simulations generally require larger computational power.}
In the domain of self-driving cars, developers resort to abstract simulation models~\cite{10.1145/3239372.3239409,DBLP:journals/tits/SontgesA18,DBLP:conf/ivs/AlthoffKM17}, rigid-body simulations~\cite{8877728,10.1007/978-3-030-60508-7_9}, and soft-body simulations~\cite{Gambi2019Police,RiccioTonella_FSE_2020} among others.

\noindent \major{\emph{Basic simulation models}, like MATLAB and Simulink models as well as abstract driving scenarios~\cite{DBLP:conf/ivs/AlthoffKM17}, have been mainly utilized for model-in-the-loop simulations, benchmarking of trajectory planners, and Hardware/Software co-design.  They implement fundamental abstractions (e.g., signals, motion primitives) but target mostly non-real-time executions and lack photo-realism, which limits their applicability for testing SDC systems.
}

\noindent \major{\emph{Rigid-body simulations} approximate the physics of bodies by modeling entities as undeformable bodies~\cite{abdessalem2018testing}. % or as compositions of a limited number of rigid three-dimensional objects such as boxes, cylinders, and convex meshes 
Rigid-body simulations implement a very coarse approximation of reality and can simulate only basic object motions and rotations.
Consequently, rigid-body simulations cannot simulate realistic and critical scenarios (e.g., car crashes, inertia) accurately, even when they are combined with rendering engines to achieve photo-realistic simulations~\cite{DosovitskiyRCLK17,BondiDKPSFDHIJT18,XuLXXYHMW19}.}
% . Nevertheless, they scale well in the number of simulated entities (e.g., vehicles) and can be effectively combined with powerful rendering engines to achieve photo-realistic simulations required by SIL testing~\cite{DosovitskiyRCLK17,BondiDKPSFDHIJT18,XuLXXYHMW19}.}

\noindent \major{\emph{Soft-body simulations} improve over rigid-body simulations and can simulate a wide range of simulation cases in addition to primitive body motions and rotations.
As stated by Dalboni and Soldati~\cite{Dalboni}, soft-body simulations can simulate body deformations, % fractures, vibrations, 
anisotropic mass distributions, and inertia, which are essential in many CPS domains.}
\major{For SDCs, soft-body simulations are a better fit for simulating safety-critical driving scenarios~\cite{Gambi2019Police} and, like rigid-body simulations, they can be coupled with powerful rendering engines to achieve photo-realism (e.g.,~\cite{beamNG}). 
Consequently, in our work, we leverage soft-body simulations for simulation-based testing of SDCs.}
% Similar to rigid-body simulations, soft-body simulations can however, because soft-body simulations are more accurate than rigid-body simulations, they are also more computationally demanding. Hence, simulating driving scenarios generally takes longer and exacerbates the need to improve SDC testing cost-effectiveness.}  
%and less scalable in the number of simulated bodies. Consequently, soft-body simulations are less suitable for complicated traffic scenarios where the movement of the simulated entities is generally unrestricted (i.e., there are no collisions between the simulated entities). In contrast, soft-body simulations are a better fit for implementing safety-critical scenarios (e.g., car crashes~\cite{DBLP:conf/sigsoft/GambiHF19}) and focused scenarios in which high simulation accuracy, even in extreme situations, matters the most (e.g., simulating an unbalanced load of trucks or driving with a flat tire).}

\major{
\subsection{Simulation-based Testing of Lane Keeping Systems}
\label{sec:cps-testing}
In this paper, we study how \framework can optimize the testing of the software that controls self-driving cars using physically accurate driving simulations. Specifically, we focus on testing Lane Keeping systems (LKS) that implement one of the fundamental features of autonomous driving.
%\alex{Check whether we have already introduced LKS and move the acronym there}

Simulation-based testing requires creating relevant testing scenarios and reifying them into concrete executions~\cite{intelligence:testing}. 
In accordance with current research on automated testing of LKS~\cite{SBST:2021}\cite{SBST2022}, we consider scenarios that take place on a sunny day on single, flat roads surrounded by plain green grass. Consequently, tests take the form of the following driving task: driving without going off the lane from a given starting position, i.e., the beginning of a road, to a target position, i.e., the end of that road.

% Test Definition - Input road + OBB oracle
The roads defining these driving tasks are obtained by interpolating \emph{road points} using cubic-splines 
% on a two-dimensional map 
to obtain a smooth \emph{road spine}, i.e., the road's center line (see Figure~\ref{fig:virtual-roads}). Driving simulators use the road spines to implement the actual driving tasks to execute.
%
 % reports two examples of virtual roads.
% In the figure, white dots represent the road points, triangles and squares represent the starting and target locations, and a yellow line represents the interpolated road spine.
%
\begin{figure}[tp]
    \centering
    \includegraphics[width=0.8\textwidth]{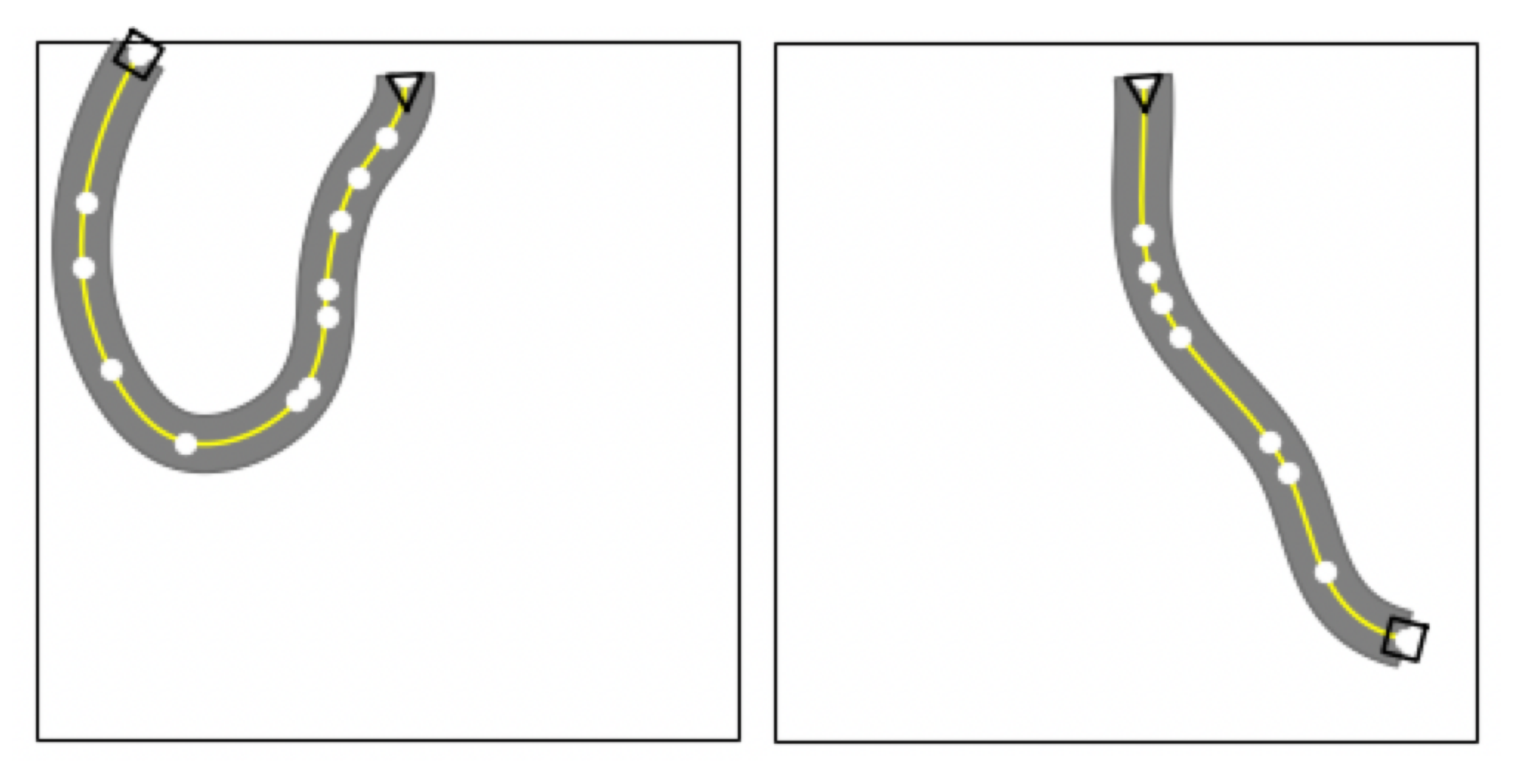}
    \caption{ \major{Virtual roads for testing Lane Keeping systems. The white dots represent the road points, the (central) yellow lines represent the interpolated road spine, the triangles represent the starting locations, and the squares represent the target locations}}
    \label{fig:virtual-roads}
\end{figure}

In this context, unsafe tests correspond to virtual roads that expose problems in the ego-vehicle while driving autonomously on them, for instance,  causing it to drive off-road or invade the opposite lane.
% , or stop in the middle of the road.
%
As discussed in the next Section, \framework extracts a set of features from the road spine and road points that enable it to predict whether the corresponding virtual road will expose a problem in the ego-vehicle before the test execution.

\framework relies on the open-source testing infrastructure developed for the CPS testing competition of the SBST (Search-Based Software Testing) workshop~\cite{SBST:2021}.
This infrastructure can automatically implement executable simulations from the road spines, execute them, and collect their results (e.g., pass/fail).
We opted for this infrastructure for two main reasons:
(1) It utilizes BeamNG.tech~\cite{beamNG} simulator; hence, it can execute physically accurate and photo-realistic driving simulations.
(2) It has already been used to benchmark several automatic test generators (see~\cite{SBST:2021}\cite{SBST2022}); hence, it enables us to study the generality of \framework.
%\seba{@christian, please after this, clarify whether the generators randomly generate test cases without using the GA. It is a comment we need to address from reviewer 1. Adding this information here would fix it.}
\framework uses Frenetic~\cite{DBLP:conf/sbst/CastellanoCTKZA21} as the main test generator, which uses a genetic algorithm for defining road points on a cartesian plane.

The open-source testing infrastructure developed for the CPS testing competition~\cite{SBST:2021} enables \textit{driving agents} to drive simulated vehicles and get programmatic control over running simulations (e.g., pause/resume simulations, move objects around).
We consider two different driving agents as test subjects for our evaluation:
%The first, \textbf{BeamNG.AI}, is the driving agent shipped with the BeamNG.tech, and the second,
\minorrevision{
The first is the driving agent shipped with the BeamNG.tech, which we refer to as \textbf{BeamNG.AI}, and the second,
}
\minorrevision{
%\textbf{Driver.AI}\footnote{\url{https://github.com/alessiogambi/AsFault/blob/asfault-deap/src/asfault/drivers.py}}, is a trajectory planner~\cite{Gambi2019}.
is an open-source trajectory planner, which we refer to as \textbf{Driver.AI}\footnote{\url{https://github.com/alessiogambi/AsFault/blob/asfault-deap/src/asfault/drivers.py}}~\cite{Gambi2019}.
}
As explained by BeamNG.tech developers, a parameter called the “risk factor” (RF) controls the driving style of BeamNG.AI: low RF values (e.g., 0.7) result in smooth driving, whereas high RF values (e.g., 1.2 and above) result in an edgy driving that may lead the ego-car to “cut corners”.
Driver.AI instead analyzes the road geometry and plans the car trajectory by computing for each turn the maximum safe driving speed ($v$) using the standard formula for centripetal force on flat roads with static friction ($\mu$)~\cite{misc:safe-speed}:
 \begin{equation}
  v = \sqrt{\mu \times r \times g}
 \end{equation}
 where $r$ is the turn radius and $g$ is the free-fall acceleration.

Driver.AI relies on the user to provide the value of the friction coefficient, as well as information about the maximum acceleration and deceleration of the ego-car. In our evaluation, we estimated those values empirically following a trial-and-error approach.
It is important to mention that, at the moment, both BeamNG.tech and Driver.AI do not have previous versions of their driving agents. This means that their behavior can only be altered or investigated by experimenting with the parameters already discussed in the context of our study. As a consequence, the target of our regression testing strategy is primarily focused on enabling SDC test selection, with the main goal of reducing the effort required to detect faults. For future work, assuming new versions of both BeamNG.tech and Driver.AI are delivered, we plan to experiment with consecutive versions of these AI agents so that it is possible to investigate the potential fault-detection capability of both of them.
}

\major{
\subsection{Article Terminology}
\label{sec:terminology}

To avoid any confusion in terminology, it is important to note that in the rest of the paper, we will refer to simulation-based test cases generated by \framework as \textbf{test cases}. 
Test cases are composed of virtual roads composed of a sequence of multiple road segments, as exemplified in Figure~\ref{fig:virtual-roads}. Formally, \textbf{road segments} refers to (parametric) portions of roads of test cases; hence, they can be straight segments (no curvature), left turns (positive curvature), or right turns (negative curvature).

We refer to test cases that have been executed and evaluated in simulation as \textbf{executed test} cases. 
Then, if a test is passed successfully, we refer to it as a \textbf{passing test}, and if it failed, potentially revealing some issues with the system under test, we refer to it as a \textbf{failing test}. 

On the other hand, as we elaborate more in the next sections, \framework automatically assigns labels to the test cases regarding them being likely to fail or pass without executing them. 
In this context, we refer to the test cases which are considered by \framework to be likely to pass as \textbf{safe test} cases and the ones that are considered likely to fail as \textbf{unsafe test} cases.

Regarding the features used in \framework, \textbf{static (road) features} refer to any test case features that can be calculated without running any simulations, \ie they are suitable for predicting test results (simulation results) before running simulation. As discussed in detail in the next section, we propose to use two different sets of road features: \textbf{road characteristics} and \textbf{diversity metrics}.

Regarding the experiments to answer RQ$_2$, we will discuss \textbf{offline experiments} that involves test selection from a previously generated (offline) pool of test cases in Section~\ref{sssec:study-round}. 
We conducted the offline experiment in two experimental setups that mimic the issues of having a limited testing budget in the context of SDCs: 1)~\textbf{\FIX}, in which the amount of total test cases that can be executed in the simulation environment is \emph{fixed} to a certain number. 2)~\textbf{\REACH} in which we continue executing the test cases until we \emph{reach} a certain number of failing tests. 

As discussed later in Section \ref{sec:study-real}, we complement RQ$_2$ evaluations with \textbf{real-time experiments}, in which we study the application of \framework to automated test generation, \ie the test pool is being generated in \emph{real-time}, and only the \emph{unsafe} tests are being kept and executed.
There, we have two experimental setups: 1)~with a \textbf{pre-trained} ML model. 2)~with an \textbf{adaptive} ML model that could be retrained with the correct labels of the generated test cases. 
}

\section{The \framework Approach}
\label{sec:sdc-scissor}

\major{
In this section, we first overview \framework's software architecture and its main usage scenarios
%(see Figure~\ref{fig:architecture} and Section \ref{sec:overview}).
(Section \ref{sec:overview}); next, we describe the %SDC Test Case Feature Sets we 
selected features used as inputs to \framework
% to classify safe and unsafe test cases % before executing them 
(Section \ref{sec:features}); finally, we 
%discuss in detail the components and the approach (see Section \ref{sec:approach-sdc}) behind \framework (explaining 
explain how \framework uses these features to classify test cases before executing them (Section \ref{sec:approach-sdc}).
% identify safe and unsafe test cases ).
}

\major{\subsection{\framework Architecture Overview}
\label{sec:overview}

\framework supports two main usage scenarios: \textit{Benchmarking} and \textit{Prediction}.
In the \textit{Benchmarking} scenario, SDC developers (or testers) leverage \framework to determine the best ML model(s) to classify SDC simulation-based tests as safe or unsafe.
In the \textit{Prediction} scenario, instead, \framework uses the most promising ML model(s) to classify newly generated test cases. % (before executing them).

\framework Software Architecture (Figure~\ref{fig:architecture}) implements these 
% \textit{Benchmarking} and \textit{Prediction} 
scenarios by means of five main software components, which have the main following responsibilities and relations:

% ALESSIO: Avoid "Widow"
\begin{enumerate}[(i)]
    \item \texttt{SDC-Test Generator} generates %random 
SDC simulation-based test cases.
    \item \texttt{SDC-Test Executor} executes the tests and stores the test results, i.e., safe or unsafe labels, to allow training of the ML models.
% (or execute any set of test cases specified as input). 
% The test case results produced by \texttt{SDC-Test Executor} are stored locally, to be used later to automatically classify test cases with the safe or unsafe labels; %) , relevant for the training and validation steps; 
    \item \texttt{SDC-Features Extractor} extracts the input features 
% of the executed
from the SDC simulation-based test cases.
% as described in Section \ref{sec:features}. 
    \item \texttt{SDC-Benchmarker} uses these features and 
% corresponding % (safe and unsafe)
collected labels %as input
to train the selected ML models and determines which ML model best predicts the tests that are more likely to detect faults.
% Finally, 
    \item \texttt{SDC-Predictor} uses the %specified 
trained ML models to classify newly generated test cases, thus achieving cost-effective SDC simulation-based testing via test selection. 
\end{enumerate}

% (i.e., without executing the new test cases).
}

\major{
\begin{figure*}[t!]
\centering 
%link to the source of the figure: https://drive.google.com/drive/folders/1ZmpP575r-o8V0y4jZHq9HwQJxQFjEAab
	\includegraphics[width=0.99\textwidth]{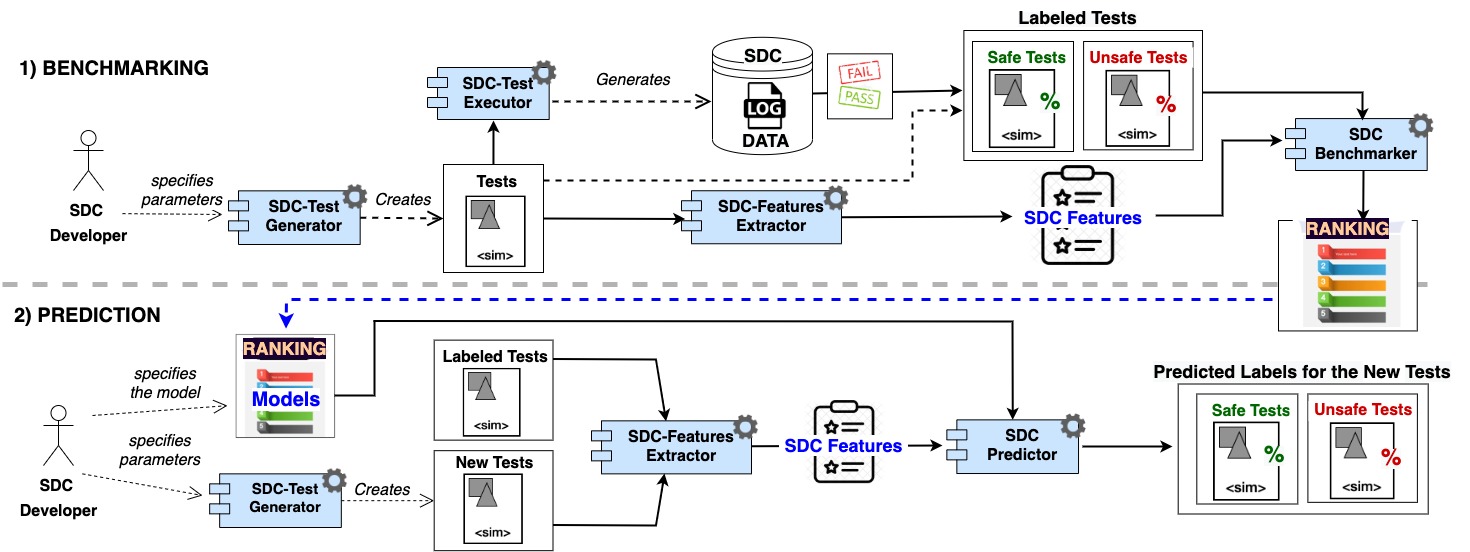} 
		%\vspace{-2mm}
	\caption{Overview of \framework's Software Architecture}
	\label{fig:architecture}
	\vspace{-2mm}
\end{figure*}
}

\major{\subsection{SDC Test Case Features}
\label{sec:features}}
%\alex{@ALL: Since the description of features is not complete, I did not review this section. When @Christian completes it, I can give it a go}
% Road Characteristics \& Road Metrics}
\major{
\textbf{SDC Test Case Road Characteristics - Features Set 1}  (Used in RQ1, RQ2, and RQ3).
To predict whether test cases
are likely to result in safe or unsafe test cases before their execution, we use a set of simple static features extracted from the \emph{global characteristics} (we refer to \emph{Road Characteristics}) of the virtual roads used as test cases.
We extract two types of Road Characteristics describing the main road \emph{attributes} (see~Table~\ref{table:road_general_feat}) and descriptive \emph{statistics} about the road composition (see Table~\ref{table:road_segment_stat_feat}).
Exemplary road attributes we consider are the total length of the virtual road, its starting and target positions on the map, and the count of left and right turns.
To calculate road statistics, instead, we adopt the following procedure:
(1) We extract the \emph{driving path} that the ego-car must follow during the test execution; this path defines the test case and contains the road segments that the ego-car must traverse to reach the target position from the starting position.
(2) We extract the metrics such as segment length, road angle, and pivot radius from the road segments.
% The definition of these segments are based on a turn angle threshold within specific road length. For instance, if the road makes an ``enough sharp'' turn in 10 meters, then this 10 meters of road belongs to a turn segment. If the next 10 meters are still doing the same type of turn then it belongs to the same turn segment as the previous 10 meters. 
% The segment end until the direction of the road changes.
(3) We compute descriptive statistics by applying standard aggregation functions (e.g., minimum, maximum, average) on the collected road segment metrics.
\begin{table}
    \centering
    \caption{\minor{Road Attributes  extracted by the \textit{SDC-Features Extractor}. In the table, we report for each feature their name, description, and range (based on the tests in the generated datasets)}}.
    \label{table:road_general_feat}
     \begin{tabular}{l l c}
     \toprule
       \textbf{Feature} & \textbf{Description} & \textbf{Range}\\
     \midrule
     Direct Distance & Euclidean distance between start and finish (Meters) & [0\hfill--\hfill489.9] \\
     Length & Total length of the driving path (Meters) & [50.6\hfill--\hfill3317.9] \\
     Num L Turns & Number of left turns on the driving path & [0\hfill--\hfill18]\\
     Num R Turns & Number of right turns on the driving path & [0\hfill--\hfill17] \\
     Num Straight & Number of straight segments on the driving path & [0\hfill--\hfill11]   \\
     Total Angle & Cumulative turn angle on the driving path & [105\hfill--\hfill6420] \\
     \bottomrule
    \end{tabular}
\end{table}
\begin{table}
\centering
\caption{\minor{Road Statistics extracted by the \textit{SDC-Features Extractor}. In the table, we report for each feature their name, description, and range (based on the tests in the generated datasets).}}
\label{table:road_segment_stat_feat}
    \begin{tabular}{l l c}
     \toprule
     \textbf{Feature}  & \textbf{Description} & \textbf{Range}\\
     \midrule
     Median Angle & Median turn angle on the driving path & [30\hfill--\hfill330]   \\
     Std Angle & Standard deviation of turn angles on the driving path & [0\hfill--\hfill150] \\
     Max Angle & Maximum turn angle on the driving path & [60\hfill--\hfill345] \\
     Min Angle & Minimum turn angle on the driving path & [15\hfill--\hfill285] \\
     Mean Angle & Average turn angle on the driving path & [52.5\hfill--\hfill307.5] \\
     \midrule
     Median Radius & Median turn radius on the driving path & [7\hfill--\hfill47] \\
     Std Radius & Standard deviation of turn radius on the driving path & [0\hfill--\hfill22.5] \\
     Max Radius & Maximum turn radius on the driving path & [7\hfill--\hfill47] \\
     Min Radius & Minimum turn radius on the driving path & [2\hfill--\hfill47]  \\
     Mean Radius & Average turn radius on the driving path & [5.3\hfill--\hfill47] \\
     \bottomrule
    \end{tabular}
\end{table}

}

\begin{figure}
    \centering
    \includegraphics[width=0.5\textwidth]{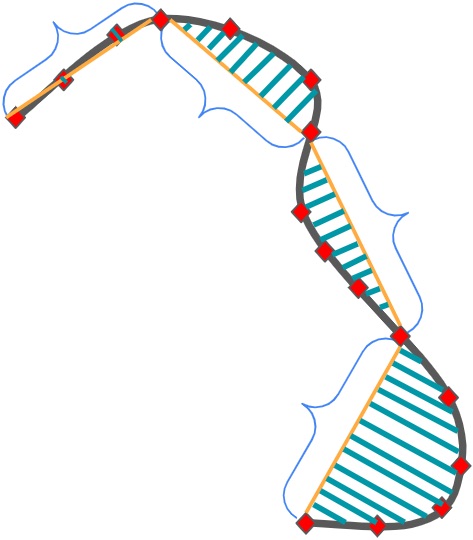}
    \caption{\major{Road diversity as area (green) between the road (black) and direct segment line (yellow).}}
    \label{fig:road-diversity}
\end{figure}

\major{
\textbf{SDC Test Case: Diversity Metrics - Features Set 2} (Used in RQ3).}
\major{
%\seba{@Christian, please describe what values such new metrics can have and how you compute them more formally, in a formula... (maybe you could have a table with name of the feature, the formula, and the range of values, as for other features...)}
To predict whether test cases are likely to result in safe or unsafe test cases before their execution, we also designed a new set of road features called \textit{Diversity Metrics}.
Specifically, we calculate per road segment the area that is spawned between the direct line of a segment (start and end of the segment) and the actual road.
The concept of the diversity feature is also explained in Figure~\ref{fig:road-diversity}, where the green area represents the diversity of a single road segment.
The curly braces indicate the segments of the road.
A segment consists of road points marked as red diamonds.
Furthermore, the yellow lines represent the direct paths between the start and end points of each segment.
Concretely, we used for the calculation of the area \texttt{Shapely}~\cite{Shapely}, an open-source library for Python to perform geometric calculations.
%\sajad{i don't think we need this much of details on implementation here}
For each identified segment, we define a Shapely \texttt{Polygon} object that includes the road points and the line representing the direct segment line.
All classes of Shapely provide a similar interface as well for calculating the area of a Shapely object.
The previously constructed Polygon has a property called \texttt{area}.
With this approach, we retrieve the area (also known as diversity in our context) of the segments.
On this basis, we calculate two additional features; (i) \texttt{Full Road Diversity}, and (ii) \texttt{Mean Road Diversity}.
As described in Table~\ref{table:diversity-feature}, the \texttt{Full Road Diversity} is computed by summing up all areas spawned by each segment of a road, whereas the \texttt{Mean Road Diversity} feature is the mean value of all areas of a single road.
The main assumption for using these new features is that the road is more diverse if the spawned area is greater and, therefore, unsafer.
}

\begin{table}
\centering
\caption{\minor{Diversity features extracted by the \textit{SDC-Features Extractor}. In the table, we report for each feature their name, description, and range (based on the tests in the generated datasets).}}
\label{table:diversity-feature}
\major{
    \begin{tabularx}{\textwidth}{l X c}
     \toprule
     \textbf{Feature}  & \textbf{Description} & \textbf{Range}\\
     \midrule
     Full Road Diversity & The cumulative diversity of the full road composed of all segments. & [0\hfill--\hfill$\infty$]   \\
     Mean Road Diversity & The mean diversity of the segments of a road. & [0\hfill--\hfill$\infty$]   \\
     \bottomrule
    \end{tabularx}
    }
\end{table}

\major{
%\alex{Pay attention to never leave header widow or orphan!}
\Needspace{5\baselineskip}
\subsection{The \framework's Workflow}
\label{sec:approach-sdc}}

\major{
% Test Generation and Text Execution
%\seba{@Christian, in the following paragraph, which generators we used? I assume AsFault, Frenetic, right? Let's explain why..}
As described in Section~\ref{sec:background}, \framework's leverages an existing, open-source, and extensible SDC testing infrastructure to execute the test cases (\texttt{SDC-Test Executor}). Likewise, it relies on existing test generation algorithms integrated with that infrastructure to automatically generate the test cases to optimize (\texttt{SDC-Test Generator}). Hence, \framework can already be used to improve the cost-effectiveness of several test generators.

During \emph{Benchmarking}, \framework utilizes \texttt{SDC-Test Generator} and \texttt{SDC-Test Executor} to collect the necessary data for training the ML Models, i.e., labeled test cases; next, it relies on \texttt{SDC-Benchmarker} to determine the ML models that best classify the SDC test cases as safe or unsafe as described below.
% that are likely to detect faults.
%
Given a set of labeled test cases and the corresponding input features extracted by \texttt{SDC-Features Extractor}, \texttt{SDC-Benchmarker} trains and evaluates an ensemble of standard ML models using the well-established \texttt{sklearn}\footnote{https://scikit-learn.org/} library. 
Next, it assesses each ML model's quality using %either 
K-fold cross-validation and the whole dataset.
% (or a training and test set methodology using different percentage splits of the dataset provided as input to the tool); and,
Finally, it identifies the best-performing ML models according to Precision, Recall, and F-score metrics~\cite{Birchler2022Cost} and outputs the best (trained) models as well as the features needed to operate them.

\framework can work with various ML models. In this study, we consider ML models that have been successfully used for defect prediction or other classification problems in Software Engineering ~\cite{Bezerra, Kaur,PanichellaSGVCG15,SorboPASVCG16,RANI2021111047,PanichellaR20}. Specifically, we consider 
Naive Bayes (that applies Bayes' theorem to train a probabilistic classifier)~\cite{Caruana06anempirical}, 
Logistic Regression (that uses a logistic function to model the probability of observing a certain class)~\cite{ref1},
J48 (that creates a decision tree following the well-known C4.5 algorithm)~\cite{Weka,UAV:2022},
and Random Forests (that uses an ensemble of decision trees)~\cite{TinKamHo1998}.

During \emph{Prediction}, \framework takes as input the (trained) ML Models and the definition of the features needed to use them.
Next, it generates new test cases using \minorrevision{\texttt{SDC-Test Generator}} and utilizes \texttt{SDC-Features Extractor} to extract the necessary features.
Finally, it invokes \texttt{SDC-Predictor} for classifying safe or unsafe test cases before executing them.
% \framework predicts whether the tests are likely to be safe or unsafe before their execution using input features extracted by \texttt{SDC-Features Extractor}. Specifically, this component extracts SDC features, as discussed in Section \ref{sec:features}
% (see~Table~\ref{table:road_general_feat} and  Table~\ref{table:road_segment_stat_feat}).

%~\cite{Bezerra, Kaur,PanichellaSGVCG15}.
% \begin{itemize}
%     \item \textbf{Naive Bayes} which train a probabilistic classifier based on the Bayes' theorem~\cite{Caruana06anempirical}. 
%     \item \textbf{Logistic} follows the well-known C4.5 algorithm to create a decision tree.
%     \item \textbf{Random Forest} uses an ensemble of decision trees~\cite{TinKamHo1998}.
%   % \item \textbf{}
%     %\item \textbf{}
% \end{itemize}

In the next section, we describe the studies we conducted to evaluate the benefits of using \framework for test selection in the context of SDCs. After that, we present and discuss the achieved results.

% see Section~\ref{sec:design}, which  describes the empirical study design) and the results and insights (see Section~\ref{sec:results} and Section Section~\ref{sec:discussion}) achieved in experimenting with \framework in the context of SDCs.
}
% , providing also a corresponding probabilities (values ranging between 0 to 1).
%\update{ALESSIO: Clarification is needed. Does the predictor requires a training set DIFFERENT than Benchmarker? That is, does the PREDICTOR needs to RETRAIN the models? If so, this must be clarified in the initial section as well !}
% designed to apply a training and test sets methodology. Specifically, it takes as input a training set of tests (previously labelled) and a new set of test cases (generated by the SDC-Test Generator), 

\section{Study Design}
\label{sec:design}

\major{In this paper, we investigate Machine Learning-based test selection techniques for improving the cost-effectiveness of simulation-based testing of SDCs. 

The first challenge (\textit{RQ$_1$}) we focus on is to investigate whether, and to what extent, it is possible to classify test cases for SDCs as safe or unsafe before executing them, i.e., only considering \emph{input features}, such as the one discussed in Section~\ref{sec:sdc-scissor} \minorrevision{by conducting offline and real-time experiments}.
Specifically, we investigate the use of ML models for classifying test cases in the context of Lane Keeping systems (see Section~\ref{sec:background}).
% , in which unsafe tests corresponds to scenarios cause the ego-vehicle to depart from its lane~\cite{Gambi2019}.
% . As explained in Section \ref{sec:background}, in this context,  

The second challenge we focus on is devising techniques that effectively leverage features extracted from SDC test cases to reduce testing costs while keeping testing effectiveness high. Hence, we investigate whether \framework improves the cost-effectiveness of simulation-based testing of SDCs, compared to baseline approaches (\textit{RQ$_2$}). 
%
% Specifically, we focus on checking whether \framework reduces the time dedicated to executing safe tests and without affecting testing effectiveness (i.e., its ability to identify unsafe tests) compared to such baselines.}

A further aspect we investigate is whether there is an upper bound on the precision and recall achieved by ML techniques in identifying SDC safe and unsafe test cases when using static SDC features (available before executing the tests). Hence, we focus on investigating whether fine-tuning the ML algorithms (e.g., calculating derived features and performing hyper-parameter tuning) improves \framework's ability to discern safe test cases from unsafe ones (\textit{RQ$_3$}). 

Finally, to investigate the practical usefulness of \framework, we integrated our tool into the context of an industrial organization in the automotive domain (details of such an investigation are reported in Section \ref{sec:integration}).
% Hence, compared to previous studies in the state-of-the-art, this research question provides to this article a practical focus, with \framework  integrated SDC-Scissor into the context of an industrial organization in the automotive domain, named AICAS \footnote{https://www.aicas.com/wp/}.

% \alex{The following paragraphs are not clear, I omit them for the moment:}
% About all experiment we have performed in the context of regression testing, one could also focus on building training data from previous versions of the AI-agent, different from the current version. Indeed, the test outcome (i.e., safe/unsafe) of the same test case may vary depending on the versions used. 
% %
% However,  at the moment, neither BeamNG.AI nor Driver.AI integrated in \framework, are implemented in multiple versions; thus, their behavior can be only influenced via the parameters already discussed in the context of our study.
% %
% We plan to experiment with different versions of these driving agents when they will be released and investigate the generalizability of our techniques.
% %
% Section~\ref{sec:threats} discusses this aspect in more details.

In the following sections, we describe the dataset used in our study and the steps we followed to address these challenges. % and investigations, including a description of the 
}

%Specifically, we investigate two alternative setups (explained later in this section): pre-trained ML models, which may find application in regression testing, and dynamically retrained ML models, which are suitable in automated test generation. 
%
%We refer to the former setup as \emph{Offline} training, whereas we refer to the latter as \emph{\Realtime} training.
%
\major{
\subsection{SDC Test Cases Dataset Preparation}
\label{sec:context}
% As discussed in Section \ref{sec:sdc-scissor},  -> Not sure we did discuss it!
%We used \framework to generate the test cases and collect the corresponding labels that form our dataset.
%Doing so, we can extract features that characterize aspects of safe and unsafe test cases before their execution and without introducing computational overhead.
%\subsubsection{Dataset of SDC Test Cases}
%\label{sec:study_datasets}
%To avoid biasing the dataset, % We reduce the risk oobtain an unbiased sampling of the SDC test space, with 
%\alex{@All Is this still the case? Did we consider experiments with multiple configurations? I understood we use only BeamNG.AI and Risk Factor 1.5... }
%we collected data using multiple test subjects in different configurations and used \framework to generate test cases randomly.
%create a dataset of several
% (required to draw conclusions about the generalizability of the proposed approach), 
% In addition, from the resulting test cases, we computed a set of SDC features concerning aspects of the created SDC test cases (discussed in Section \ref{sec:sdc-scissor}), which represents the input features to train the ML models integrated in \framework (see Section \ref{sec:sdc-scissor}).
% and obtained the required (safe or unsafe) label for each test case, which is required to train and assess the ML models (as described in Section \ref{sec:sdc-scissor}).
% As result of this process, 
%\seba{@Christian, please (or check) update the following numbers..}
To enable the prediction of safe and unsafe SDC test cases, we used \framework for executing the generated test cases and collected labels (safe/unsafe) from the test results (pass/fail).
%As reported in Table~\ref{table:datasets}, we generated  $8,500$ test cases and collected labels from $14,175$ simulations using two driving agents and four configurations. 
\minorrevision{
%\alex{This is confusing and incomplete: in the table 4 we never mention 8500 test cases. Here in the description we do not mention the 148K points of road segments. Please clarify that in each test case/simulation there are multiple road-segments, therefore from the 14 we collected data about 148K road segments. }
As reported in Table~\ref{table:datasets}, we generated
%$8,500$ test cases and collected
 a dataset with $14,175$ data rows with full road features
 %data points
 that are obtained from simulations of 8,500 tests using two driving agents and four configurations.
%\alex{R3 is confused here: maybe we can say we generated 8500 virtual roads and executed them as test cases against the two test subjects, which resulted in collecting 14K data points (labels?).}
}
%In total, we collected approximately $14,175$ road data points. 
What can be observed from the table is that \framework takes AI engines' inputs to generate the test cases, this lead to test cases having different configurations of roads and, as a consequence, different sets of road segments composing them. 
\minorrevision{The test cases, their labels, and the SDC features characterizing them are the main data used for conducting our experiments. An overview of the data is reported in Table~\ref{table:datasets}.}

\begin{table}
\caption{\minorrevision{Dataset Summary of SDC test cases on segment level and full road level (composed by segments).}}
\label{table:datasets}
\begin{tabular}{l l r r r}
\toprule
% Header 2
\multicolumn{1}{c}{\textbf{Test Subject}} &
\multicolumn{1}{c}{\textbf{Feature Set}} &
\multicolumn{3}{c}{\textbf{Data Points}}\\
% Header 2
&  & \textbf{Unsafe} & \textbf{Safe} & \textbf{Total} \\
\midrule
BeamNG.AI cautious & Full Road & 312 (26\%) & 866 (74\%) & 1’178 \\     
BeamNG.AI moderate & Full Road & 2'543 (45\%) & 3'095 (55\%)  & 5’638 \\  
BeamNG.AI reckless & Full Road & 1'655 (96\%) & 74 (4\%)      & 1’729 \\ 
Driver.AI & Full Road & 1'045 (19\%) & 4'585 (81\%) & 5’630 \\ \cline{5-5}
& & & & \minorrevision{\textbf{14'175}} \\
\midrule
BeamNG.AI moderate & Road Segment & 2'543 (3\%) & 72'433 (97\%) & 74’976 \\
Driver.AI & Road Segment & 2'494 (3\%) & 71'145 (97\%) & 73’639 \\ \cline{5-5}
& & & & \minorrevision{\textbf{148'615}} \\
\bottomrule
\end{tabular}
\end{table}
}

\major{
\subsection{Research Method}
% With the usage of \framework, 
We designed a set of experiments to answer our research questions:
\begin{itemize}
    \item \textit{Machine Learning-based Experiments (RQ$_1$)}: The first set of experiments investigates whether ML models trained with the selected SDC test case features %about \textit{Road Characteristics}, as input features for the various models, 
    can identify safe and unsafe test cases before their execution. 
    \item \textit{Offline Experiments (RQ$_2$)}: 
    %\alex{This is incomplete, we need to include here also the real-time experiments. Maybe we can introduce RQ\_\{2.1\} and RQ\_\{2.2\} to reduce the confusion?}
    The second set of experiments 
    % are used to 
    investigates if and how much \framework improves the cost-effectiveness of SDC simulation-based testing compared to baseline approaches.
    \minorrevision{
    \item \textit{Real-Time Experiments (RQ$_2$)}: In these experiments, we train an adaptive model based on data observed while executing the tests and compare it with a pre-trained model.
    }
    \item \textit{Optimization Experiments (RQ$_3$)}: The third set of experiments 
    % are used to 
    investigates how \framework performance improves by adding new SDC features and 
    % experimenting with 
    tuning ML Models hyperparameters. % tuning strategies. 
    Specifically, in RQ$_3$, we focus on investigating whether there is an actual upper bound on the precision and recall achieved by the ML techniques in identifying SDC safe and unsafe test cases when using static SDC features (available before executing the tests).
  %  \item \textit{Experiments on Industrial use Case (RQ$_4$)}: The last set of experiments %are used to  investigates the extent to which \framework can be integrated into the context of industrial organizations in the automotive domain.
\end{itemize}
}

\major{
\subsubsection{Machine Learning-based Experiments (RQ$_1$)} 
\label{sec:study-model}
In the context of RQ$_1$, we study whether ML models can be used to predict safe or unsafe test cases and which combinations of features allow us to achieve more accurate predictions. 
% results in the SDCs context. %Therefore, we experiment with various types of ML models and classify the test cases generated by \framework.
%\seba{@Christian, check the following part out.. if you do not have J48, I can run the experiment for you..}
%\alex{This is something we already mentioned. This part might be omitted to save space and reduce repetitions}
As discussed in Section \ref{sec:sdc-scissor}, we integrated into \framework several ML models, and in the context of our work, we experimented with \textit{Logistic Regression}~\cite{Tolles2016}, the \textit{J48}~\cite{Weka}, the \textit{Random Forest}~\cite{TinKamHo1998}, and the \textit{Naive Bayes}~\cite{Caruana06anempirical} as ML models.} 

%\alex{Possible repetition}
\major{
We trained the ML models mentioned above using a training and test sets split strategy for each of the configurations listed in Table~\ref{table:datasets} separately. 
%That is, we did not mix data generated while testing the various test subjects and configurations. 
We evaluated the performance of each ML model 
%using as the test set the remaining data points in the same configuration. Specifically, we evaluate models' performance
by computing the standard metrics of precision, recall, and F-score~\cite{Baeza-YatesR2011,Bezerra, Ceylan, Kaur,CanforaLPOPP13,PanichellaSGVCG15}.}
%, computed as follows:
% \begin{align*}
%     \textit{Precision} &= \frac{TP}{TP + FP} \\
%    \\
%    \textit{Recall} &= \frac{TP}{TP + FP} \\
%     \\
%     Accuracy &= \frac{TP + TN}{TP + TN + FP + FN}
% \end{align*}
%In the formula, we refer with \textit{TP} the true positive cases (i.e., unsafe tests correctly identified), while with \textit{FP}, the cases in which safe tests have been miss-classified as unsafe tests.
%Vice versa, in the formula, we refer with \textit{FP} the true negative cases (i.e., safe tests correctly identified), while with \textit{FN}, the cases in which unsafe tests have been miss-classified as safe tests.

\major{
\textbf{Rebalancing of training data}. Since unsafe scenarios are an exception --not the norm-- when generating random tests, the raw data we collected with \framework is unbalanced toward safe cases. Therefore, we re-balanced the training data (in the case of the training and test sets split strategy) to avoid skewed distributions that would otherwise bias the ML models towards one specific class. Specifically, we adopted random oversampling, a re-balancing technique proven to be robust~\cite{10.5555/3000292.3000304}, to supplement the training data with multiple copies of some of the minority classes. 
% Consequently, we first randomly sample the entire data set to form an initial training set; next, we identify the less represented class (unsafe tests) and randomly select and duplicate data of such minority class, this until both classes have roughly the same size.
}

\major{
\textbf{Size of the training dataset}. To study how the training set size affects the ML models' performance, we created balanced training  datasets of increasing size (Table~\ref{tab:training_dimensions}).
However, we generated the test datasets to evaluate the ML models by randomly sampling the data point not included in the training datasets. Notably, we did not re-balance the test datasets to preserve the underlying distribution classes in the data. 
}

\begin{table}
\centering
\caption{Model Training Dimensions}
\label{tab:training_dimensions}
\begin{tabular}{p{0.15\linewidth} p{0.35\linewidth}  p{0.4\linewidth}}
\toprule
\textbf{Dimension} & \textbf{Description} & \textbf{Dimension Configurations}\\
\midrule
 Dataset & Using different datasets to train the model & BeamNG.AI (RF 1,1.5,2), Driver.AI, and Combined Datasets\\
% Features&Changing the features used for the model&Full Road Features, Road Segment Features \\
 Training Set & Changing training set size by using different percentage split for training and test sets &
 \begin{tabular}[t]{@{}l@{}} 
 40\% training set \& 60\% test set;\\
 50\% training set \& 50\% test set;\\
 60\% training set \& 40\% test set;\\
 80\%  training set \& 20\% test set.
 \end{tabular}\\
\bottomrule
\end{tabular}
\end{table}

\major{We also study the effects of different training strategies on each ML model's performance. To do so, 
% instead of creating a balanced dataset for each configuration, 
we evaluated the ML models using standard K-fold cross-validation~\cite{Refaeilzadeh2009}.
In particular, we set $K=10$ (i.e., 10-fold cross-validation) and utilize all the available data in each configuration.}

\major{
\subsubsection{Offline Experiments (RQ$_2$)} \label{sssec:study-round}
To answer RQ$_2$, we investigate whether \framework improves the cost-effectiveness of simulation-based testing of SDCs, compared to baseline approaches. The quality focus is to understand whether \framework reduces the time dedicated to executing safe (irrelevant) tests without affecting testing effectiveness (i.e., its ability to identify unsafe tests) compared to such baselines.

\framework can use pre-trained models to classify safe and unsafe test cases. 
%However, it can also retrain the ML models on the fly, as new data are collected from test executions.
Therefore, we designed experiments to analyze how using pre-trained ML models for selecting (existing) test cases improves regression testing. 
%Likewise, we plan experiments to assess the effectiveness of \framework as a means to dynamically add new test cases to automatically generated test suites (see Section~\ref{sec:study-real}).
For those experiments, we consider the combinations of ML models and features that achieve the best results in the context of RQ$_1$ (see Section~\ref{sec:results_rq1}).
In addition, we contextualize the results achieved by \framework using a baseline approach that performs a random selection of test cases. Notably, random selection is considered one of the standard baselines for evaluating test selection strategies~\cite{Shin2018, Yoo:2010}. 
%\seba{@christian, check the following} 
Finally, we also compare \framework against a slightly more intelligent baseline approach that selects test cases by ordering the test to be executed considering their road length (in decreasing order). The conjecture of this second baseline is that the longer the road, the higher the probability of observing a fault.
}

% Experiment description
\begin{table}
\centering
\setlength{\tabcolsep}{18pt}
\caption{Offline Experiment Dataset
%\alex{data in the table does not correspond to data in the text. We say we have 2225 safe and 1334 failing, but the entire data set is reported as 3095 and 2543}
}
\resizebox{12cm}{!}{
\begin{tabular}{ @{}lcc@{} }
\toprule
     \textbf{Dataset} & \textbf{Number of Safe Tests} & \textbf{Number of Unsafe Tests}   \\ 
\midrule
    Complete Set & 3095 & 2543 \\
    Training Set & 2034 & 2034 \\
    Test Pool (95/5) & 1061 &  55  \\
    Test Pool (80/20) & 1061 & 265  \\
    Test Pool (60/40) & 763& 509  \\
    Test Pool (30/70) & 218 & 509 \\
\bottomrule
\end{tabular}
}
\label{tab:overview_pool_comp}
\end{table}

%\chris{@Seba: Please check the non-highlighted text. We used another dataset because in the previous data of Bill we don't have the road points to extract also the new features.}
Studying the effectiveness of \framework offline requires test cases and executions;
therefore, we used a dataset with known test execution times.
\minorrevision{
%\seba{explain which one in the next sentence..}
Due to the lack of backward compatibility of BeamNG.tech, we generated a new dataset for complementing our evaluation (see Table~\ref{tab:cost-effectiveness}) involving the usage of the most recent version of BeamNG.tech.
For all other evaluations, we used the data as reported in Table~\ref{tab:overview_pool_comp}.
In summary, the separated new dataset consists of $3'559$ with $2'225$ safe and $1'334$ failing tests labeled with the BeamNG.AI (RF 1.5).
}
%therefore, we used the data previously collected in the context of RQ$_1$.
%Specifically, we consider the data generated while testing BeamNG.AI in the moderate configuration (RF 1.5).
%We decided to do so because this configuration provides the largest number of test cases and executions.
\minor{As reported in Table~\ref{tab:overview_pool_comp}, we created a \emph{Training Set}, accounting for 80\% of the whole data set, and we used the remaining 20\% of data for testing. We created a balanced Training Set, but we purposely created four unbalanced \minorrevision{\textbf{Test Pools} with different distributions of unsafe cases, ranging from few (5\% of the testing data) to many (70\% of the testing data).
In creating our test pools, we under-sampled safe test cases (e.g., Test Pool (30/70)) since the number of unsafe test cases was inferior to the total amount of test cases in our complete dataset.
}
Our conjecture is that using different Test Pool compositions allows us to assess \framework's performance in various settings.}

% nr_tests: 3559, nr_pass: 2225, nr_fails: 1334

% ALESSIO: The figures are a bit confusing and do not really add anything to the paper, IMHO. But I leave the here for the moment

\begin{figure}
\centering
\includegraphics[width=\textwidth]{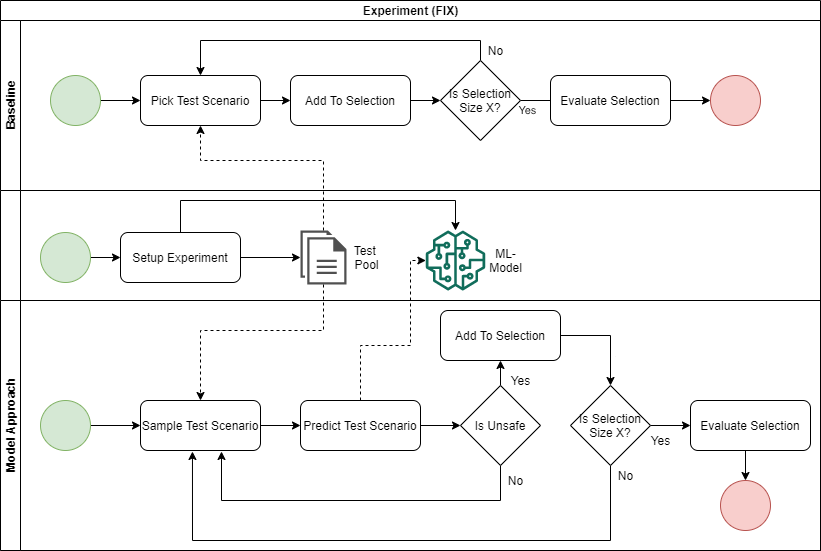}
\caption{FIX Experiment Overview. 
%\alex{there is now a YES after add to selection that shall be removed}
}
\label{fig:exp_fix}
\end{figure}

\major{
\textbf{Experimental setups of offline experiments}.
We conducted the offline experiment in two experimental setups, referred to as \FIX and \REACH. Since they mimic the issues of having a limited testing budget in the context of SDCs, We believe they are representative. 
% in industrial contexts 
We repeated the experiments in both setups $30$ times to increase the confidence in the achieved results.}

%\seba{@christian, I have changed the following, but if we do not have the second baseline approach, we have to modify it}
The \FIX setup investigates the benefits of using \framework when the resources allocated for testing are limited, i.e., the amount of test cases that can be executed in the simulation environment is \textbf{fixed} \major{to a value $S$ (e.g., $S = 5, 6, etc.$)}. 
\major{The process we followed to experiment with the \FIX setup is illustrated in Figure \ref{fig:exp_fix} alongside the baseline processes. The baseline approach draws tests from the test pool (randomly or by considering their road length) and adds them to the test suite until the test suite reaches the target size $S$. 
% \FIX approach of 
\framework, instead, samples the tests from the test pool but adds them to the test suite only if the ML model predicts that they are unsafe; as before, the process ends when the test suite reaches the target size $S$.
In this setup, more effective techniques select larger portions of unsafe tests; therefore, we evaluate the performance of \framework using the ratio of unsafe to safe test cases in the final test suites compared to the baseline approaches.
}

\begin{figure}
\centering
\includegraphics[width=\textwidth]{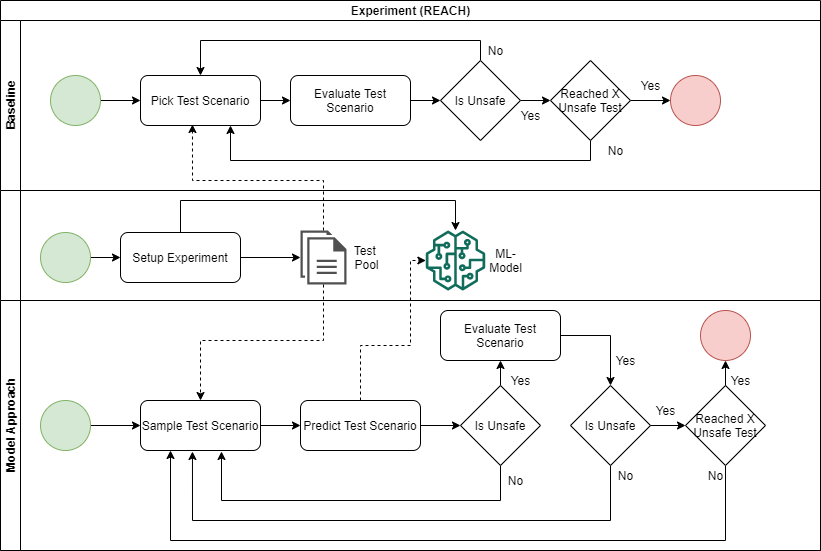}
\caption{REACH Experiment Overview.}
\label{fig:exp_reach}
\end{figure}

%\seba{@christian, I did not change the following since I think we cannot compare the second baseline in the REACH type of experiments.}
The \REACH experiment, instead, investigates the ability of \framework to reduce the time to identify at least $N$ unsafe test scenarios. 
% We conjecture that testing time should be spent on executing unsafe test cases, as those help developers expose problems in SDCs. earlier. 
In our experiment, we set $N=10$ since the time to identify that many unsafe test cases potentially requires the execution of many more (safe) test cases.
The process we followed to experiment with the \REACH setup is illustrated in Figure \ref{fig:exp_reach} alongside the random baseline approach. As before, the baseline randomly samples tests from the test pool and executes them until $N$ unsafe tests have been identified. \REACH, instead, 
% follows a similar process but 
executes only those tests that are predicted to be unsafe by the ML models.
In this setup, more effective techniques identify $N$ unsafe tests sooner;
therefore, we consider the number of true positives (TP),\footnote{True positives are tests predicted as unsafe and verified to be so; conversely, true negatives are tests predicted and verified to be safe.} true negatives (TN), false positives (FP), and false negatives (FN) predicted by the ML models. Having information about TP, TN, FP, and FN enables us to count how many tests were needed to reach the goal, how long it took to do so, and how much time was wasted in evaluating safe test cases.

\minorrevision{
\subsubsection{Real-Time Experiments (\texorpdfstring{RQ\textsubscript{2}}{RQ2})}
We complement the previous \textit{Offline Experiments} to answer RQ$_2$, which focuses on applying \framework to regression test case selection, with \textit{Real-Time Experiments} in which we study the application of \framework to automated test generation.
}

\minorrevision{
We conducted the Real-Time Experiments according to the following procedure:
\begin{inparaenum}[(i)]
\item \framework to generate random test cases;
\item for each newly generated test case,  \framework classifies it as safe/unsafe; and,
\item we filter out test cases classified as safe before generating the next test case, whereas we executed the test cases classified as unsafe.
\end{inparaenum}
As the test subject, we used BeamNG.AI in the moderate configuration (RF equal to 1.5) as this configuration is a compromise between overly conservative and overly aggressive driving styles.
}
%}

\begin{figure}
    \centering
    \includegraphics[width=\textwidth]{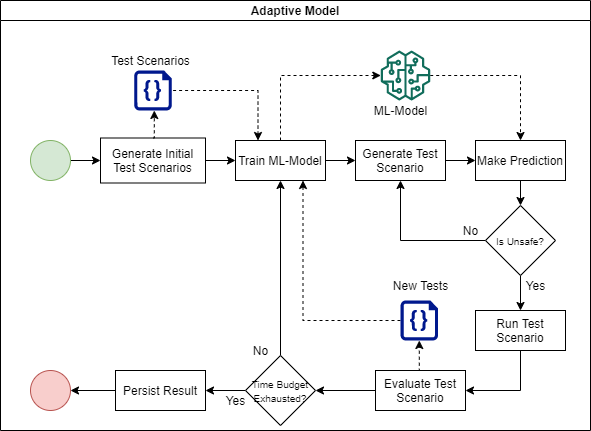}
    \caption{Overview of the Adaptive Model configuration for the Real-Time Experiments.}
    \label{fig:adaptive_prc}
\end{figure}

\minorrevision{
A cost-effective test generator devotes more time to executing (likely) unsafe tests that can expose defects rather than executing safe test cases, which might not contribute any additional insight into the behavior of the SDC under test.
Correctly identifying unsafe test cases, therefore, is paramount and depends on the quality of the ML model used as a classifier which, in turn, depends on the technique employed by the ML models and the data used to train them.
Particularly relevant in this context is whether the ML model is predefined and fixed or allowed to be updated online as new data become available.
The trade-off between these two configurations is that ML models have little operational costs once trained but may miss relevant behaviors; on the contrary, dynamically retrained ML models can cope with missing training data but at the cost of additional time spent in retraining them.
%\major{
Therefore, we compare the following two approaches:
\begin{itemize}
    \item \textbf{Pre-trained Model} in which we used the best performing model identified during the Machine Learning-based Experiments (Section~\ref{sec:results_rq1}). We trained this model using the re-balanced dataset for the case of BeamNG.AI RF 1.5, as this is the configuration of the test subject used for this experiment.
    \item \textbf{Adaptive Model} in which we also used the best performing model identified during the Machine Learning-based Experiments (Section~\ref{sec:results_rq1} but trained with only $60$ randomly generated test cases. After this initial training, we retrain the ML model after executing the predicted unsafe test cases using the newly collected ground truth labels for those test cases. Figure \ref{fig:adaptive_prc} illustrates this process. Notably, since the ML model may be inaccurate, this process collects both positive and negative labels.
\end{itemize}
%}
}

\minorrevision{
As before, we contextualize the results achieved by \framework using a baseline approach that implements plain vanilla random generation, i.e., it does not filter the test cases.
}

\minorrevision{
We ran each configuration on a dedicated machine equipped with an Intel Core i5-6600K~(3.5~GHz), 16~GB RAM, and an NVIDIA GeForce GTX 1070 GPU and set the test generation time budget to six hours.
}

\minorrevision{
%\major{
During each execution of the experiment, we stored all the tests generated by \framework so we could execute the test cases filtered out by \framework \emph{post-mortem} to calculate metrics such as accuracy, precision, and recall.
%}
}

\minorrevision{
Table \ref{tab:evaluation_met} provides an overview of the metrics used for the evaluation of \framework across the various configurations.
Those metrics include the count of unsafe tests found during each experiment (true positives), true negatives, false positives, and false negatives. Additionally, we consider how \framework allocated the time budget to run safe and unsafe test cases, generate test cases, and rebuild the ML models.
}

\begin{table}
\minorrevision{
\caption{Evaluation metrics for the Real-Time Experiments.}
\centering
\renewcommand{\arraystretch}{1.5}
\resizebox{12cm}{!}{
\begin{tabular}{p{5cm}p{5.5cm}p{2cm}}
\toprule
 \textbf{Metric} & \textbf{Description}& \textbf{Range}\\
\toprule
Number of Unsafe Test Execution & The number of unsafe tests the approach simulated during the experiment & 0-N\\
Number of Safe Tests Execution &The number of safe tests the approach simulated during the experiment & 0-N\\
Time Allocation & How much time relative to the total time was spent with an action & 0-1\\
True Positives/Negatives & Number of correct predictions for categories safe and unsafe & 0-Number of Predictions\\
False Positives/Negatives & Number of incorrect predictions for categories safe and unsafe & 0-Number of Predictions\\
\bottomrule
\end{tabular}
}
}
\label{tab:evaluation_met}
\end{table}

% REAL TIME PARTS (RQ2)
\minorrevision{
In the second study, \framework leverages real-time data (i.e., dynamically generated tests) and continuously (re-)trained ML models; this setup lets us evaluate the application of the proposed technique for automated test generation.
As described before, in both setups, we compared the time-saving ability of \framework with respect to the random selection strategy as well as its ability to detect more faults while allocating lower test execution costs.
}

\major{
\subsubsection{Optimization Experiments (RQ$_3$)} \label{sec:optimization}
}

\major{
RQ$_3$ investigates whether there is an upper bound on the precision and recall of ML techniques in identifying SDC safe and unsafe test cases when using SDC test case features available before executing the tests. 
A range of different optimization algorithms can be used to achieve potentially better results with respect to the default configuration of parameters of the ML models. Two of the most common hyperparameter tuning methods are Random Search and Grid Search \cite{2986459.2986743,BergstraB12,AdnanAUR22}.
Grid search performs better for spot-checking combinations that are known to perform well. 
Therefore, we experiment with Grid search as a hyperparameter optimization approach and investigate how \framework's performance improves when it employs fine-tuned ML models.
% potential optimal combinations of parameters for the ML models integrated with .
Specifically, with Grid Search, we experimented with several parameter combinations for the best ML models using a 10-fold validation setting, as summarized below.
%\alex{Do we have a reason to chose those parameters?}

For the \textbf{Decision Tree (J48)}  we covered all possible combinations of the following parameters:
    \begin{itemize}
       \item \textbf{C (confidenceFactor)}: Is the confidence factor, and we experimented with values  $[0.001, 
       %0.002, 0.005, 
       0.01, 
       %0.02, 
       0.05, 
       0.1, 
       %0.2, 
       0.5]$ 
       \item \textbf{M (minNumObj)}: Is the minimum number of instances in a leaf, and we experimented with values  $[1, 
        %5, 
        10, 20, 
        %30, 
        %40, 
        50, 
        %70, 
        100]$
       \item \textbf{R (reducedErrorPruning)}: Reduced error pruning is an alternative algorithm for pruning that focuses on minimizing the statistical error of the tree.  We  experimented with the following values  $[yes, no]$
       \item \textbf{S (subtreeRaising)}: This is a specific method of pruning whereby a whole set of branches further down the tree are moved up to replace branches that were grown above it. We experimented with  the following values  of it $[yes, no]$
    \end{itemize}

%For the \textbf{Logistic Regression} model   we covered all possible combinations of the following parameters:
%    \begin{itemize}
%        \item \textbf{R (ridge)}: Is the Ridge in the log-likelihood, and we experimented with values  $logspace(2,-9, num=25)$
%    \end{itemize}

%\seba{@christian, did we experiment with other models?}

For the \textbf{Random Forest}, we covered all possible combinations of the following parameters:
\begin{itemize}
        \item \textbf{I (numIterations)}: Is the number of trees in the forest, and we experimented with values  $[5, 
        10, 
        %50, 
        %80, 
        100, 
        %200, 
        %500, 
        %800, 
        1000, 
        %1500, 
        2000]$ \item \textbf{K (numFeatures)}: Is the max number of features considered for splitting a node, and we experimented with values  $[0, 
        %1, 
        %5, 
        10, 
        %20, 50, 
        %70, 
        100, 
        %200, 
        500, 
        %700, 
        1000]$  
        \item \textbf{depth}: Is the maximum depth of the tree (0 unlimited), and we experimented with values  $[
        0, 
        %1,
        %2, 3, 4, 
        5, 
        %6, 7, 8, 
        10, 
        %11, 12, 13, 14, 
        %15, 
        %16, 17, 18, 19, 
        20]$ 
        \item \textbf{M (minNumObj)}: Is the minimum number of instances in a leaf , and we experimented with values  $[1, 
        %5, 
        10, 20, 
        %30, 
        %40, 
        50, 
        %70, 
        100]$
    \end{itemize}

%For the \textbf{Naive Bayes} we covered all possible combinations of the following parameters:
%\begin{itemize}
%    \item No parameters \seba{@Christian, if we do not}
%\end{itemize}

For the \textbf{Gradient Boosting}, we covered all possible combinations of the following parameters:
\begin{itemize}
    \item 'loss' = ['log\_loss', 'deviance', 'exponential']
    \item 'learning\_rate' = [0.01, 0.1, 0.2, 0.4]
    \item n\_estimators' = [10, 100, 1000]
    \item 'criterion' = ['friedman\_mse', 'squared\_error', 'mse']
\end{itemize}

For the \textbf{Logistic Regression}, we covered all possible combinations of the following parameters:
\begin{itemize}
    \item 'penalty' = ['l1', 'l2', 'elasticnet', 'none']
    \item 'dual' = [True, False]
    \item 'max\_iter' = [10, 100, 1000]
    \item 'solver' = ['newton-cg', 'lbfgs', 'liblinear', 'sag', 'saga']
\end{itemize}

For the \textbf{Support Vector Machine}, we covered all possible combinations of the following parameters:
\begin{itemize}
    \item 'penalty' = ['l1', 'l2']
    \item 'loss' = ['hinge', 'squared\_hinge']
    \item 'dual' = [True, False]
\end{itemize}

It is important to note that we perform Grid Search %setting results in the application of a
(with a 10-fold cross-validation strategy) over all experiments (for a total of over 700
% 2*2*2+4*2*3*5+3*4*3+5*5*4*5+5*5*2 = 714
experimented combinations of parameters) and use the best combination of features and ML model from Section~\ref{sec:study-model}.
% that achieved the best classification performance when using the default hyper-parameters 
% configurations of the ML models (as it is possible to see in 
% (Section~\ref{sec:study-model}).
}

\major{Section~\ref{sec:results} elaborates on the achieved experimental results for all research questions, while Section~\ref{sec:discussion} reflects on the results reported in such section, providing complementary insights, findings, and implications.\\}

\section{Results}
\label{sec:results}

\newcounter{finding}
\newenvironment{finding}%
{
    \refstepcounter{finding}\par\medskip
    \begin{framed}%
    \textbf{\textit{Finding~\thefinding. }}%
}%
{
    \end{framed}
} %

%This section reports, for each research question, the obtained results and the main findings.

\major{This section presents the achieved results organized by research questions, while Section~\ref{sec:discussion} 
discusses them in depth.}

\major{
\subsection{Machine Learning-based Experiments (\texorpdfstring{RQ\textsubscript{1}}{RQ1})}
\label{sec:results_rq1}
}

\major{
In this section, we discuss the results of RQ$_1$. Specifically, we describe the results achieved using the \emph{Road Characteristics} 
listed in Section \ref{sec:features} as input features to build the ML models. %(Section~\ref{sec:full-road_results}). 
%Then, we discuss the contribution of the input features in classifying test cases (Section~\ref{sec:feature_analysis_results}).
%\alex{Not clear what does this mean:}
%Then, we discuss the contribution of the input features in classifying test cases (Section~\ref{sec:feature_analysis_results}).
}
%\alex{Check the consistency: do we refer to Road Characteristics? Are there some other features to list?}
\major{
\subsubsection{Machine Learning-based  Experiments with \emph{Road Characteristics}}
\label{sec:full-road_results}}
% In this section, we will discuss the result of the full road approach comparing the different dimensions discussed previously in Section \ref{sec:design} (see Table \ref{tab:training_dimensions}).

\major{
% \paragraph{BeamNG.AI Datasets.}
% An important difference with the setting of experiments with the initial dataset is that
We evaluated the ML models trained using \emph{Road Characteristics}  as the main SDC features
with four splits of training and test data, as summarized in Table~\ref{tab:training_dimensions}.
However, for the sake of readability, we report here only the results achieved by the best-performing configuration, i.e., 80\% training and 20\% for testing. The full results can be found in our replication package~\cite{RP2021}.
Table~\ref{tab:rq1} reports Precision, Recall, and F-score for both unsafe and safe labels separately to study how the ML models can classify each case (i.e., the experiments summarized in Table~\ref{tab:training_dimensions}). It is important to note that in all experiments reported in Table~\ref{tab:training_dimensions}, we rebalanced the training data (as discussed in Section \ref{sec:study-model}).

\begin{table}[h!tp]
\caption{Performance of the ML models trained using road features. The results refer to the split of 80/20 between training and test data. The best results are shown in boldface.}
\label{tab:rq1}
\centering
\begin{tabular}{llllllll}
\toprule
\textbf{Model} 
%& \textbf{Acc.} 
& \multicolumn{3}{c}{\textbf{Unsafe Test Cases}} & \multicolumn{3}{c}{\textbf{Safe Test Cases}} \\
\cmidrule{2-7}
 %&
 & \textbf{Prec.} & \textbf{Recall} & \textbf{F$_1$} & \textbf{Prec.} & \textbf{Recall} & \textbf{F$_1$} \\
\midrule
\textbf{BeamNG RF 1.5} 
%&
&  &  &  &  &  &  \\
J48 
%& 65.6\% 
& 69.2\% & \textbf{67.4\% }& 68.2\% & 61.5\% & 63.5\% & 62.5\% \\
Naïve Bayes 
%& 66.7\% 
& \textbf{79.3\%} & 53.2\% & 63.6\% & 59.3\% & \textbf{83.1\%} & 69.2\% \\
Logistic 
%& \textbf{70.9\%} 
& 78.1\% & 65.3\% & \textbf{71.1\%} & \textbf{64.8\%} & 77.8\% & \textbf{70.7\%} \\
Random Forest 
%& 68.5\% 
& 75.8\% & 62.7\% & 68.6\% & 62.5\% & 75.6\% & 68.4\% \\
\midrule
\textbf{Driver.AI} 
%&  
&  &  &  &  &  &  \\
J48 
%& 44.1\% 
& 19.5\% & 64.1\% & 29.9\% & 82.9\% & 39.6\% & 53.6\% \\
Naïve Bayes 
%& 38.8\% 
& 20.3\% & \textbf{78.5\%} & 32.3\% & \textbf{85.8\%} & 29.8\% & 44.2\% \\
Logistic 
%& 56.3\% 
& \textbf{22.7\%} & 56.5\% & \textbf{32.4\%} & 85.0\% & 56.3\% & 67.7\% \\
Random Forest 
%& \textbf{57.2\%} 
& 22.3\% & 52.6\% & 31.3\% & 84.4\% & \textbf{58.2\%} & \textbf{68.9\%} \\
\bottomrule
\end{tabular}
\end{table}
}

%\alex{Check the consistency of the names of Dataset. Did we introduce these datasets by name?}
\major{
% BeamNG.AI RESULTS
Regarding the BeamNG.AI dataset, with Risk Factor 1.5, the ML model performing the best in terms of F-score is Logistic (with 71\% for both labels), followed by Random Forest (between 68\%-69\% for both labels). The other models, instead, achieved lower F-score values.
% On the other hand, with more biased datasets towards either safe or unsafe classes (AF1 and AF2), we observe different performance for the classes, e.g, F$_1$ of 57.9\% (unsafe) vs. 71.8\%(safe) on RF 1 dataset, and 90.4\% (unsafe) vs. 32.6\% (safe) on RF 2 dataset for the Logistic model.
}

\major{
% \paragraph{Driver.AI Dataset.}
Regarding the Driving.AI dataset, we observe that the ML models achieved lower accuracy (49.1\%) than the BeamNG.AI dataset. 
%
%\alex{Didn't we balanced the datasets?}
\minorrevision{
This result can be explained by looking at how \textit{unbalanced} the Driver.AI dataset is %\alex{The training or testing dataset?}.
since Driver.AI drives carefully, its dataset comprises mainly safe scenarios, and the predictions of the ML models tested on it are biased toward safe predictions. 
}

Comparing the F-score achieved by the ML models against the Driver.AI and BeamNG.AI datasets shows this problem more evidently: the ML models performed comparably well for safe and unsafe classes against the BeamNG.AI dataset, whereas they performed well only for the safe test class in the case of Driver.AI.
However, we can observe some similarities between all ML models in terms of F-score values when trained on the Driving.AI dataset and the BeamNG.AI dataset.
For instance, for both datasets, Logistic and Random Forest tend to achieve better results.
In both cases, and especially in the case of Driver.AI, most ML models struggle to classify safe test cases when compared to  unsafe test cases. 

\begin{finding}
\framework is able to classify safe and unsafe test cases in both the BeamNG.AI 
 and the Driving.AI datasets, with the Logistic  and Random Forest models achieving the most reliable results in terms of F-score. 
% values for labels. 
However, all ML models achieved very poor results  on the Driving.AI dataset (49.1\%) when compared to the BeamNG.AI dataset. A result that we can justify by looking at the \textit{unbalance set of test cases} in the Driver.AI dataset. 
\end{finding}
}

\major{
\subsubsection{Analysis of Relevant Features.}
\label{sec:feature_analysis_results}
% In our study, we considered two sets of features, full road, and road segment features.
Although the ML models trained using the road features can effectively classify the test cases as safe or unsafe, it is crucial to know the level of contribution of each of these features. We analyzed the road features for the  BeamNG dataset discussed in Table \ref{tab:rq1} using two popular feature evaluation methods: \textit{information gain} and \textit{correlation}}.
While the detailed analysis results are reported in Appendix \ref{app:feature_analysis_results}, we summarise the main findings here. 

\major{
\begin{finding}
The \emph{Road Characteristics} extracted by \framework contribute differently to identifying the safe and unsafe test cases.
The \emph{Road Characteristics} concerning the pivot radius (min, mean, std, median), the sum of the turn angles, the number of left and right turns, and the total length of the road are among the most important features, which are all belonging to the set of road features.
\end{finding}
}

\major{
\subsubsection{Impact of Risk Factor (RF)}
To make it more clear how \framework's performance is affected by varying RF values, we compared its performance on BeamNG datasets with RF 1, 1.5, and 2 separately. While we report the details in Appendix \ref{app:rf}, here we summarise the main findings. 
}
\major{
\begin{finding}
The accuracy of \framework is influenced by their driving style and the diversity of datasets. For example, for more aggressive driving agents, the accuracy achieved by the ML models was higher than for cautious driving agents. Hence, predicting unsafe test cases is harder for cautious drivers than for reckless ones. Consequently, improving the testing of SDCs is more challenging for less aggressive driving agents.
\end{finding}
}

\major{
\subsubsection{Knowledge Transfer Between Different Driving Agents}
We also studied the ability of the ML models to transfer knowledge from a driving agent to another one by training ML models with one AI's dataset (BeamNG RF 1.5) and testing it with another AI's dataset (Driver.AI) and \emph{vice versa}. While we report the details in Appendix \ref{app:knowledge_transfer}, here we summarise the main findings. 
}
\major{
\begin{finding}
Our results show that the knowledge is not transferable from one driving agent to another, i.e., \framework performed worse when training ML models on data from a specific driving agent and testing them on data from a different one. 
However, ML models trained on the BeamNG data performed only slightly worse when evaluated on the Driver.AI data.
\end{finding}
}

% \major{In Section \ref{sec:discussion}, we discuss further results of RQ1, with particular attention to aspects that can impact the results of \framework (e.g., different risk factors in the case of BeamNG, etc.).}

\major{
\subsection{Offline Experiments (\texorpdfstring{RQ\textsubscript{2}}{RQ2})}
\label{sec:results_rq2}
%\seba{@christian, check if we included the second baseline and change the following part}
In this section, we discuss the results of RQ2.
Specifically, we focus on devising techniques that effectively leverage features extracted from SDC test cases to minimize testing costs while keeping testing effectiveness high. 
For this reason, we investigate whether \framework improves the cost-effectiveness of simulation-based testing of SDCs, compared to baseline approaches (\textit{RQ2}).
Hence, we report the results of the \FIX and \REACH experiments (detailed in Section \ref{sssec:study-round}).
Additionally, we report the results of the comparison between various ML models against the baseline approaches (described in Section \ref{sssec:study-round}) by considering different test pool compositions.
}

%\seba{@christian, check if we included the second baseline and change the following part}
\subsubsection{\FIX Experiment results}
\begin{figure*}
\centering
\includegraphics[keepaspectratio=true,width=0.9\textwidth]{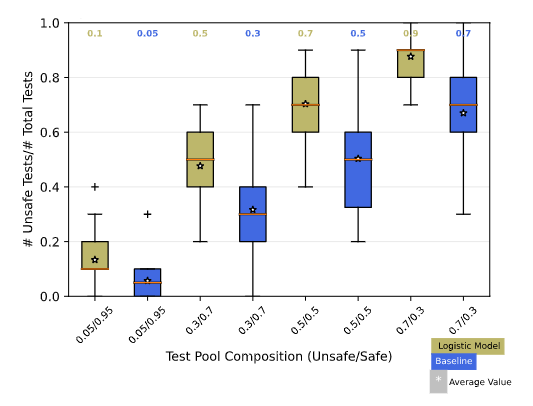}
\caption[Comparison Logistic Model and Baseline]{Comparison Logistic Model and Baseline across different Test Pool Compositions.}
\label{fig:pool_comp}
\end{figure*}

\begin{figure}
\centering
\caption{ Number of executed unsafe scenarios during the experiments on 
a)Test Pool(0.05/0.95) %: Baseline(mean=0.57), Logistic(mean=1.33), J48(mean=1.00), Random Forest(mean=1.20). 
b) Test Pool(0.3/0.7) %: Baseline(mean=3.17), Logistic(mean=4.76), J48(mean=4.17), Random Forest(mean=5.27). 
c)Test Pool(0.7/0.3) %: Baseline(mean=6.70), Logistic(mean=8.77), J48(mean=8.33), Random Forest(mean=7.83).
}
\label{fig:pool_005}
\begin{minipage}{0.48\textwidth}
\centering
a
\includegraphics[keepaspectratio=true,width=\textwidth]{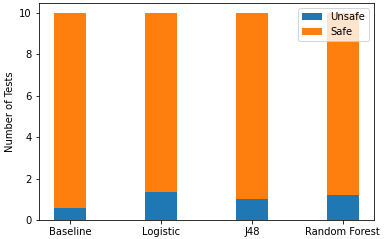}
\end{minipage}\hfill
\begin{minipage}{0.48\textwidth}
\centering
b
\includegraphics[keepaspectratio=true,width=\textwidth]{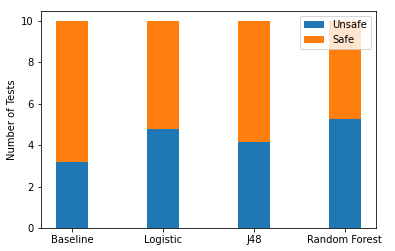}
\end{minipage}\hfill
\begin{minipage} {0.48\textwidth}
\centering
c
\includegraphics[keepaspectratio=true,scale=0.45]{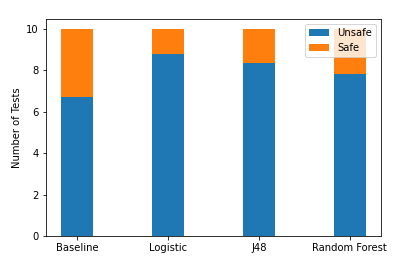}
\end{minipage}
\end{figure}

The goal of this experiment is to optimize the usage of the available resource in terms of testing execution time and effectiveness.
%We evaluated the results based on the number of unsafe scenarios. 
Figure \ref{fig:pool_comp} compares the ratio of unsafe tests selected for execution using different ML models against the first baseline approach (random selection) across different test pool compositions.
As can be observed from the figure, the Logistic model outperformed the baseline in all different test pool compositions  (described in Section \ref{sec:design}).
%The relative difference between the baseline and the models grows relative to the test pool composition.
Figure \ref{fig:pool_005} illustrates that with fewer unsafe test cases in the pool, we observe improvements in the number of selected unsafe tests using ML models over the baseline.
In the pool with the least unsafe tests, the Logistic model finds 133\% more unsafe tests compared to the baseline approach. In the more balanced testing pool, Logistic finds 50\% more unsafe tests, while with the pool with more unsafe than safe tests, it identifies 30\% more unsafe tests.
%Further, the different models show consistent results over the different compositions.
The Logistic model performs slightly better than the other models in all compositions except one (0.3/0.7), where Random Forest performed the best.

The confusion matrices in Table \ref{tab:conf_pool_r005} further illustrate the concrete results in terms of effectiveness with the various pool compositions.
In the pool with only 0.05 unsafe tests (Table \ref{tab:conf_pool_r005}-a), the Logistic model achieved 10 false negatives and 260 true negatives; this means that the model avoided the execution of 549 safe tests (considering that safe test cases take around 24 seconds in average to be executed), thus potentially reducing cost by more than 200 minutes in total on the less critical scenarios.
However, the false-positive number is still high, with a cumulative 263 false-positives identified.
As can be observed in Table \ref{tab:conf_pool_r005})-b, for the Test Pool 0.7/0.3, the Logistic model achieved over 260 true positives and only 37 false positives.
We observe that the precision correlates with the dataset composition; indeed, for datasets having more unsafe tests, the precision for unsafe tests is higher. 
For datasets having fewer unsafe tests, we obtain the opposite effect in the results. 
Figure \ref{fig:pool_comp} shows that the ML model performance and the baseline depend on the test compositions. 
The baseline and ML models perform better in test pools with more unsafe tests. 
Thus, according to our results, designing an appropriate test pool composition is of critical importance to achieving accurate prediction results.
%Having test pool compositions with fewer unsafe scenarios, increases the benefit of having some predictive power.
%In test pool with a large number of unsafe test scenarios, the benefit of having some predictive power is diminished, since even a random selection has a high chance of finding an unsafe scenario.

\major{
\begin{finding}
\framework outperforms the random baseline approach in selecting unsafe tests across all test pool compositions, which is critical for more effective testing practices.
In the test pool composition 0.3/0.7 (safe to unsafe), \framework found 30\% more unsafe tests; in the test pool composition 0.95/0.05 (safe to unsafe), instead, it found 133\% more unsafe tests.
\end{finding}
}

\begin{table}
    \caption{Confusion Matrix for Logistic Model, cumulative over 30 rounds for a) Test Pool (0.05/0.95), b) Test Pool (0.7/0.3)}
    \centering
    \parbox{.48\linewidth}{
    \centering
    a
    \begin{tabular}{l|l|c|c|c}
    \multicolumn{2}{c}{}&\multicolumn{2}{c}{Predicted class}&\\
    \cline{3-4}
    \multicolumn{2}{c|}{}&Unsafe&Safe&\multicolumn{1}{c}{}\\
    \cline{2-4}
    \multirow{2}{*}{Actual Class}& Unsafe & 40 & 10\\
    \cline{2-4}
    & Safe & 260 & 549 \\
    \cline{2-4}
    \multicolumn{1}{c}{} & \multicolumn{1}{c}{} & \multicolumn{1}{c}{} & \multicolumn{1}{c}{} & 
    \end{tabular}
    }
    \parbox{.48\linewidth}{
    \centering
    b
    \begin{tabular}{l|l|c|c|c}
    \multicolumn{2}{c}{}&\multicolumn{2}{c}{Predicted class}&\\
    \cline{3-4}
    \multicolumn{2}{c|}{}&Unsafe&Safe&\multicolumn{1}{c}{}\\
    \cline{2-4}
    \multirow{2}{*}{Actual Class}& Unsafe & 263 & 48\\
    \cline{2-4}
    & Safe & 37 & 81 \\
    \cline{2-4}
    \multicolumn{1}{c}{} & \multicolumn{1}{c}{} & \multicolumn{1}{c}{} & \multicolumn{1}{c}{} & 
    \end{tabular}
    }
    \label{tab:conf_pool_r005}
\end{table}

%\seba{@Christian, at the end of this section we should add the part concerning the second baseline (ordering test cases based on their road length)}
\major{
We assessed the cost-effectiveness of \framework also against a second baseline whose selection strategy is based on the road length.
The assumption is that the longer the road is the more likely it will be unsafe.
In contrast to the random baseline, which selects the tests randomly from the test set, the second baseline orders the tests according to the road length and selects the longest ones.
In Table~\ref{tab:cost-effectiveness}, the cost-effectiveness of \framework is compared to both baselines.
The Random Forest and Logistic models have the best cost-effectiveness compared to both baselines with a selection of $80\%$ unsafe tests.
On the other hand, the SVM and Naive Bayes have a worse selection than both baselines selecting only $40\%$ unsafe tests each, whereas the random and RL baselines select an average $42.6\%$ and $60\%$ unsafe tests, respectively.
}

%        random_forest: ce_sdc_scissor=4.0000, ce_random_baseline=0.7061, ce_rl_baseline=1.5000
%    gradient_boosting: ce_sdc_scissor=1.5000, ce_random_baseline=0.6656, ce_rl_baseline=1.5000
%                  SVM: ce_sdc_scissor=0.6667, ce_random_baseline=0.7861, ce_rl_baseline=1.5000
% gaussian_naive_bayes: ce_sdc_scissor=0.6667, ce_random_baseline=0.7626, ce_rl_baseline=1.5000
%  logistic_regression: ce_sdc_scissor=4.0000, ce_random_baseline=0.7942, ce_rl_baseline=1.5000
%        decision_tree: ce_sdc_scissor=0.4286, ce_random_baseline=0.7369, ce_rl_baseline=1.5000
% 1/9=0.1111	2/8=0.2500	3/7=0.4286	4/6=0.6667	5/5=1.000	6/4=1.5000	7/3=2.3333	8/2=4.0000	9/1=9.0000
\begin{table}
    \centering
    \major{
    \caption{Cost-effectiveness ($\frac{\#failing}{\#passing}$) of \framework against a random baseline and a road length-dependent baseline.}
    \begin{tabularx}{\textwidth}{cXXX} \toprule
        \multirow{2}{*}{\textbf{Model}} & \multicolumn{3}{c}{\textbf{Cost-effectiveness (Percentage of failing tests)}}     \\ \cline{2-4}
                                        & \textbf{\framework}   & \textbf{Random Baseline}  & \textbf{RL Baseline}  \\ \midrule
        Random Forest                   & $4.0$ ($80\%$)        & $0.7419$ ($42.6\%$)       & $1.5$ ($60\%$)        \\
        Gradient Boosting               & $1.5$ ($60\%$)        & $0.7419$ ($42.6\%$)       & $1.5$ ($60\%$)        \\
        SVM                             & $0.6667$ ($40\%$)     & $0.7419$ ($42.6\%$)       & $1.5$ ($60\%$)        \\
        Naive Bayes                     & $0.6667$ ($40\%$)     & $0.7419$ ($42.6\%$)       & $1.5$ ($60\%$)        \\
        Logistic Regression             & $4.0$ ($80\%$)        & $0.7419$ ($42.6\%$)       & $1.5$ ($60\%$)        \\
        Decision Tree                   & $0.4286$ ($30\%$)     & $0.7419$ ($42.6\%$)       & $1.5$ ($60\%$)        \\ \bottomrule
    \end{tabularx}
    \label{tab:cost-effectiveness}
    }
\end{table}

\major{
\begin{finding}
\framework outperforms a baseline approach that selects test cases based on their road length.
The baseline has a cost-effetiveness of $1.5$ whereas the Random Forest and Logistic provide a cost-effectiveness of $4.0$ each.
%\seba{@Christian, complete this finding if needed (or possible)}
\end{finding}
}

\subsubsection{\REACH Experiment}
The goal of this experiment is to investigate whether the usage of ML models allows for reducing the total test execution time. 
By reducing the total test execution costs, a testing pipeline would be able to spend more testing time on more safety-critical test cases.
The task in this experiment was to identify, as early as possible, ten unsafe tests while minimizing the number of total executed test cases. 
To perform the various comparisons, for each experimented strategy, we collected information about the number of test cases required to reach ten unsafe cases as well as the cumulative cost (i.e., the execution time) to run all the test cases (i.e., till the final unsafe scenario was identified). 
Further, we collected information concerning the execution time for both safe and unsafe test cases.
The conjecture behind this analysis is that the testing cost concerning safe cases should as limited much as possible, whereas the test cost dedicated to unsafe cases is beneficial to identify flaws of SDC in virtual environments. 

\begin{figure*}
    \centering
    \includegraphics[keepaspectratio=true,scale=0.6]{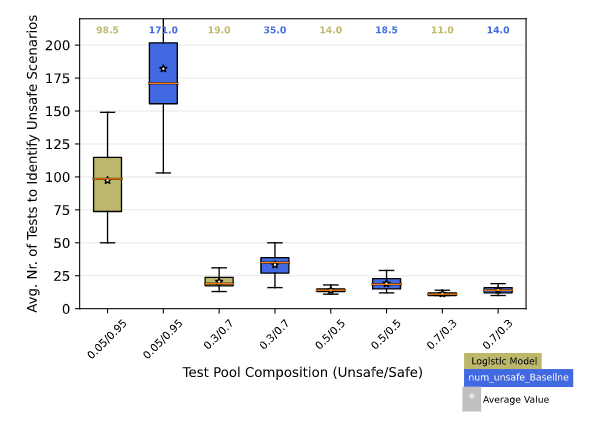}
    \caption[Comparing the Logistic model and baseline]{Comparing the Logistic model with the baseline across the different test pools.}
    \label{fig:test_num_across}
\end{figure*}

\begin{figure*}
    \centering
    \includegraphics[keepaspectratio=true,width=0.8\textwidth]{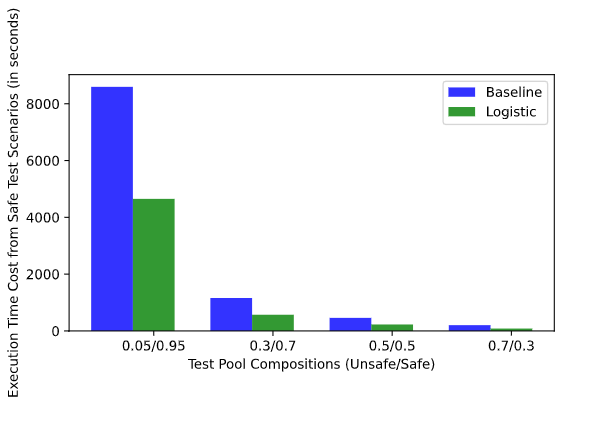}
    \caption[Comparison across different test pools]{Time spent for the execution of safe tests, Logistics vs. Baseline across different test pools 
    %(0.05/0.95 means=8595/4650, 0.3/0.7 means=1159/569, means=459/227, 0.7/0.3 means=203/83).
    }
    \label{fig:test_cost_across}
\end{figure*}

\begin{figure}
    \centering
    \begin{minipage}{0.49\textwidth}
        \centering
        a
        \includegraphics[keepaspectratio=true,width=\textwidth]{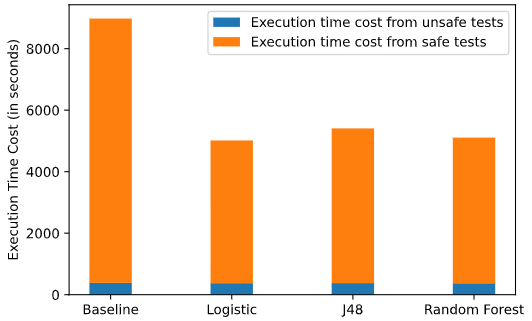}
    \end{minipage}
    \begin{minipage}{0.49\textwidth}
        \centering
        b
        \includegraphics[keepaspectratio=true,width=\textwidth]{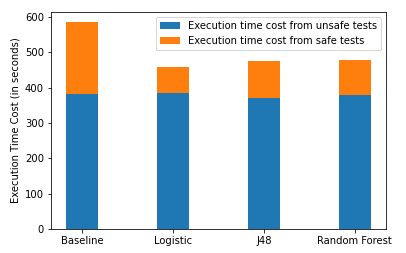}
    \end{minipage}
    \caption{Time spent on executing each safe and unsafe test case for different models in
    a) test pool (0.7/0.3) % cost from safe(baseline mean=203, Logistic mean=73, J48 mean=104, means=104, random forest means=100).
    b) test pool (0.05/0.95) % cost from safe(baseline mean=8595,Logistic mean=4640,J48 mean=5029, random forest mean=4736).
    }
    \label{fig:test_reach_005_pool}

\end{figure}

Figure \ref{fig:test_num_across} and Figure \ref{fig:test_cost_across} provide an overview of the performance of the baseline compared to the Logistic model (the best-performing model in previous experiments) across different test pool compositions.
Table \ref{tab:results_reach} summarizes the results of the \REACH experiment.
We observed that the Logistic model performed better across all test pool compositions. 
The test costs strictly depend on the required numbered of tests to be executed before identifying the minimum set of 10 unsafe tests. Although the difference in the number of required tests tends to be higher in the pool with fewer unsafe tests (in the 0.05/0.95 pool between 171 to 98.5 tests, in the 0.7/0.3 between 14 to 11 tests), \framework allows for reducing test execution time dedicated to less critical tests when the test pool presents more unsafe tests. 
Figures \ref{fig:test_reach_005_pool} show that in the smaller unsafe pool it is higher the test execution time dedicated to less critical tests. The test execution time for these less critical tests is 85\% higher in the baseline than in the Logistic model. 
%\seba{@Christian, fix the sentence below, as for other similar sentences (the reviewer asked it, we had a similar issue in abstract - tentatively fixed by me I think.)}
In the larger pool, the Logistic model selects 80\% unsafe tests, whereas the baselines only have 42.6\% and 60\%, respectively.
%Further, we observe that the cost for executing unsafe test scenarios is consistent among overall models and test pools; this is expected since in all cases the goal is to find 10 unsafe cases, which should have similar run-times.

\major{
\begin{finding}
We investigate whether \framework can reduce the number of executed tests required to find at least N unsafe tests. 
%\seba{@Christian, fix the sentence below, as for other similar sentences (the reviewer asked it, we had a similar issue in abstract - tentatively fixed by me I think.)}
Our results show that \framework outperformed the baselines across all test pools, with the Logistic model reducing the unnecessary execution time dedicated to safe tests by selecting 80\% unsafe tests, whereas the baselines select 42.6\% and 60\% unsafe tests, respectively. 
%Further we have seen that depending on the metric the models performed better compared to baseline in different test pool compositions. 
 \framework performed better compared to the baseline when test pools are characterized by fewer unsafe tests. 
\end{finding}
}

\begin{table*}
\centering
\caption{\minor{Results of the \REACH experiments comparing the Logistic model and the baseline in various test pool compositions (safe/unsafe test ratio). Execution time is reported in seconds, and the values are averaged across the experiment repetitions.}}
% \newcolumntype{P}[1]{>{\centering\arraybackslash}p{#1}}
\begin{tabular}{@{}llll@{}}
\toprule
      \multirow{2}{*}{\textbf{Model/Pool}}& \multirow{2}{*}{\textbf{Tests \#}} & \multicolumn{2}{c}{\textbf{Execution Time}} \\
     & & \textbf{Safe} & \textbf{Unsafe}\\
\midrule
 \textbf{Smart Selector} &  &  &\\
    Test Pool (0.05/0.95) & \textbf{98.5} & 4664 & 375\\ 
    Test Pool (0.3/0.7) & 19 & 475 & 376 \\
    Test Pool (0.5/0.5) & 14 & 214 & 389 \\
    Test Pool (0.7/0.3) & 11 & \textbf{54} &	379\\ 
  \midrule
     \textbf{Baseline} &  &  & \\
    Test Pool (0.05/0.95) & \textbf{171} & 8079 & 382 \\ 
    Test Pool (0.3/0.7) & 35 & 1243 & 383 \\ 
    Test Pool (0.5/0.5) & 18.5 & 439 & 391 \\ 
    Test Pool (0.7/0.3) & 14 &	\textbf{193} &	387 \\ 
  \bottomrule

\end{tabular}
\label{tab:results_reach}
\end{table*}

\major{
In Section \ref{sec:discussion}, we discuss further results of RQ2, providing additional insights on this research question.
}

\minorrevision{
\subsection{Real-Time Experiments (\texorpdfstring{RQ\textsubscript{2}}{RQ2})}
} \label{sec:study-real}

\minorrevision{
%\textbf{\textit{Results of Real-time Experiment}}.
In this section, we
%discuss
present
the results of the real-time experiments, where we compare the results of a pre-trained model and a real-time model with the baseline approach.
%}
}

\begin{figure}
    \centering
    
    \begin{minipage}[t]{0.9\textwidth}
    \centering
    a
    \includegraphics[keepaspectratio=true,width=\textwidth]{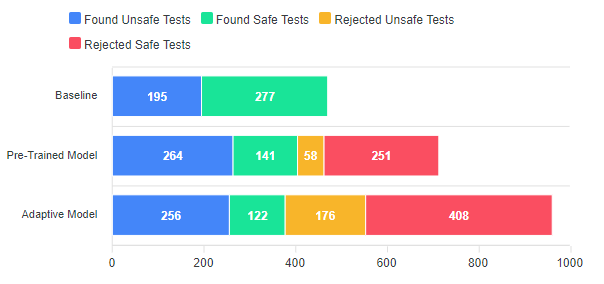}
    \end{minipage}
    \begin{minipage}[t]{0.9\textwidth}
    \centering
    b
    \includegraphics[keepaspectratio=true,width=\textwidth]{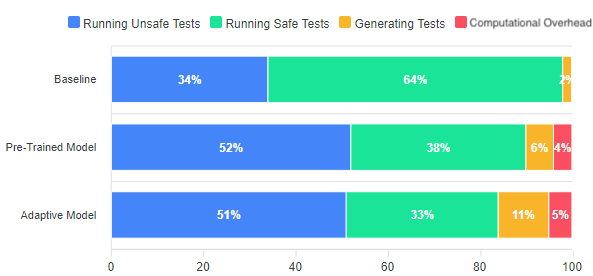}
    \end{minipage}
    \caption{\minorrevision{Comparison of the metrics for different real-time approaches in a 6-hour run
    a) generated test cases distribution. % Baseline (total=472), Pre-trained Model (total=714), real-time Model(total=962)
    b) spent time distribution across different tasks.
    }}
    \label{fig:generated_tests_rt}
    \label{fig:time_allocation}
\end{figure}

\minorrevision{
\textit{Baseline vs. Pre-trained and Adaptive Models.}
Figure \ref{fig:time_allocation} gives an overview of the results achieved by the experimented models.
We observe that the baseline executed a higher number of test cases (472).
The pre-trained model runs more test cases (405) than the real-time approach (378).
Figure \ref{fig:time_allocation} summarizes our main observations, as elaborated in the next paragraphs. 
}

\minorrevision{
The pre-trained and real-time models apply a machine learning-based test selection, which leads to numerous rejected (i.e., non-executed) test cases: real-time and pre-trained experienced 588 and 309 rejected tests, respectively.
The baseline uses 98\% of the time to execute test cases; only 2\% is dedicated to generating test cases. The pre-trained and real-time approaches use more time for test generation (6\% pre-trained, 11\% real-time approach). 
In addition to the longer test generation process, these two approaches allocate time for predictions and evaluation of tests (pre-trained 4\%, real-time 5\%), which the baseline does not need to perform.
Compared to the pre-trained approach, the real-time approach continuously trains the machine learning model with new tests.
}

\minorrevision{
Interestingly, although the baseline executes more test cases, both pre-trained and real-time approaches found more unsafe test cases (baseline 195, pre-trained 265, real-time 256).
The pre-trained model was able to find 35\% more unsafe test cases, executing only 49\% of safe tests.
In Figure \ref{fig:time_allocation}, we can observe that the baseline only spends 34\% of the time running unsafe tests, while 64\% of the test time was spent on executing safe test tests.
In contrast, our proposed approaches dedicated more than 50\% of the time to unsafe tests, which is positive since, in a testing environment, the goal is to find more errors in less time (in our case, it corresponds to exposing more weakness in SDC).
}

\minorrevision{
\begin{finding}
Our results show that even though the baseline approach executes more test cases, both the real-time and the pre-trained (i.e., offline) models integrated into \framework are able to find more unsafe tests than the baseline.
The time investment of predicting the outcome of test cases and generating more tests is beneficial for testing purposes. The pre-trained model was able to find 35\% more unsafe tests than the baseline, with the baseline only dedicating 34\% of the time budget to assessing unsafe tests.
The offline model spends 52\% running unsafe and only 38\% on safe test cases.
\end{finding}
}

\begin{table}
\centering
\minorrevision{
\caption{Comparison between pre-trained and real-time models.}
\begin{tabular}{lccccc}
\toprule
\multirow{2}{*}{\textbf{Model}} & \multirow{2}{*}{\textbf{Acc.}} & \multicolumn{2}{c}{\textbf{Unsafe}} & \multicolumn{2}{c}{\textbf{Safe}} \\
 &  & \textbf{Prec.} & \textbf{Recall} & \textbf{Prec.} & \textbf{Recall} \\ \midrule
    Pre-trained Model & 72.1\% &  65.2\% & 82\%  & 81.2\% & 64\% \\
    Real-time Model & 69\%  & 67.7\% & 59.3\%  & 69.9\% & 77\% \\ \bottomrule                          
\end{tabular}
}
\label{tab:pre_adaptive_compare}
\end{table}

\minorrevision{
\textit{Adaptive vs. Pre-trained Model.}
Figure \ref{fig:time_allocation} shows that the testing time allocation for the pre-trained and real-time models is similar, but the real-time model spends more time on test generation (11\%) than the pre-trained one (6\%).
The pre-trained model is based on the previously generated dataset with
%\alex{Before we mentioned 8000 something, are those different test cases}
5,643 (consisting of 3,559 valid test descriptions as described in Section~\ref{sec:design}) test cases, whereas the real-time model started with generating an initial dataset of 60 test cases as described in Section \ref{sec:design}.
Table \ref{tab:pre_adaptive_compare} shows that the pre-trained model achieved a higher accuracy (72.1\%) than the real-time model (69\%). 
The lower accuracy explains the higher number of test cases generated by the pre-trained model (tests generated; real-time 962, pre-trained 714). 
%The confusion matrices \ref{tab:conf_adaptive}, \ref{tab:conf_pre-trained} shows that the number of false positives is lower for the real-time approach (122), compared to the pre-trained 141 approaches, but the false negative is significantly higher (real-time 176, pre-trained 56). 
%The high number of false-negative is reflected in the unsafe recall, where the pre-trained model is significantly higher (0.82) than the recall of the real-time model (0.593). 
Although the pre-trained model has higher accuracy in general and higher unsafe recall, it only found 3.13\% more unsafe tests than the real-time model.
}

\minorrevision{
\begin{finding}
%We observe whether using a pre-trained and an real-time allow to achieve similar results.
The offline model achieved an accuracy of 72.1\%, which is higher than the real-time model (69\%).
%The higher accuracy is reflected in the number of unsafe test scenarios found (real-time 256, pre-trained 264).
A real-time approach can achieve similar results compared to an offline model, with the real-time model finding only 3.13\% fewer unsafe tests than the offline model. In achieving such results, the real-time model only used an initial set of 60 test cases, whereas the offline model leveraged 5,643 tests. 
\end{finding}
}

\minorrevision{
%\major{
\textbf{Training costs: Pre-trained and Adaptive Models v.s. Random Baseline}.
From a qualitative point of view, the cost of the training dataset is about 0 for the random baseline, while it is $>0$ for the pre-trained and adaptive Models.
It is important to mention that, for all results discussed in Section \ref{sec:study-real} and for the adaptive and pre-trained models, we did not include the cost required for training the ML models on the training data. 
This choice was made since  the cost of training the best ML model can be considered negligible compared to the cumulative cost of generating all tests and executing them. Indeed, the average cost to train the Logistic Regression model (i.e., the best ML model) on 
%\seba{@christian, can you estimate the time required to train Logistic on 60 test cases and 5,643 test cases. About this, just compute the training not the classification process. Consider that here we are not talking about 10-fold, but simple training on 60 test cases and 5,643 test cases.}
60 test cases is of about 0.139 seconds, whereas the cost to train the same ML model on 5,643 tests (for the offline model) is of about 0.685 seconds.
However, since for other ML models or particular settings of the same ML model (e.g., different from its standard configuration), we could achieve rather higher training costs, we discuss this topic in the threat to validity.
}

\minorrevision{
%\seba{@Christian, check and complete the following..}
\textbf{Training dataset preparation: Pre-trained and Adaptive Models v.s. Random Baseline}.
It is important to report that the comparison of \framework and the random baseline does not take into account the time (i.e., the cost) required for the training dataset preparation in the real-time experiments. 
From a qualitative point of view, the cost of the preparation of the training data is about 0 for the random baseline (since no training is needed), while for the pre-trained and adaptive models, this has a non-negligible cost. 
The preparation of the training data includes: (i) the time required for the design, implementation, and testing of the road characteristics (i.e., one week of full-time work) into \framework; (ii) and the cost for the automated extraction of such features from all test cases 
%\seba{@Christian, can you complete the following number..} 
(158 seconds). 
In total, this required us (i.e., to the first author of this work around one week of work). 
Hence, while  both the pre-trained and adaptive models are more cost-effective than a random baseline when selecting test cases, the training data preparation cost represents a very high cost to be sustained upfront, which becomes beneficial only over a long period of test execution time. 
%}
}

\begin{table*}[H]
    \caption{Pre-trained Model, Confusion Matrix}
    \parbox{.45\linewidth}{
    \begin{tabular}{l|l|c|c|c}
    \multicolumn{2}{c}{}&\multicolumn{2}{c}{Predicted class}&\\
    \cline{3-4}
    \multicolumn{2}{c|}{}&Unsafe&Safe&\multicolumn{1}{c}{}\\
    \cline{2-4}
    \multirow{2}{*}{Actual Class}& Unsafe & 264 & 58\\
    \cline{2-4}
    & Safe & 141 & 251 \\
    \cline{2-4}
    \multicolumn{1}{c}{} & \multicolumn{1}{c}{} & \multicolumn{1}{c}{} & \multicolumn{1}{c}{} & 
    \end{tabular}
    \label{tab:conf_pre-trained}
    }
    \caption{real-time Model, Confusion Matrix}
    \parbox{.45\linewidth}{
    \begin{tabular}{l|l|c|c|c}
    \multicolumn{2}{c}{}&\multicolumn{2}{c}{Predicted class}&\\
    \cline{3-4}
    \multicolumn{2}{c|}{}&Unsafe&Safe&\multicolumn{1}{c}{}\\
    \cline{2-4}
    \multirow{2}{*}{Actual Class}& Unsafe & 256 & 176\\
    \cline{2-4}
    & Safe & 122 & 403 \\
    \cline{2-4}
    \multicolumn{1}{c}{} & \multicolumn{1}{c}{} & \multicolumn{1}{c}{} & \multicolumn{1}{c}{} & 
    \end{tabular}
    \label{tab:conf_adaptive}
    }
\end{table*}

\begin{figure}
\centering
\begin{minipage}{.5\textwidth}
  \centering
  \includegraphics[width=\linewidth]{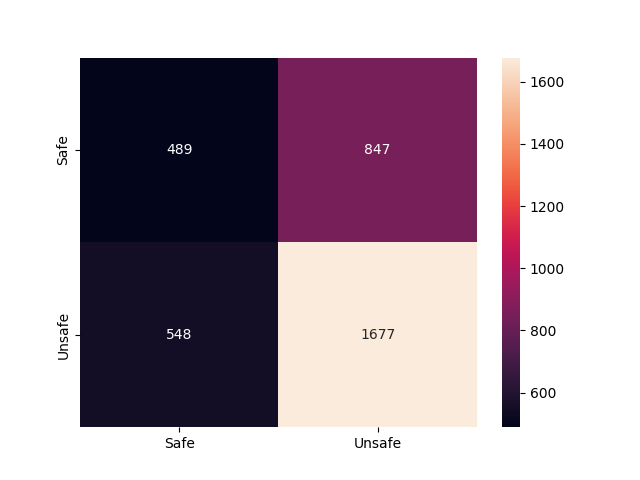}
  \caption{Confusion matrix for the Gaussian Naive Bayes model.}
  \label{fig:cm-random-forest}
\end{minipage}%
\begin{minipage}{.5\textwidth}
  \centering
  \includegraphics[width=\linewidth]{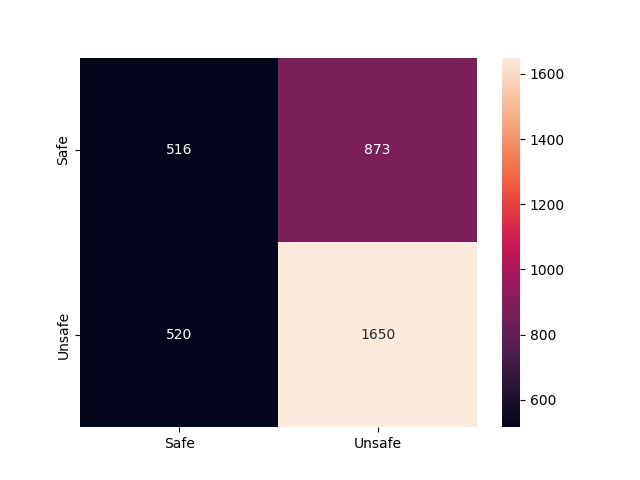}
  \caption{Confusion matrix for the J48 Decision Tree model.}
  \label{fig:cm-j48}
\end{minipage}
\end{figure}

\major{
\subsection{Optimization Experiments (\texorpdfstring{RQ\textsubscript{3}}{RQ3})}
\label{sec:results-rq3}
In RQ3, we focus on investigating whether there is an actual upper bound of ML techniques in identifying SDC safe and unsafe test cases when using static SDC features (available before executing the tests).
%\seba{hence, @Christian, here if results are not so great, after so many optimizations (see work from drone issues as reference for describing results of grid search), we outline that these types of features have probably an upper bound of what they can actually predict, in section discussion you could elaborate what types of metrics could be designed for future work (use the paper from sajad). however, before this, try to identify the best model looking at the confusion matrix on FAIL. If the best model in this regard is J48, we could actually show and discuss the tree (see work from wontfix paper).}
We performed Grid Search for the Random Forest, J48, Gradient Boosting, Logistic, Naive-Bayes, and Support Vector Classifier to identify the best hyper-parameters for each model.
Table~\ref{tab:grid-search} summarizes the results of Grid Search by showing the F-score (F$_1$) for safe and unsafe test cases as well as the averaged F-score.

The best two models regarding the averaged F-score are the Gaussian Naive Bayes (F$_1=60.0\%$) and the J48 Decision Tree classifiers (F$_1=59.5\%$).
Although these two models have similar averaged F-scores, they are distinct among the classes.
Among the unsafe tests, the J48 Decision Tree achieved an F-score of  $70.3\%$, but for the safe tests, it achieved $42.6\%$.
In the case of the Naive Bayes model, we have among the unsafe tests an F-score of $41.0\%$ and $71.0\%$ for the safe tests.

\begin{table}
\major{
\centering
\caption{Best ML models with recall, precision, and F-score.}
\label{tab:best-ml-models-scores}
%\resizebox{0.75\linewidth}{!}{
\begin{tabularx}{\textwidth}{lcccccc}
	\toprule
	\multirow{2}{*}{\textbf{ML Technique}}  & \multicolumn{2}{c}{\textbf{Precision}} & \multicolumn{2}{c}{\textbf{Recall}}  & \multicolumn{2}{c}{\textbf{F$_1$}} \\ \cline{2-7}
	                                        & Safe & Unsafe & Safe & Unsafe & Safe & Unsafe \\ \midrule
	J48                                     & $49.8\%$ & $65.4\%$ & $76.0\%$ & $37.1\%$ & $42.6\%$ & $70.3\%$ \\
	Naive Bayes                             & $66.0\%$ & $47.0\%$ & $75.0\%$ & $37.0\%$ & $71.0\%$ & $41.0\%$ \\ \bottomrule
\end{tabularx}
}
\end{table}

For the best two models according to their averaged F-score, we show their corresponding confusion matrices in Figure~\ref{fig:cm-random-forest} and Figure~\ref{fig:cm-j48}.
Furthermore, a detailed overview of their precision, recall, and F-scores among the classes are reported in Table~\ref{tab:best-ml-models-scores}.

%\seba{@Christian, before the following I would add a table reporting, only for the best models (RF and J48) all values of precision, reall and F-score for all labels. Hence a specific table only on these to models. You will remove column "Param. Config.", since it is already reported in table 10. This is important, since looking mainly on the F-score it is difficult to grasp the low-level results.. }
Both confusion matrices show a similar distribution.
The models identify most of the true unsafe test scenarios with 1'677 and 1'650 cases, but in predicting the safe tests, the models have a low true positive rate with 409 and 516 correct predicted safe tests.

%In the case of the J48 Decision Tree model, the corresponding tree is illustrated in Figure~\ref{fig:j48-pruned-tree}.
%The tree shows that the model makes its predictions based only on the \textit{Road Distance} feature.
%The tree of the J48 classifier is small and has two leaves and has a size of three.
%In general, this classifier has a simple structure and the way how it classifies the test scenarios is based only on the \textit{Road Distance} feature.

\major{
\begin{finding}
%\seba{@Christian, add a finding on this RQ3...}
The Naive Bayes and J48 models reach a weighted average F$_1$ of 60\%.
Their confusion matrices show that they are able to predict unsafe tests, whereas their performance on detecting safe tests is less accurate.
\end{finding}
}

}

\begin{table}
\major{
\centering
\caption{Best ML model configurations after a Grid search}
\label{tab:grid-search}
%\resizebox{0.75\linewidth}{!}{
\begin{tabularx}{\textwidth}{lXccc}
	\toprule
	\multirow{2}{*}{\textbf{ML Technique}}  & \multirow{2}{*}{\textbf{Param. Config.}}  & \multicolumn{2}{c}{\textbf{F$_1$}}                      & \multirow{2}{*}{\textbf{Weighted avg. F$_1$}} \\ \cline{3-4}
	                                        &                                           & \textbf{Safe}             & \textbf{Unsafe}           & \\ \midrule
    \multirow{4}{*}{Random Forest}          & I=5,                                      & \multirow{4}{*}{35.1\%}   & \multirow{4}{*}{72.4\%}   & \multirow{4}{*}{57.8\%} \\
                                            & K=10,                                     &                           &                           & \\
                                            & depth=10,                                 &                           &                           & \\
                                            & M=50                                      &                           &                           & \\ \midrule
    \multirow{2}{*}{J48}                    & C=0.5,                                    & \multirow{2}{*}{42.6\%}   & \multirow{2}{*}{70.3\%}   & \multirow{2}{*}{\textbf{59.5\%}} \\
                                            & M=20                                      &                           &                           & \\ \midrule
    \multirow{4}{*}{Gradient Boosting}      & criterion=friedman\_mse,                  & \multirow{4}{*}{77.0\%}   & \multirow{4}{*}{0.0\%}    & \multirow{4}{*}{48.0\%} \\
                                            & learning\_rate=0.01,                      &                           &                           & \\
                                            & loss=log\_loss,                           &                           &                           & \\
                                            & n\_estimators=10                          &                           &                           & \\ \midrule
    \multirow{4}{*}{Logistic}               & dual=False,                               & \multirow{4}{*}{76.0\%}   & \multirow{4}{*}{12.0\%}   & \multirow{4}{*}{52.0\%} \\
                                            & max\_iter=10,                             &                           &                           & \\
                                            & penalty=none,                             &                           &                           & \\
                                            & solver=saga                               &                           &                           & \\ \midrule
    Naive-Bayes                             & No parameters                             & 71.0\%                    & 41.0\%                    & \textbf{60.0\%} \\ \midrule
    \multirow{3}{*}{SVC}                    & dual=False,                               & \multirow{3}{*}{76.0\%}   & \multirow{3}{*}{28.0\%}   & \multirow{3}{*}{58.0\%} \\
                                            & loss=squared\_hinge,                      &                           &                           & \\
                                            & penalty=l2                                &                           &                           & \\ \bottomrule
\end{tabularx}
}
\end{table}

%\begin{figure}
%    \centering
%    \includegraphics[width=0.6\textwidth]{imgs/j48-pruned-tree.png}
%    \caption{J48 pruned tree.}
%    \label{fig:j48-pruned-tree}
%\end{figure}

\major{
\section{Integration of \framework in the Industrial Use Case}
\label{sec:integration}

\major{
\subsection{Experiments involving an Industrial use Case (AICAS)} 
\label{sec:use-case}
}

\major{
% \textbf{The AICAS Automotive Use Case}.
%\& its Signal-based Protocol} % Experimental Settings?
% As mentioned in previous sections, we focus (in 
We investigate the extent to which \framework can be integrated into the context of industrial organizations in the automotive domain, addressing one of the open questions in simulation-based testing~\cite{3533818,Birchler2022Cost,Gambi2019,abdessalem2018testing} for SDCs.
%
% This investigation is needed since an open question in the context of recent research on simulation-based testing~\cite{3533818,Birchler2022Cost,Gambi2019,abdessalem2018testing} for SDCs, is: \textit{
% To what extent can simulation-based testing be integrated in an industrial organization in the automotive domain?}. 
%
We identified the AICAS company\footnote{\major{https://www.aicas.com/wp/}} as an ideal use case for this investigation. 
%
% Specifically, 
AICAS develops JamaicaCAR, an OSGi-based technology for the automotive sector, currently running in more than five million cars worldwide.
% offering a powerful application framework for complex, interactive applications in Car Head Units and in-vehicle infotainment systems. JamaicaCAR, with its FCA UConnect Infotainment technology, used in the IIC Security testbed, is running in more than five million of cars worldwide.
% Delivering applications via this platform is complex due to the heterogeneity that arises in this context—both in terms of the hardware that the system must run on and the diversity of application requirements.
%This use case will focus on delivery of sample applications undergoing the V&V process within
%the OSGi development pipeline to the JamaicaCAR platform.
A pressing challenge for AICAS concerns the need to combine simulations and HiL testing protocols to optimize the testing costs. Specifically, AICAS aims to reduce testing costs by automatically generating inputs, i.e., signals, 
compatible with the Controller Area Network (CAN) Bus protocol\cite{canbus-history} in simulated environments.

\begin{figure}[t]
    \centering
    \includegraphics[width=0.9\textwidth]{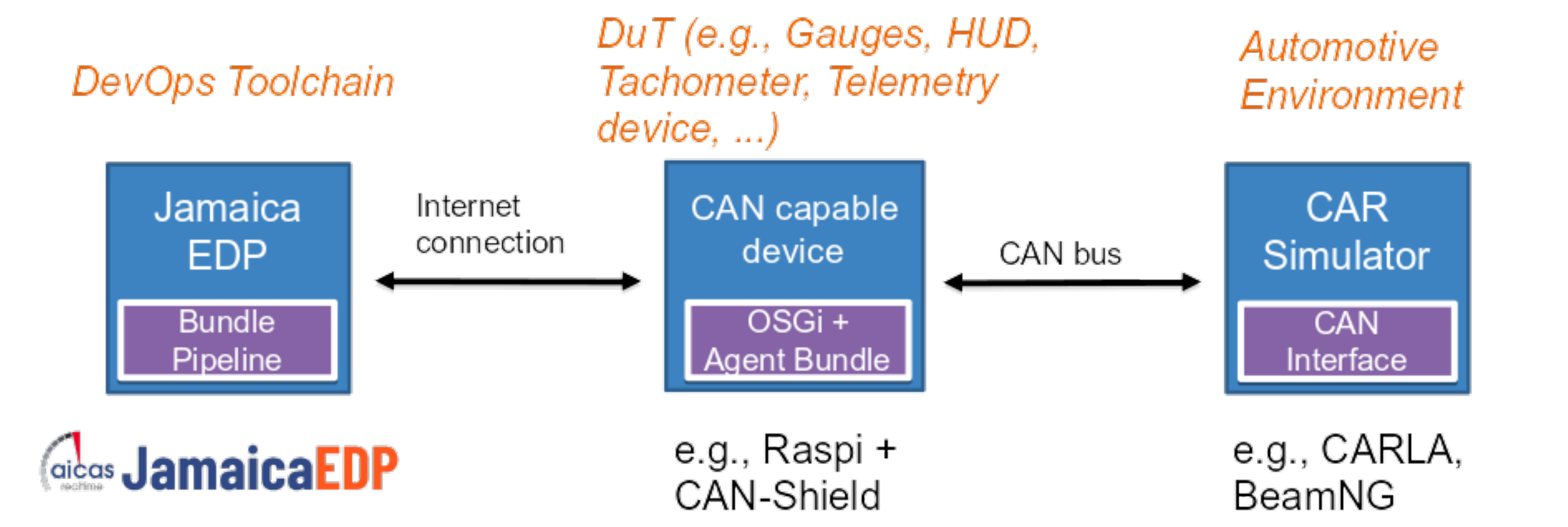}
    \caption{AICAS's Jamaica EDP validation setup.}
    \label{fig:canbus}
\end{figure}

\major{
\begin{figure}[t]
    \centering
    \includegraphics[width=0.9\textwidth]{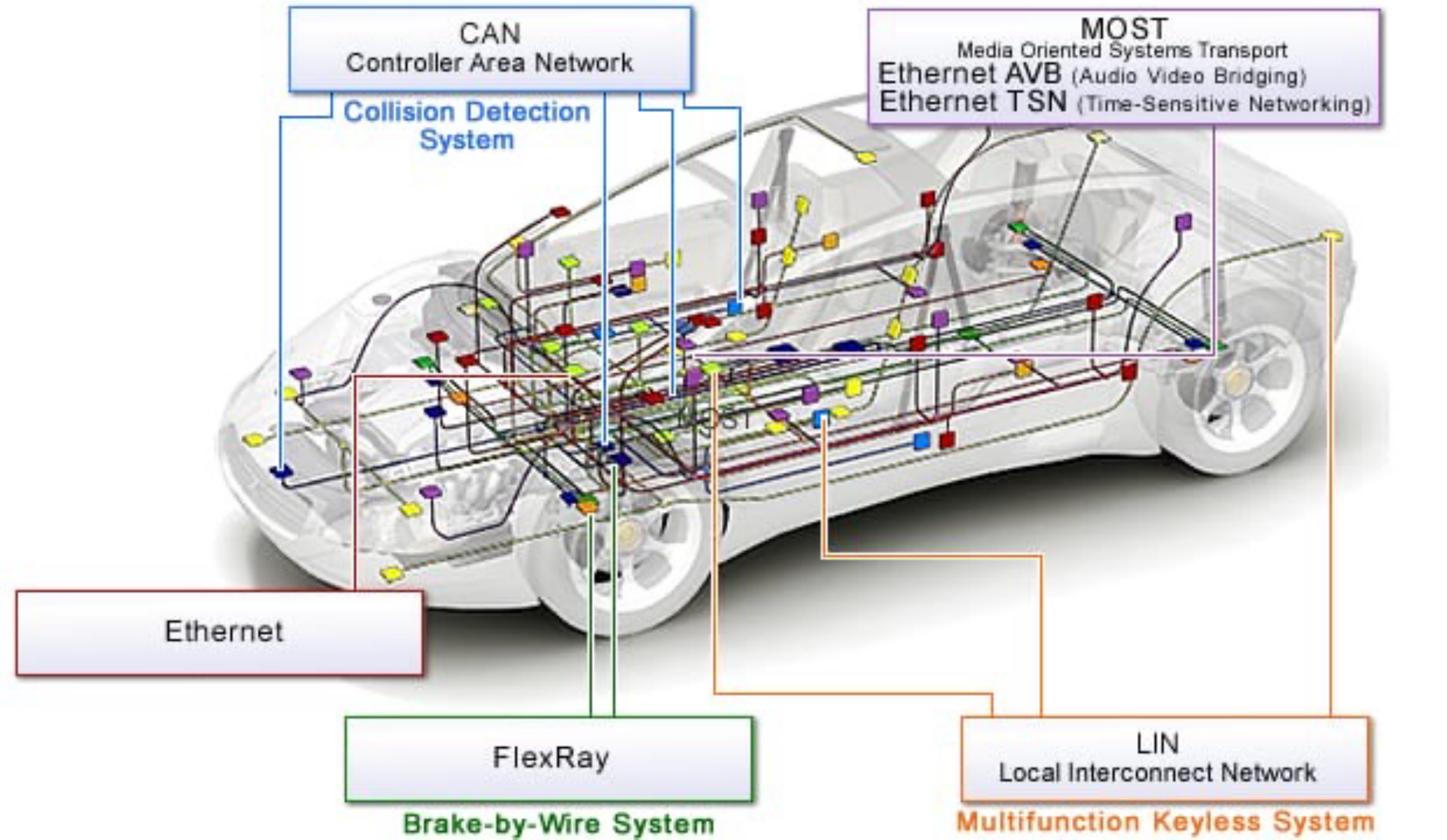}
    \caption{Can Bus in the context of an SDC.}
    \label{fig:canbus-SDC}
\end{figure}
}
}

\major{
% \textbf{The Can Bus protocol in the AICAS use case}. 
Based on the trajectories planned by the planning module of SDC, the control module of SDC typically takes charge of the longitudinal and lateral control of the vehicle and 
% By using control algorithms (e.g., proportional integral derivative (PID) control \cite{X9} and model predictive control (MPC) \cite{X10}), this module 
generates appropriate control commands (e.g., steering, acceleration, brake) that it sends to the related hardware component of the SDC via the CAN Bus 
% \cite{canbus-history,8923315,GunduM22} 
(see Figure \ref{fig:canbus-SDC}). 
% Specifically, in this context, the CAN Bus module consists is a message-based protocol which enables ECUs (Electronic Control Units) of automotive systems, as well as other devices, reliable and well-pririotized communication with each other. 
% CAN is supported by a rich set of international standards under ISO 11898, which makes this module critical for several functionalities provided by SDCs, including Lane Keeping Assistance (LKA), Adaptive Cruise Control (ACC), and Automatic Emergency Braking (AEB).
}
\major{
\begin{figure}[t]
    \label{fig:canbuspipeline}
    \centering
    \includegraphics[width=0.98\textwidth]{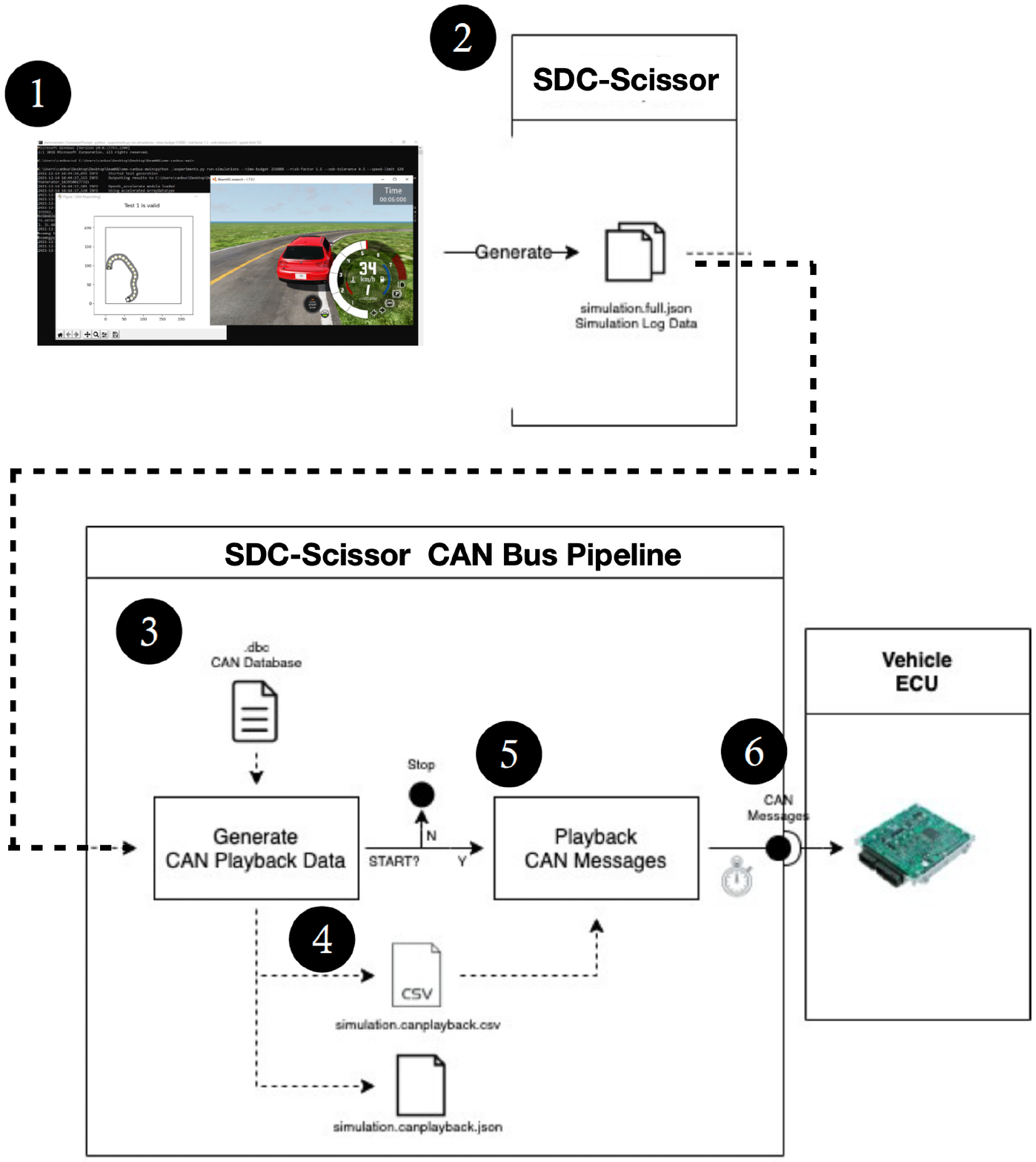}
    \vspace{-14mm}
\caption{CAN Bus code pipeline integrated into  \framework}
\end{figure}
}

\major{
\begin{figure}[t]
    \centering
    \includegraphics[width=0.98\textwidth]{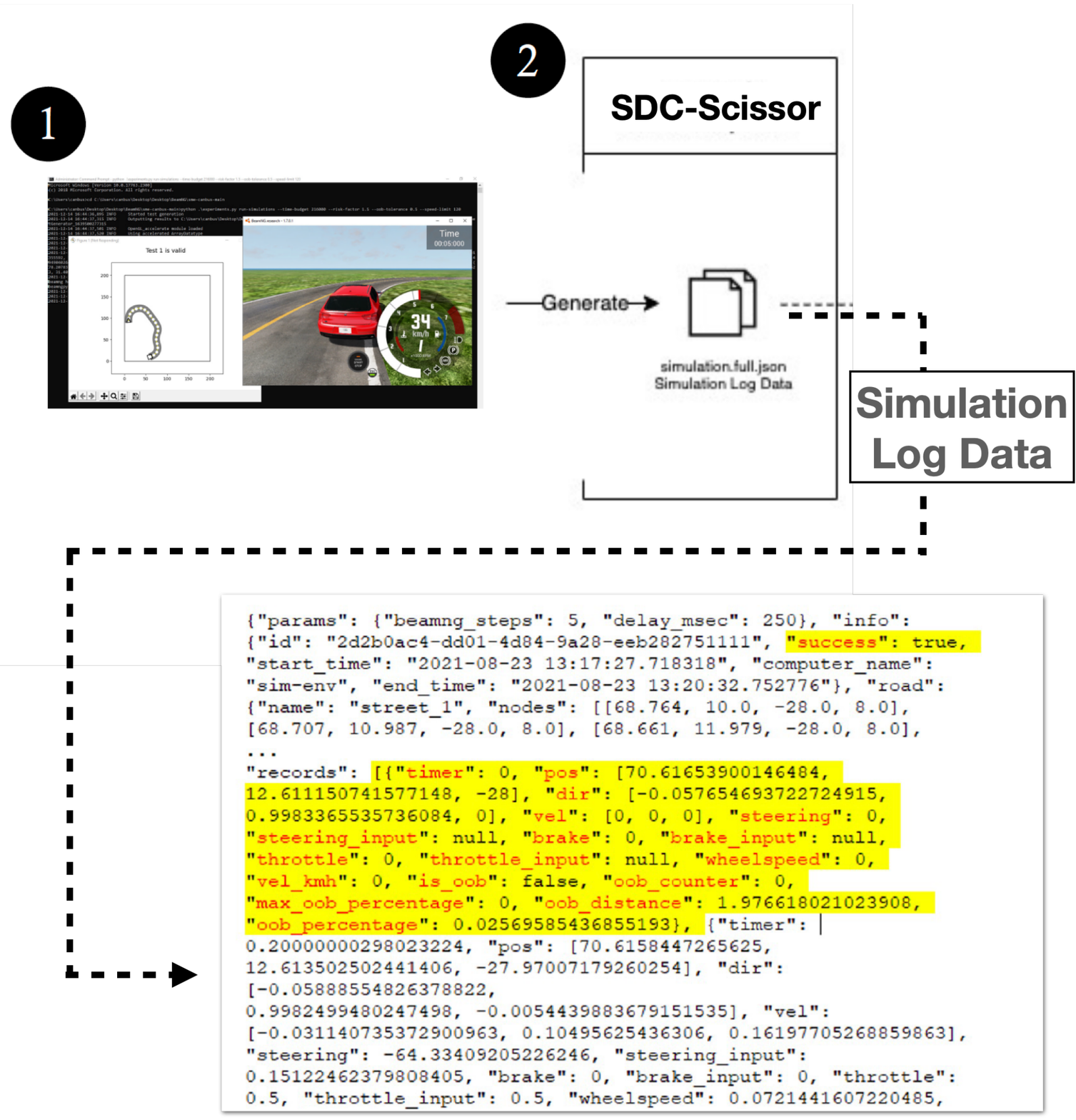}
    \vspace{-14mm}
    \caption{\framework's CAN Bus code pipeline: SDC Test Case Generation and Storage.}
        \label{fig:canbus-test-generation}
\end{figure}
}

\major{
\begin{figure}[t]
    \centering
    \includegraphics[width=0.98\textwidth]{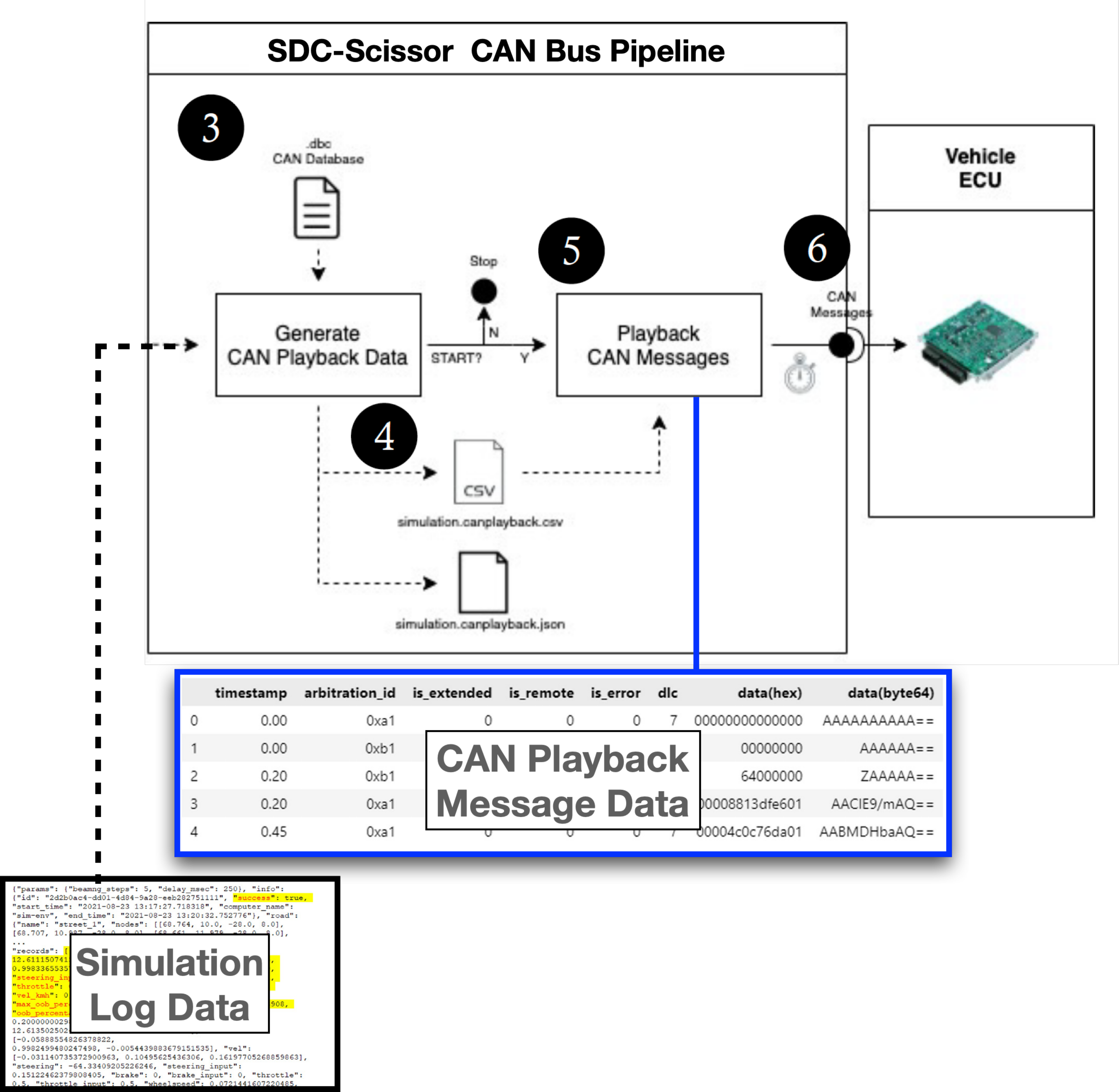}
    %\vspace{-4mm}
    \caption{\framework's CAN Bus code pipeline: SDC Test Data Conversion \& Generation of CAN Playback Data.}
    \label{fig:canbus-conversion}
\end{figure}
}

\major{
% \textbf{Validation Setups}. 
To allow validation of the described scenarios, AICAS provided us with 
% a setup that extends Jamaica CAR by a number of CAN bus equipped 
devices under test (DuT) equipped to communicate via the CAN Bus. 
%
% We connected
% The devices can be accessed by Jamaica CAR and is 
We connected the devices to the CAN bus and the CAN bus to a driving simulator that allowed us to generate the appropriate signals (see Figure \ref{fig:canbus}).
The devices act as a validation context for the described automotive scenarios.
% As it is possible to see from Figure \ref{fig:canbus} compatible signals for the Can Bus protocol can be potentially generated by simulation environments, including CARLA \cite{Dosovitskiy17} and BeamNG.

There are several main advantages of integrating test cases generated by \framework in the testing workflow of AICAS:
\begin{itemize}
    \item \textit{Increased level of test automation}: Currently, AICAS inputs are manually generated or designed by testers and developers in its organization. The usage of an integrated framework such as \framework can enable the generation of test cases automatically, increasing automation and diversity of generated SDC scenarios.
    %\alex{One might argue this is thanks to BeamNG, not SDC Scissor}
    \item \textit{Increased level of realism}:  Most of the manually entered signals inserted in the Can Bus protocol by the testers and developers of the AICAS organization do not reflect a real driving set of signals (e.g., the provided acceleration and steering angle of the vehicle are not reflecting a real driving test scenario, which makes the used inputs in most cases too random or unrealistic).
   % \alex{Same as before. One might argue this is thanks to BeamNG, not SDC Scissor}
\end{itemize}
}

\major{
\textbf{Integration steps}.
To investigate the extent to which \framework can be integrated into the context of AICAS, we extended \framework with a CAN Bus code pipeline (see the full pipeline in Figure \ref{fig:canbuspipeline}), which automates the following steps:
\begin{itemize}
    \item \textit{SDC Test Case Generation and Storage (Steps 1-2)}: As visualized in Figure \ref{fig:canbus-test-generation}, we first use  \framework to generate 3,559 SDC test cases (with BeamNG, with RF 1.5 - moderate driving), execute them, and store the corresponding execution log in a JSON file (i.e., the actual \emph{simulation.full.json} containing all information concerning the generated and executed tests by \framework, see Figure \ref{fig:canbus-test-generation}), which constitutes the dataset of our experiments.
    \item \textit{SDC Test Data Conversion \& Generation of CAN Playback Data  (Steps 3-5)}:  In this stage, we   
    convert (and visualized in Figure \ref{fig:canbus-conversion}) the execution log from the JSON file (i.e., \emph{simulation.full.json} generated by \framework to CAN Playback Data (i.e., the file \emph{simulation.canplayback.*}).
 \item \textit{Transmission of CAN-based Signals  (Steps 6)}: The messages (i.e., the CAN Playback Data) generated in the previous step are then transmitted to the CAN Device according to defined timestamps, consistent with the one generated by \framework while executing SDC test cases. Specifically, referring to the specified used CAN database (i.e.,  \textit{<.dbc>}), we converted  \framework test case data (i.e.,  \textit{<simulation.file.json>}) to CAN messages (i.e.,  \textit{<simulation.canplayback.csv>}).
Using a specified CAN interface device, logged CAN frames are played back to external CAN bus devices.
 These final steps allow us to finally send realistic SDC signals concerning the driving scenarios to the CAN Device (i.e., SDC test cases generated by \framework) in an automated fashion).  
\end{itemize}

From a technological point of view, the definition and implementation of the pipeline in Figure \ref{fig:canbuspipeline} required us to leverage the following libraries: (i) \textit{Python-CAN}\footnote{\major{https://python-can.readthedocs.io/en/master/}}, which allows controlling various CAN interface devices in the Python environment; (ii) the \textit{cantools}\footnote{\major{https://cantools.readthedocs.io/en/latest/}}, which support CAN database encoding and decoding actions (from the device to the Simulator, and vice versa).
%AICAS \footnote{https://www.aicas.com/wp/}
%Specifically, to perform such investigation, we generate new test cases to asses the ability of \framework to generate signals compatible with the Can Bus protocol of the AICAS organization 
%In Section \ref{sec:results}, we discuss the  results of the study conducted to answer RQ4.
} 

\major{
\subsection{Industrial Use Case (AICAS): Integration results } 
\label{sec:use-case-integration}
}

\major{
To investigate the extent to which \framework can be integrated into the context of AICAS, we extended \framework with a CAN Bus code pipeline described in Section \ref{sec:use-case} and shown in Figure \ref{fig:canbuspipeline}.
The development and integration of this pipeline in the AICAS context required around five months of work: considering the time to design the pipeline till its implementation and integration, including the time for running all the required experiments reported in this article (this includes the generations of test cases by \framework, their execution, the analysis of the data, etc.).

Table \ref{table:datasets-use-case} reports the details of the test cases generated by \framework.
Specifically, we generated around 3,600 test cases, which required a total execution time of 12h, 17m, and 11s, with an average simulation time of 12.428 seconds for each test case and a max. observed simulation time of 21.4 seconds.

\major{
\begin{table}
\centering
\caption{Dataset Summary}
\label{table:datasets-use-case}
\major{
\begin{tabular}{p{5cm} c}
\toprule
% Header 2
\multicolumn{1}{c}{\textbf{Property}} &
\multicolumn{1}{c}{\textbf{Value}} \\
\midrule
\textit{Nr. SDC test cases Generated by \framework} (BeamNG RF 1.5)  & 3,559\\
\midrule
\textit{Total Simulation Time}  & 12h 17m and 11s\\
\midrule
\textit{Average Simulation Time}  & 12.428 s\\
\midrule
\textit{Max. Simulation Time}  & 21.4 s  \\
\bottomrule
\end{tabular}
}
\end{table}
}

\major{
The most challenging steps of the  integration of \framework into the context of AICAS are represented by the \textit{SDC Test Data Conversion \& Generation of CAN Playback Data} (Steps 3-5, shown in Figure \ref{fig:canbuspipeline}) and the \textit{Transmission of CAN-based Signals  (Steps 6, shown in Figure \ref{fig:canbuspipeline})}.
The main aspect that made this task challenging was the need for signal conversions and \textit{mapping} between \framework's signals and CAN Playback Data. As shown in Figure \ref{fig:mapping}, for each signal generated by \framework, we had to generate a corresponding value mapped with the CAN Playback module. 

\minorrevision{
Based on the simulation-based signals generated by the implemented \framework pipeline, we were able to 
generate appropriate control commands (e.g., steering, acceleration, brake), and send them to the related hardware component of the SDC via the CAN Bus.
}
Table \ref{table:integration} reports the details of \framework's integration process.
Specifically, for all  3,600  generated test cases, which required a total execution time of 12h, 17m, and 11s, it required a total of 52.391 seconds to \framework for enabling the automated signal conversions,  \textit{mapping}, and  transmission of CAN messages. 

As visualized in Figure \ref{fig:conversion}, it requires 14.721 ms on average to \framework to translate simulation-based signals into CAN-compatible signals. 
In comparison with the current manual signal generation process, it requires on average 1-2 days for AICAS developers and testers to design and then generate a sequence of CAN signals corresponding to 10-15 test cases generated by \framework (according to the qualitative assessment of our main contact people within AICAS). In addition to the test automation enabled by \framework in the context of AICAS, the generation of  a more realistic sequence of SDC signals (corresponding to signals of a realistic SDC car driving in a virtual test case) is vital for the identification of safety-critical scenarios to be executed and tested via the CAN Bus protocol.
}

\major{
\begin{finding}
The main challenging aspect of integrating \framework into the context of AICAS was the need to systematically automate the \textit{mapping} between \framework's signals and CAN Playback Data.
Considering 3,600 generated test cases as input, it required a total of 52.391 seconds to \framework for enabling the automated signal conversions, consisting of about 14.7 ms on average to \framework to translate simulation-based signals in CAN-compatible signals of a single test case. The corresponding process of manually designing and generating the sequence of CAN signals corresponding to 10-15 test cases generated by \framework takes between 1-2 days for AICAS developers and testers.
\end{finding}
}

\major{
\begin{figure}[t]
    \centering
    \includegraphics[width=0.9\textwidth]{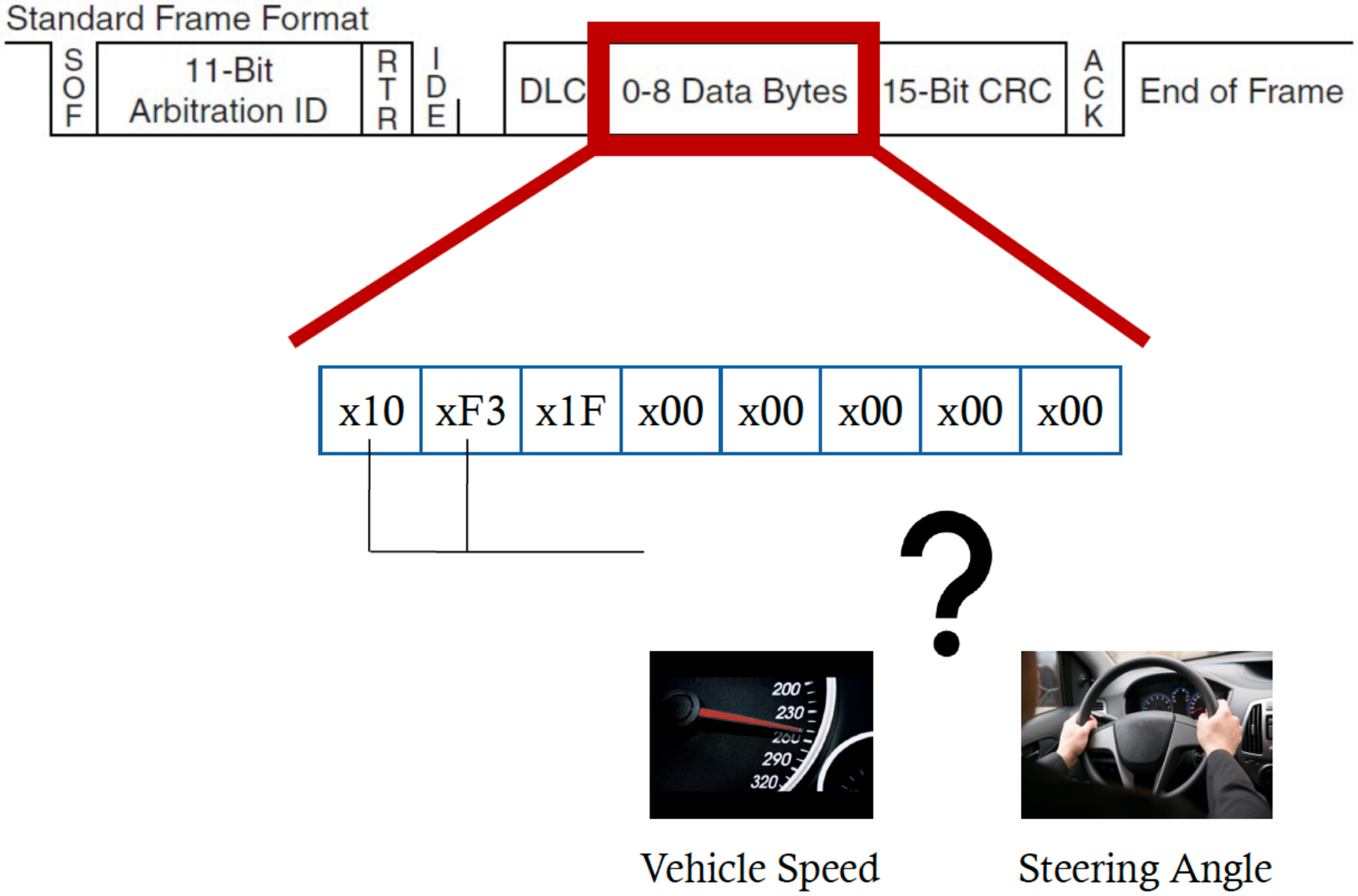}
    \major{\caption{Mapping between \framework's signals and CAN Playback Data}}
    \label{fig:mapping}
\end{figure}
}

\major{
\begin{table}
\centering
\caption{Results of the Integration Process}
\label{table:integration}
\major{
\begin{tabular}{p{7cm} c}
\toprule
% Header 2
\multicolumn{1}{c}{\textbf{Property}} &
\multicolumn{1}{c}{\textbf{Value}} \\
\midrule
\textit{Nr. SDC test cases Generated by \framework} (BeamNG RF 1.5)  & 3,559\\
\midrule
\textit{Total Conversion of Messages + Transmission of CAN signals}  & 52.391 s\\
\midrule
\textit{Mean Time for Conversion of Messages +  Transmission of CAN signals (per each SDC test case)}  & 14.721 ms\\
\midrule
\textit{Min Time for Conversion of Messages +  Transmission of CAN signals (per each SDC test case)}  & 7.892 ms  \\
\bottomrule
\textit{Min Time for Conversion of Messages +  Transmission of CAN signals (per each SDC test case)}  & 30.006 ms  \\
\bottomrule
\end{tabular}
}
\end{table}
}

\begin{figure}[t]
    \centering
    \includegraphics[width=0.99\textwidth]{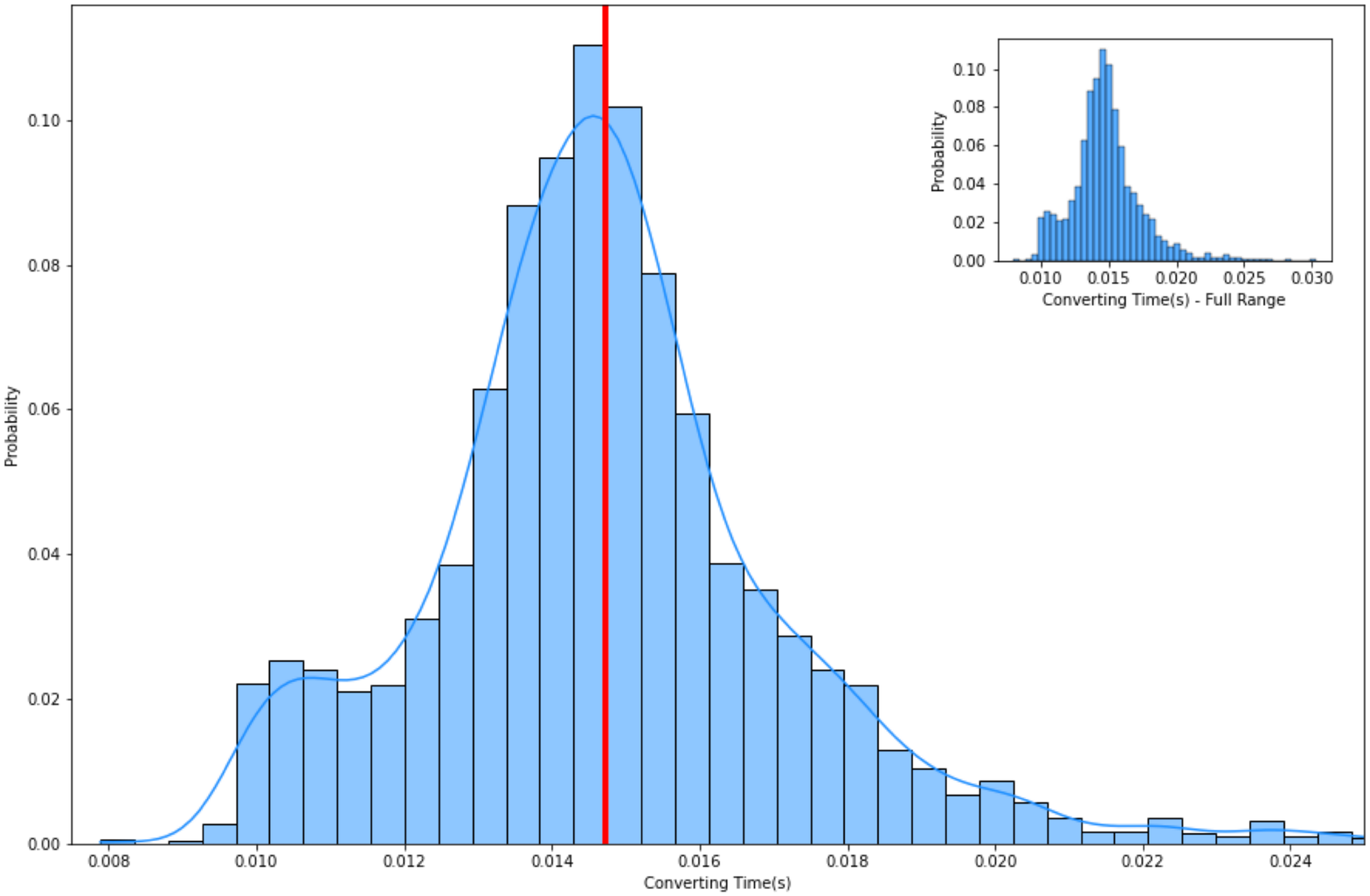}
    \major{\caption{Performance of Conversion and Transmission time.}}
    \label{fig:conversion}
\end{figure}

}
}

\section{Discussion}
\label{sec:discussion}
\major{This section discusses additional factors that can influence the results of the various research questions, providing more insights and findings about them.
Moreover, it also provides a concrete discussion on directions for future research in the field. \\
}

%RQ1 MAJOR TAKE AWAY MESSAGES

 \major{
 \subsection{Discussion of Experiments Using  \textit{Road Characteristics} as Input Features to the ML Models} % (RQ$_1$ and RQ$_2$) }
 \label{sec:discussion_rq1}
 }

 \major{
 As we have observed from the conducted experiments in RQ$_1$, \framework is able to classify safe and unsafe test cases in both the Driving.AI dataset and the BeamNG.AI dataset, with the Logistic  and Random Forest models achieving the most reliable results in terms of F-score values for labels. 
 Moreover, we also observed that the \emph{Road Characteristics} extracted by \framework contribute differently to identifying the safe and unsafe test cases. The \emph{Road Characteristics} concerning the pivot radius (min, mean, std, median), the sum of the turn angles, the number of left and right turns, and the total length of the road are among the most important features, which are all belonging to the set of road features.

 In the context of RQ$_1$, there are other factors that can impact the results of \framework, such as (i) the risk factor (RF) of the SDCs; (ii) the ability of the ML models to transfer knowledge from a driving agent to another one (i.e., between BeamNG RF 1.5 dataset and the Driver.AI dataset); finally, (iii) we complement the previous \textit{Offline Experiments}, which focus on applying \framework to regression test case selection, with \textit{Real-Time Experiments} in which we study the application of \framework to automated test generation.\\
 }

\major{
\subsection{Further Remarks and Future Directions}
 \major{This work can have relevant implications for developers and researchers. Hence, this final discussion reflects further remarks on the results of all questions, with a specific focus on  future directions of RQ$_3$ and RQ$_4$ for  developers and researchers.
} 

 \major{For what concerns \textbf{developers}, the designed tool allows identifying specific problems that need to be carefully monitored in simulation environments at the time of testing. These include, for instance, the need for coping with testing multiple hardware versions and diversified test inputs to verify correctness with realistic test inputs. Also, it is of paramount importance to be able to generate inputs that lead to a different safety-critical situation in a safe manner (i.e., without harming humans).  \framework allows to generate and identify test cases that can cause the SDC to fail by using different safety criteria (in the context of this work, we focus on the line-keeping feature as the main safety criterion, but further criteria can be easily integrated and tested). 
}
 
 \major{The integration of  \framework into the AICAS use case allows us to demonstrate that the proposed approach can automate the testing process of such a large automotive company, coping with the need to complement their hardware-based simulation (based on the Can Bus protocol) with simulation-based testing automation.  Specifically,  \framework allows addressing two pressing challenges of AICAS such as  the need for (i) an \textit{Increased level of test automation} (e.g., AICAS inputs are manually generated or designed by testers and developers in its organization) with test cases automatically generated to increase the diversity of generated SDC scenarios; (ii) and the need of \textit{Increase level of realism}, since most of the manually entered signals inserted in the Can Bus protocol by the testers and developers of the AICAS organization do not reflect a real driving set of signals (e.g., the provided acceleration and steering angle of the vehicle are not reflecting a real driving test scenario, which makes the used inputs in most cases too random or unrealistic).
} 
 
 \major{
To enable the detection and fixing of SDC bugs during the evolution of SDCs, developers can focus on configuring  \framework to test different combinations of simulators, and AI agents in diversified testing cases, to identify faults in the AI engine and the connected hardware of the system. Of course, we expect that test cases for assessing and detecting SDC bugs can vary between different organizations. To perform such new experiments, \framework  can be used to generate new test cases by increasing the level of realism of the generated simulation by including  obstacles in the generated tests. This is to observe the behavior of the SDCs as well as the ability of \framework to identify safe and unsafe test cases in the context of more articulated test cases.
}

 \major{
From the discussion of the results of RQ$_3$, we identified that there is an upper bound of the extent to which static SDC features (i.e., features available before executing the tests) can be used to predict SDC testing outcomes. This represents a relevant topic for both \textbf{developers and researchers} for future investigation. From one side, we may argue that novel static SDC features need to be designed to achieve better results (in terms of precision, recall, and F-score). On the other side, we also observed in RQ$_3$ how the usage of different SDC features and hyperparameter optimization strategies do not lead to drastically better results. Given the complexity of the simulation environment and its simulated physics, we argue that to cope with the upper bound of static SDC features, better results can be achieved by combining static metrics and runtime SDC metrics (i.e., metrics available during the execution of SDC test). The rationale of such implication is that there is limited information that can be used to derive if SDC test cases will fail or not before their execution, and achieving better results requires designing metrics that are available during the execution of test cases. For instance, one could consider using the average distance, speed, and steering angle in the proximity of an SDC failure (namely, a crash or a violation of the safety criterion, such as the lane-keeping feature).
}
 
 \major{For what concerns \textbf{researchers}, this work triggers activities towards better testing and analysis of SDCs. First and foremost, given the identified safe and unsafe test cases, it can be used to derive higher-order \cite{JiaH09} SDC-specific mutation operators. 
 For example, the integration of obstacles and different fault detection strategies related to other safety criteria (different from the lane-keeping feature) during the execution of test cases could lead to mutants that change the test case outcome towards more faulty SDC behaviors. More complicated would be dealing with runtime adjustments of SDC test cases, which may require to be instantiated by perturbing the SDC behavior during the testing process. %Interface-related mutants could be produced as higher-order mutants from already existing mutants---\eg those from object-oriented language mutants  \cite{KimCM01}---by modifying the communication between CPS and devices, for example altering the ordering of method calls and/or passed parameters.  
} 
 
 \major{Also, the work could foster  the development of specific static analysis tools for SDC, looking for SDC-specific recurring problems observed in failing test cases. 
 Complementary empirical research could be directed to investigate the difficulty (\eg duration) of fixing SDC-specific bugs and developing tools guiding developers in allocating the appropriate development effort to various types of SDC bugs.
In the context of SDCs, the usage of \framework can help researchers (and developers) have a deep knowledge of SDC bugs and their root causes, which is potentially facilitated by their high reproducibility. 
Specifically, being able to reproduce a bug is crucial during bug triaging and debugging tasks but not always possible in field testing~\cite{BettenburgJSWPZ07,HuangZD13,ZimmermannPBJSW10,7332519}. 
 }

\major{
Fixing or addressing SDC-specific bugs and automatically assessing the correctness of the SDC behavior represent a critical challenge for  developers and researchers. Hence, future studies should look at further safety-related bugs due to the uncertainty of SDC behavior, concerning, for instance, the effect of different SDC initializations in the SDC test case outcomes.  %\seba{here @Christian, you could provide some examples of flaky tests, to motivate future research.}
During our experiments, we also noticed a non-deterministic behavior of the test outcomes, also known as flaky tests.
Concretely, depending on the definition of a failing test for \framework, we observed 1\% to 5\% flaky test cases, which we discarded when creating our dataset.
Future research should address the concern of having flaky tests in virtual environments since they lower the reliability of simulation-based tests of safety-critical systems such as SDCs.
}

\major{
Finally, SDC developers heavily rely on different experts (they need to have both software and hardware knowledge) to assess the correctness of SDC test outcomes. 
As the judgment of the experts highly depends on their experience and domain knowledge, such human oracles may not be reliable or can be considered subjective. 
This human-based assessment can be supported by reproducible SDC test regression frameworks, such as \framework, to mitigate the effect of subjective assessments of the correctness of SDC test outcomes.
}
}

\section{Related Work}     
\label{sec:related}
% What is our work about?
\major{
SDC-Scissor improves CPS testing cost-effectiveness by identifying and discarding likely irrelevant (i.e., safe) tests. Therefore, SDC-Scissor's main application areas are (automated) test generation and test regression selection.
Specifically, SDC-Scissor employs Machine Learning models to classify tests as safe or unsafe before their execution. Research has yielded many approaches to reduce testing efforts~\cite{DBLP:journals/infsof/ElberzhagerRME12,9251094}. 
These approaches can be classified into the following categories: \textit{test case selection}~\cite{Chen:1996}, \textit{test suite reduction, test case minimization}~\cite{Rothermel:icsm1998}, and \textit{test case prioritization}~\cite{Rothermel:icsm1999}.
Test case selection identifies subsets of available tests relevant (or necessary) for testing a given change in the code; test suite reduction removes redundant test cases from existing test suites, thus leading to smaller test suites that can execute faster; test case minimization removes irrelevant statements from the tests, reducing their size; finally, test case prioritization approaches rank test cases by the likelihood of detecting faults such that their execution can lead to finding faults soon.

Most of the available approaches focus on regression testing and do not employ Machine Learning~\cite{Yoo:stvr2010}. Only recently~\cite{DBLP:journals/ese/PanBGB22}, we observed a positive increment in the number of proposed approaches that rely on ML to select and prioritize test cases; however, those approaches focus mostly on traditional software systems (e.g.,~\cite{DBLP:conf/aitest/Roper19}), and the problem of reducing testing effort for Cyber-Physical Systems remains open~\cite{DBLP:journals/stvr/Sadri-Moshkenani22}.
In particular, compared to traditional software systems, CPS face additional challenges due to their continuous interactions with the environment and the tight coupling between the hardware and software components comprising them. Hence, standard testing approaches are ineffective, inefficient, or inapplicable~\cite{Briand2016}.

Testing of CPSs typically follows the X-in-the-loop paradigms~\cite{matinnejad2013automated} which involves a great deal of simulation
and takes the form of the model in the loop (MiL), software in the loop (SiL), and hardware in the loop (HiL), depending on the level of abstraction adopted to represent the CPS's software and hardware components and the relevant environmental elements.
Considering the specific requirements of X-in-the-loop testing, researchers proposed various optimization techniques tailored for CPSs. We discuss the most relevant examples in the following and point interested readers to Sadri-Moshkenan's survey for a more detailed discussion~\cite{DBLP:journals/stvr/Sadri-Moshkenani22}.

% Automated Test Generation
Effective CPS testing requires the generation of test cases that effectively stress the system under tests to systematically find critical and challenging test cases~\cite{Gambi2019}. However, many of the proposed approaches (e.g.,~\cite{SBST:2021,SBST2022,Gambi2019,li2020av}) rely on randomization to generate tests and require the execution of all the generated tests.
As we showed in our evaluation, without proper support (e.g., SDC-Scissor), those approaches struggle to efficiently identify relevant scenarios.
Abdessalem and co-authors, instead, augmented traditional evolutionary search algorithms commonly used for automated test generation with Machine Learning models to improve the cost-effectiveness of CPS testing. They evaluated their approaches on SDC collision avoidance.
Specifically, Abdessalem~\etal~\cite{DBLP:conf/kbse/AbdessalemNBS16} used Artificial Neural Networks to predict test cases' fitness without executing them.
By doing so, They could avoid the lengthy execution of test cases that might not contribute much towards achieving testing goals (i.e., finding problems in the system under test).
More recently, Abdessalem~\etal~\cite{Abdessalem2018} employed a Decision Tree to guide the test generation. In particular, during the test generation, Abdessalem~\etal train a Decision Tree that can identify regions of the test input space that likely lead to generating critical test cases.
Compared to Abdessalem~\etal's work, we adopt a similar approach but investigate the use of different Machine Learning models to classify tests as safe or unsafe. Additionally, we apply SDC-Scissor to a different problem, i.e., testing the SDC Lane Keeping system.

% Test selection
In traditional settings, test selection and prioritization are performed by computing test similarity or test adequacy (i.e., code coverage). However, given the complexity of test inputs for CPSs (e.g., simulated environments), computing those metrics is technically challenging. Consequently, new similarity metrics and procedures to compute them have been proposed. For instance, Arrieta \etal~\cite{arrieta2018multi, DBLP:conf/splc/ArrietaWSE16} proposed to measure the similarity between the test cases based on the so-called \textit{signal values} of all the states for the simulation-based test cases. 
Traditional test adequacy metrics may not be adequate for CPSs that are based on Artificial Intelligence and Deep Learning. Because of this, current research efforts focus on identifying domain-specific heuristics to select test cases. For instance, Arrieta \etal \cite{DBLP:journals/tii/ArrietaWMSE18} and Shin \etal \cite{Shin2018} proposed to select the test cases based on high-level objectives such as requirement coverage, the risks of damaging CPS Hardware components, and test execution times.

Compared to those studies, we investigate a different CPS domain and different test selection objectives.

Regarding test selection objectives, we focus on improving the cost-effectiveness of simulation-based tests to assess safety requirements. In contrast, previous studies prioritized the execution of tests based on their fault-detection capability~\cite{DBLP:journals/jss/ArrietaWSE19}, or selected tests based on signals diversity~\cite{DBLP:journals/tii/ArrietaWMSE18, arrieta2018multi, DBLP:conf/splc/ArrietaWSE16}, that require test execution. Since, in the SDC domain, executing simulation-based tests is prohibitive, we face the challenge of selecting test cases before their execution. Consequently, our techniques consider only the initial state of the car and the road features (e.g., geometry, lane markings), as those features are available without executing the tests in the simulator.
}

\section{Threats to Validity}
\label{sec:threats}

Threats to \emph{internal validity} may concern, as for previous work \cite{DBLP:conf/issta/GambiMF19,3533818,Birchler2022Cost}, the cause-effect relationships between the technologies used to generate the scenarios and their elements and the corresponding results, which strictly depends on the realism of our scenarios. Indeed, we did not recreate all the elements that can be found on real roads (e.g., weather conditions, etc.). However, to increase our internal validity, we used both BeamNG.AI and Driver.AI as test subjects. 
They both leverage a good knowledge of the roads, which means that they do not suffer from the limitations of vision-based lane-keeping systems. 
For future work, we plan to leverage the new BeamNG features, which allow experimenting with test cases composed of traffic lights as well as other cars and static objects.
\minorrevision{
%\major{
Moreover, we plan to experiment with consecutive versions of BeamNG.AI and Driver.AI (when they are available), so that it is possible to investigate the potential fault-detection capability of both of them.
}
Currently, this is not possible since both \minor{BeamNG.AI} and Driver.AI do not have previous versions of their driving agents.
%} 
\minorrevision{
Furthermore, since testing involves an underlying assumption that there will be no malicious attack on the system, future work should be conducted on more cautious driving AIs. The goal should also be to detect unsafe scenarios with a lower risk factor. A reckless driving style can be considered malicious behavior, which is, to a certain extent, provoked by the configuration RF2.
}

\minorrevision{
The current implementation of the diversity feature does not take into account the actual length of the road.
Theoretically, it is possible that a short road can have a higher diversity than a longer one, which also contradicts an assumption that a long road is generally unsafer since there is more space to encounter an unsafe state of the vehicle.
}

\major{Given the performances of the ML techniques used in our experiments may depend on the setting of their hyper-parameters. We initially leveraged their default settings, knowing that the obtained results could represent a lower bound for the classification performances. 
Then, we experimented with Grid search as a hyperparameter optimization approach (RQ$_3$) to investigate potential optimal combinations of parameters for the selected ML models.
}

\major{Finally, threats to \emph{external validity} concern the generalization of our findings. Although the (i) number of experimented test cases in our study is relatively larger \cite{DBLP:conf/issta/GambiMF19}; and (ii) we experimented with different AI engines (i.e., BeamNG.AI and Driver.AI) and integrated \framework into the development context of the AICAS use case (demonstrating that the proposed tool can automate the testing process of such a large automotive company) compared to previous studies; we cannot claim that our results can be generalized to the universe of general open-source CPS simulation environments in other domains. Therefore, further replications are desirable, and so are further studies considering more data as well as other CPS domains.

As discussed in Section \ref{sec:discussion}, for all results in Section \ref{sec:study-real}, for both the Adaptive and Pre-trained Models, we did not include the cost required for training the ML models on the training data. This choice was made since  the cost of training the best ML model can be considered negligible compared to the cumulative cost of generating all tests and executing them. 
However, this could be a threat to the external validity of our results, since for other ML models or particular settings of the same ML model (e.g., different from its standard configuration), we could achieve rather higher training costs.
Another threat could be related to the evaluation metrics used in our study, which could provide biased performance measures such as precision, recall, and F-score. Hence, for future work, we plan to leverage additional metrics such as the MCC (Matthews Correlation Coefficient), being reported as a well-known measure for unbiased performance measurements. 
}

\major{To minimize potential external validity, in conducting our experimental evaluation, we followed the guidelines by Arcuri {\em et al.} \cite{ArcuriB14} that suggest comparing results with randomized test generation algorithms (our baseline approach in RQ$_2$) and repeated the experiments several times.} 
%\seba{@christian, the following line need to be commented if no additional baseline is added} 
\major{In addition, we considered an additional baseline approach that selects test cases by ordering the test to be executed considering their road length (in decreasing order). 
}
%\sajad{add some text about the need to do run the adaptive approach multiple times in the future}

\section{Conclusions and Future Work}
\label{sec:conclusions}
%\seba{Continue from here...}

Regression testing for SDCs is particularly challenging due to the cost of running many driving scenarios in simulation. To improve the cost-effectiveness of regression testing, we introduced a test case selection approach, called \framework, that relies on a set of SDC road features extracted from driving scenarios prior to running the tests in the context of the BeamNG SDC simulation environment. Then, \framework uses ML approaches to select the test cases having a higher likelihood of experiencing unsafe situations.

\major{  
We empirically investigated the performance of \framework and compared it with baseline approaches (RQ$_1$).
Our assessment of \framework shows that \framework successfully selects test cases independently from the AI engine used or different risk levels (i.e., different driving styles), with the 
%\seba{@Christian, please check the following..} 
Logistic model providing the most stable results.
Interestingly, our results also show that the knowledge is not transferable from one AI engine to another one, i.e., \framework performed worse when training ML models on data from a specific AI engine and testing on data from a different AI engine.
% Moreover, among the defined features to train the ML models, the one that contribute the most in the accuracy of \framework are the concerning the set of full road features.

%\seba{@Christian, please change the following..} 
Our findings also suggest that \framework can reduce the number of executed tests required to find at least 10 unsafe tests (RQ$_2$).
%Specifically, \framework outperformed the baseline across all test pools, with the Logistic model selecting 80\% unsafe tests whereas the baseline select only 42.6\% and 60\% unsafe tests respectively. 
\minorrevision{Specifically, \framework outperformed the baseline across all test pools. It selected unsafe cases using the Logistic model with an accuracy of 70\%, a precision of 65\%, and a recall of 80\%.}
In terms of running time, we observed that  \framework is able to select test scenarios  in a cost-effective manner compared to two random baseline approaches  (RQ$_2$).
We experimented with Grid search as a hyperparameter optimization approach (RQ$_3$) to investigate potential optimal combinations of parameters for the selected ML models (RQ$_3$). 
%\seba{@Christian, please complete the following..} 
Our results show that there is an upper bound of an average F-score of 60\% with the J48 and Naive Bayes classifiers.
Complementary, compared to previous studies, we integrated \framework into the development context of the AICAS use case, demonstrating that the proposed tool can automate the testing process of such a large automotive company.

As future work, we plan to replicate our study on further SDC datasets, AI engines, and SDC features. Moreover, we plan to perform new empirical studies on further CPS domains to investigate how \framework performs when safety criteria concern new types of safety-critical faults different from those investigated in this study. Finally, we want to investigate different meta-heuristics and multi-objective approaches \cite{canfora2013multi,canfora2015defect} to enable test case generation based on the designed feature sets.
}

\section*{Acknowledgment}
Sebastiano Panichella, Sajad Khatiri, and Christian Birchler gratefully acknowledge the Horizon 2020 (EU Commission) support for the project \textit{COSMOS} (DevOps for Complex Cyber-physical Systems), Project No. 957254-COSMOS).
This work was partially supported by the DFG project STUNT (DFG Grant Agreement n. FR 2955/4-1).
We finally thank Kim Hyeongkyun for his valuable help in the research conducted and detailed in Section \ref{sec:integration}.

\section*{Data Availability}
The datasets generated during and/or analysed during the current study are available in the Zenodo repository~\cite{RP2021}. To foster the replicability of our study, we built a large dataset of labeled test cases~\cite{RP2021} that can be used for replicating our results and promoting further research. 
Furthermore, \framework is publicly available on GitHub~\footnote{\url{https://github.com/ChristianBirchler/sdc-scissor}}, which can be used with the data to replicate our results.

\section*{Conflict of Interest}
The authors declared that they have no conflict of interest.

\appendices
\label{sec:appendix}
\major{
\section{Analysis of Relevant Features (\texorpdfstring{RQ\textsubscript{1}}{RQ1})}
\label{app:feature_analysis_results}
% In our study, we considered two sets of features, full road, and road segment features.
Although the ML models trained using the road features can effectively classify the test cases as safe or unsafe, it is crucial to know the contribution of each of these features. For instance, more profound knowledge of the features may help to define better-suited feature sets. Hence, we analyzed in detail the road features for the BeamNG dataset discussed in Table \ref{tab:rq1}.}
Table \ref{tab:feature-analysis} reports the results of using two popular feature evaluation methods: \textit{information gain} and \textit{correlation}.
We order the features based on their evaluation scores and set a threshold (0.01 for information gain and 0.1 for correlation) for each evaluation method to select only the features with the highest contribution. It can be seen from Table \ref{tab:feature-analysis}-A and Table \ref{tab:feature-analysis}-B that the ordering and the relative score of the features are similar in most of the top cases among the two methods. 
Specifically, the top eight features are precisely the same in both methods, with a slight change in the order between ranks $2$ to $4$. Additionally, we note that the remaining features above the thresholds differ in just one feature, i.e., \textit{"std angle"}, which ranked in correlation score lower than the information gain (rank $14$ vs. $10$). 

Overall, we observe that almost all road features contributed to distinguishing safe versus unsafe test cases.  
Also, among the statistical features that we reported in Table \ref{table:road_segment_stat_feat}, features concerning the pivot radius tend to be more critical and relevant for the distinction of the classes. 
The \textit{minimum} and \textit{average radius} of the pivots are among the most contributing features, while the statistics concerning the turn angles start appearing only from rank $10$. 
\begin{table}
\caption{Feature Selection Rankings according to A) Information Gain Analysis, B) Correlation Analysis}
\centering
\parbox{.48\linewidth}{
    \centering
    \begin{tabular}{clc}
      & A & \\
    \toprule
    Rank & Feature & Inf. Gain \\
    \midrule
    1 & min pivot off & 0.140 \\
    2 & mean pivot off & 0.087 \\
    3 & total angle & 0.085 \\
    4 & num l turns & 0.084 \\
    5 & num r turns & 0.077 \\
    6 & std pivot off & 0.067 \\
    7 & median pivot off & 0.050 \\
    8 & length & 0.039 \\
    \midrule
    9 & num straights & 0.013 \\
    10 & std angle & 0.011 \\
    11 & max angle & 0.011 \\
    12 & min angle & 0.010 \\
    \midrule
    13 & max pivot off & 0.003 \\
    14 & direct distance & 0.003 \\
    15 & median angle & 0.002 \\
    16 & mean angle & 0.000\\
    \bottomrule
    \end{tabular}
    }
\parbox{.48\linewidth}{
    \centering
    \begin{tabular}{clc}
    & B & \\
    \toprule
    Rank & Feature & Correlation \\
    \midrule
    1 & min pivot off & 0.342 \\
    2 & total angle & 0.332 \\
    3 & num l turns & 0.330 \\
    4 & mean pivot off & 0.326 \\
    5 & num r turns & 0.316 \\
    6 & std pivot off & 0.270 \\
    7 & median pivot off & 0.257 \\
    8 & length & 0.222 \\
    \midrule
    9 & num straights & 0.138 \\
    10 & max angle & 0.109 \\
    11 & min angle & 0.104 \\
    \midrule
    12 & max pivot off & 0.063 \\
    13 & direct distance & 0.053 \\
    14 & std angle & 0.048 \\
    15 & median angle & 0.025 \\
    16 & mean angle & 0.017\\
    \bottomrule
    \end{tabular}
    }
\label{tab:feature-analysis}
\end{table}

% \major{
% \begin{finding}
% The \emph{Road Characteristics} extracted by \framework contribute differently to identifying the safe and unsafe test cases.
% The \emph{Road Characteristics} concerning the pivot radius (min, mean, std, median), the sum of the turn angles, the number of left and right turns, and the total length of the road are among the most important features, which are all belonging to the set of road features.
% \end{finding}
% }

%PART ON RISK FACTORS (RQ1)
\major{
\section{Impact of risk factor (RF) on Classification Performance (\texorpdfstring{RQ\textsubscript{1}}{RQ1})}
\label{app:rf}
\minorrevision{In Table \ref{tab:risk_factor2}, we report 
%an aggregate value of accuracy (column \emph{Acc.}), but we 
the precision, recall, and F-score for unsafe and safe labels regarding the BeamNG.AI datasets (with different risk factors), to make it more clear how \framework ability to classify tests is accurate on both labels, with varying
%AF
RF.}
With different risk factors, we can observe that the ML models' accuracy improved for increasing RF levels.
\minorrevision{For instance, with 
%AF2
RF 2 \framework reached a precision of 99.7\% for unsafe predicated tests.}
} 
The dataset composition seems to be the key factor explaining this result since setting the risk factor to higher values resulted in significantly more unsafe cases. 
Conversely, a small number of safe cases improved accuracy and precision for unsafe cases, counterbalanced by a decrease in the precision of safe predictions.
Finally, we can observe a similarity between the ML models' F-scores for safe and unsafe classes for the BeamNG.AI RF 1.5 case. This result can be explained by looking at how evenly distributed the safe and unsafe classes are, which illustrates the importance of having unbiased datasets for training and testing the models. 

This result supports the observation that the more the SDC under test drives safely, the harder it becomes to predict unsafe test cases.

% \major{
% \begin{finding}
% The accuracy of \framework is influenced by their driving style and the diversity of datasets. For example, for more aggressive driving agents, the accuracy achieved by the ML models was higher than for cautious driving agents. Hence, predicting unsafe test cases is harder for cautious drivers than reckless ones. Consequently, improving the testing of SDCs is more challenging for less aggressive driving agents.
% \end{finding}
% }

\begin{table}[h!tp]
\caption{Performance of the ML models trained using road features. The results refer to the split of 80/20 between training and test data. The best results are shown in boldface.}
\label{tab:risk_factor2}
\centering
\begin{tabular}{llllllll}
\toprule
\textbf{Model} 
%& \textbf{Acc.} 
& \multicolumn{3}{c}{\textbf{Unsafe Test Cases}} & \multicolumn{3}{c}{\textbf{Safe Test Cases}} \\
\cmidrule{2-7}
 %&
 & \textbf{Prec.} & \textbf{Recall} & \textbf{F$_1$} & \textbf{Prec.} & \textbf{Recall} & \textbf{F$_1$} \\
\midrule
\textbf{BeamNG RF 1} 
%&
&  &  &  &  &  &  \\
J48 
%& 61.2\% 
& 37.6\% & 69.8\% & 48.9\% & 84.2\% & 58.0\% & 68.7\% \\
Naïve Bayes 
%& 55.7\% 
& 36.7\% & \textbf{92.1\%} & 52.5\% & \textbf{93.7\%} & 42.5\% & 58.5\% \\
Logistic 
%& \textbf{66.2\%} 
& \textbf{43.3\%} & 87.3\% & \textbf{57.9\%} & 92.7\% & \textbf{58.6\%} & \textbf{71.8\%} \\
Random Forest 
%& 63.7\% 
& 40.7\% & 79.4\% & 53.8\% & 88.6\% & 58.0\% & 70.1\% \\
\midrule
\textbf{BeamNG RF 1.5} 
%&
&  &  &  &  &  &  \\
J48 
%& 65.6\% 
& 69.2\% & \textbf{67.4\% }& 68.2\% & 61.5\% & 63.5\% & 62.5\% \\
Naïve Bayes 
%& 66.7\% 
& \textbf{79.3\%} & 53.2\% & 63.6\% & 59.3\% & \textbf{83.1\%} & 69.2\% \\
Logistic 
%& \textbf{70.9\%} 
& 78.1\% & 65.3\% & \textbf{71.1\%} & \textbf{64.8\%} & 77.8\% & \textbf{70.7\%} \\
Random Forest 
%& 68.5\% 
& 75.8\% & 62.7\% & 68.6\% & 62.5\% & 75.6\% & 68.4\% \\
\midrule
\textbf{BeamNG RF 2} 
%&  
&  &  &  &  &  &  \\
J48 
%& 90.8\% 
& 98.7\% & 91.5\% & 95.0\% & 28.2\% & 73.3\% & 40.7\% \\
Naïve Bayes 
%& \textbf{93.4\%} 
& 98.7\% & \textbf{94.3\%} & \textbf{96.4\%} & 36.7\% & 73.3\% & 48.9\% \\
Logistic 
%& 83.2\% 
& 99.6\% & 82.8\% & 90.4\% & 19.7\% & \textbf{93.3\%} & 32.6\% \\
Random Forest 
%& 92.8\% 
& \textbf{99.7\%} & 92.7\% & 96.1\% & \textbf{36.8\%} & \textbf{93.3\%} & \textbf{52.8\%} \\
\bottomrule
\end{tabular}
\end{table}

% PART on transfer knowledge from a driving agent to another 
\major{
\section{Transfer Knowledge of ML models when using different Driving Agents (\texorpdfstring{RQ\textsubscript{1}}{RQ1})}
\label{app:knowledge_transfer}

We also studied the ability of the ML models to transfer knowledge from a driving agent to another by training ML models with one AI's dataset and testing it with another AI's dataset.
Specifically, we used BeamNG RF 1.5 dataset to train the ML models and used the Driver.AI test set, generated from the same set of virtual roads, to evaluate them, and \emph{vice versa}.}

\major{We considered three RF values ranging from cautious (RF 1.0) to moderate (RF 1.5) to reckless (RF 2.0). Using different values for the risk factor enables us to study the effectiveness of \framework concerning various SDCs' driving styles. 
To study the generality of our techniques, instead, we consider a second test subject, Driver.AI. Specifically, we tested Driver.AI with the same test cases used for testing BeamNG.AI in the moderate configuration. This way, we can directly compare the results achieved by both test subjects.
%
%\seba{@Cristian, please check the following paragraph, and eventually update it in the revision letter}
%From the data in Table \ref{table:datasets}, we make the following observations. First, we noticed that AsFault takes the AI engines' inputs to generate the harder scenarios, this lead to test cases having different configurations of roads and, as consequence, different sets of road segments composing them. Hence, while Table \ref{table:datasets} report a diversified set of test cases, the resulting number of road segments is almost identical. 
%As second observation, the number of unsafe tests increased with increasingly large values of BeamNG.AI's aggression factor. Second, testing Driver.AI resulted in fewer unsafe cases than testing BeamNG.AI in the moderate configuration. The above observations suggest that the aggression factor strongly influences the safety of BeamNG.AI; hence, changing its value likely results in different driving styles. At the same time, Driver.AI drives more cautiously than BeamNG.AI in the moderate configuration. Therefore, different test subjects indeed drive differently on the same roads. 
}

As is possible to observe in Table \ref{tab:mixed_accuracy}
the knowledge from one driving agent is not transferable to another one. Table \ref{tab:mixed_accuracy} shows that the ML models trained on Driver.AI and evaluated on BeamNG performed significantly worse than the same models trained on BeamNG exclusively (from 67.9\% to 41\% on average). However, when training the ML models on the BeamNG.AI dataset and evaluating them using the Driver.AI datasets, the ML models performed only slightly worse (between 49.1\% and 47.8\% on average). Interestingly, when using both datasets together, the results show a compromised solution between the accuracy achieved when training on the different AI engines separately: BeamNG 67.9\%, Driver.AI 49.1\%, and Combined datasets 55.5\%. 

\begin{table*}
\centering
\caption{ML Models' accuracy on mixed datasets.}
% \resizebox{12cm}{!}{
\begin{tabular}{@{}lccc@{}}
\toprule
     \textbf{Model} & \textbf{Training Acc.} & \textbf{Test Acc.} &  \\ 
\midrule
 \textbf{\major{BeamNG (Training)/Driver.AI (Test)}}\\
    J48  & 87\% & 46\%    \\
    Naive Bayes & 67\%     &  \textbf{56\%}        \\
    Logistic & 72\% & 45\%          \\
    Random Forest   & 100\%     & 44\%          \\
  \midrule
     \textbf{\major{Driver.AI (Training)/BeamNG (Test)}}\\
    J48  & 84\% & 44\%    \\
    Naive Bayes & 66\% &  35\%         \\
    Logistic & 81\% & \textbf{45\%}        \\
    Random Forest   & 100\%     & 43\%       \\
  \midrule
     \textbf{Driver.AI \& BeamNG Combined}\\
        J48  & 71\% & 53\%     \\
        Naive Bayes & 61\% &  49\%         \\
        Logistic & 64\% & 60\%          \\
        Random Forest   & 87\%     & \textbf{56\%}       \\
  \bottomrule                          
\end{tabular}
% }
\label{tab:mixed_accuracy}
\end{table*}

% \major{
% \begin{finding}
% Our results show that the knowledge is not transferable from one driving agent to another, i.e., \framework performed worse when training ML models on data from a specific driving agent and testing them on data from a different one. 
% However, ML models trained on the BeamNG data performed only slightly worse when evaluated on the Driver.AI data.
% % Alessio: This finding box is literally the repetition of the previous paragraph...
% % Additionally, we observed that the results correspond to a compromise between the results achieved when models are trained with data exclusively from the corresponding AI engines. When combining AI engines' datasets, 
% \end{finding}
% }
%\vspace{-5mm}

%\balance
%\newpage
\balance
\bibliographystyle{spmpsci}
\bibliography{biblio.bib}

\end{document}